\documentclass[10pt,reqno]{article}
\usepackage{amsfonts}
\usepackage{amsmath}
\usepackage{amsbsy}
\usepackage{graphicx}
\usepackage{amssymb,latexsym}
\usepackage{graphicx}
\usepackage{caption}
\usepackage{subcaption}
\usepackage{float}
\numberwithin{equation}{section}

\begin{document}

\title{Motion of the charged test particles in Kerr-Newman-Taub-NUT spacetime and analytical solutions}

\begin{titlepage}
\author{Hakan Cebeci\footnote{E.mail:
hcebeci@anadolu.edu.tr} \\
{\small Department of Physics, Anadolu University, 26470 Eski\c{s}ehir, Turkey} \\ \\
N\"{u}lifer \"{O}zdemir\footnote{E.mail:
nozdemir@anadolu.edu.tr}\\{\small Department of Mathematics,
Anadolu University, 26470 Eski\c{s}ehir, Turkey} \\ \\
Se\c{c}il \c{S}entorun\footnote{E.mail:
secilo@anadolu.edu.tr} \\
{\small Department of Physics, Anadolu University, 26470 Eski\c{s}ehir, Turkey}}

\date{ }

\maketitle

\bigskip

\begin{abstract}
\noindent In this work, we study the motion of charged test particles in Kerr-Newman-Taub-NUT spacetime. We analyze the angular and the radial parts of the orbit equations and examine the possible orbit types. We also investigate the spherical orbits and their stabilities. Furthermore, we obtain the analytical solutions of the equations of motion and express them in terms of Jacobian and Weierstrass elliptic functions. Finally, we discuss the observables of the bound motion and calculate the perihelion shift and Lense-Thirring effect for the bound orbits.
\vspace{1cm}

\end{abstract}
\end{titlepage}

\section{Introduction}

The Kerr-Newman-Taub-NUT spacetime is known as a stationary axially symmetric solution of Einstein-Maxwell field equations. The spacetime represents a rotating electrically charged source equipped with a gravitomagnetic monopole moment which is also identified as the NUT charge \cite{newman,demianski}. The solution contains four physical parameters: The gravitational mass, which is also called gravitoelectric charge; the gravitomagnetic mass (also known as the NUT charge); the rotation parameter that is the angular speed per unit mass and electric charge associated with the Maxwell field. As is well known, the NUT charge produces an asymptotically non-flat spacetime in contrast to Kerr geometry that is asymptotically flat \cite{misner}. Although the Kerr-Newman-Taub-NUT spacetime has no curvature singularities, there exist conical singularities on the axis of symmetry as in its uncharged version (namely the Kerr-Taub-NUT spacetime) \cite{miller}. One can get rid off conical singularities by taking a periodicity condition over the time coordinate. But, this leads to the emergence of closed time-like curves in the spacetime as in its uncharged version. It means that, in contrast to Kerr and Kerr-Newman solutions interpreted as regular rotating black holes, the Kerr-Newman-Taub-NUT solution cannot be identified as a regular black hole solution due to its singularity structure. Although charged and uncharged spacetimes with NUT parameter have unpleasing physical properties, it is worth to investigate such spacetimes in general relativity due to their asymptotically non-flat spacetime structures. To explore their various physical phenomena the spacetimes with NUT charge have been vastly studied in the works \cite{bonnor,bini,jantzen,aliev1,esmer,liu,bell,gamal} where in \cite{bonnor}, an alternative physical interpretation of the NUT parameter is also illustrated.

One way to explore the properties of Kerr-Newman-Taub-NUT spacetime is to study the motion of (charged) test particles in this background. In fact, the geodesic equations have been investigated in the spacetimes with Taub-NUT charge in the works \cite{gibbons, horvathy, abdujabbarov1, abdujabbarov2, pradhan}. Especially in \cite{gibbons, horvathy} dynamical symmetries of the geodesic motion in the spacetime with NUT parameter (which describes four dimensional Bogomol'nyi-Prasad-Sommerfield (BPS) monopoles and five dimensional Kaluza-Klein monopoles respectively) have been illustrated. In this sense, our aim is to obtain the analytical solutions of the equations of motion for a charged test particle in the background of Kerr-Newman-Taub-NUT spacetime and examine the effect of the NUT parameter and the charge of the test particle. Moreover, by integrating equations of motion for bound orbits, one can also calculate the precession of the orbital motion and Lense-Thirring effect.

The geodesic motion of test particles were first examined analytically in Schwarzchild spacetime in \cite{hagihara}. Later on, the motion of test particles was extensively investigated in Kerr spacetimes where the circular geodesics has also been examined \cite{wilkins,bardeen,chandrasekhar,oneil}. Recently, in \cite{fujita}, analytic solutions of the bound timelike geodesics of test particles in Kerr spacetime have been presented in terms of elliptic integrals using Mino time. In \cite{hackmann1} and \cite{hackmann2}, the geodesic equations are analytically solved in the background of Schwarzschild-(anti) de Sitter spacetimes, where the solutions are expressed in terms of Kleinian sigma functions. In \cite{hackmann3}, the investigation of the analytic solutions has been extended to Kerr-(anti) de Sitter spacetimes where in this case the solutions are presented in terms of Weierstrass elliptic functions. In a similar fashion, geodesic equations are solved in the spacetime of Kerr black hole pierced by a cosmic string \cite{hackmann4}, where the perihelion shift and the Lense-Thirring effect have also been investigated for bound orbits. In \cite{grunau1}, the geodesic equations are analytically examined in the background of Einstein-Maxwell-dilaton-axion black hole, where the effect of dilaton charge is investigated. Similarly, the analytic solutions of geodesic equations are given in higher dimensional static spherically symmetric spacetimes \cite{hackmann5}. Likewise, in \cite{grunau2} and \cite{grunau3}, the equations of geodesic motion have been examined in singly and doubly spinning black ring spacetimes respectively. In addition, the orbital motion of electrically and magnetically charged test particles have been studied in the background of Reissner-Nordstr\"{o}m \cite{grunau4} and Kerr-Newman spacetimes \cite{hackmann6}, where the effect of the charge of the test particle has been observed.

In this work, using Hamilton-Jacobi method, we derive the equations of motion for a charged test particle in the background of Kerr-Newman-Taub-NUT spacetime. By making a transformation on the time variable, we decouple the radial and angular ($\theta-$)part of the equations of motion and express all the differential equations in terms of Mino time \cite{mino}. We analyze the angular and radial part of the orbit equations and examine the possible orbit types including the special spherical orbits. Furthermore, we obtain the energy and the orbital angular momentum of the test particle for a spherical orbit. We also examine the stability of spherical orbits with respect to NUT parameter. Next, we present the analytical solutions of the equations of the motion by expressing them in terms of Weierstrass $\wp$, $\sigma$, and $\zeta$ functions. Additionally, we calculate the angular frequencies for the bounded radial and angular motions and examine the perihelion precision and the Lense-Thirring effect.

Organization of the paper is as follows: In section 2, we introduce Kerr-Newman-Taub-NUT spacetime. In section 3, we derive the equations of motion of the test particles by expressing them in terms of Mino time as well. In section 4, we make a comprehensive analysis of the angular and radial motion where we also examine the spherical orbits as a subsection. In chapter 5, we give detailed analytical solutions of all the equations that describe the orbital motion. Next, using the analytical solutions, we illustrate the plots of some orbit types for fixed spacetime parameters, the charge, the energy and the orbital angular momentum of the test particle. We also calculate and examine perihellion precision and the Lense-Thirring effect for the bound orbits. We end up with some comments and conclusions.

\section{Kerr-Newman-Taub-NUT spacetime}

The Kerr-Newman-Taub-NUT spacetime is a stationary solution of the Einstein-Maxwell field equations that is asymptotically non-flat. The metric describes a rotating electrically charged source that also includes a NUT charge which is also known as the gravitomagnetic monopole moment. In Boyer-Lindquist coordinates, Kerr-Newman-Taub-NUT spacetime can be described by the metric with asymptotically non-flat
structure,
\begin{equation}
g = - \frac{\Delta}{\Sigma} (dt- \chi d \varphi )^{2} + \Sigma
\left(\frac{d r^{2} }{\Delta} + d \theta ^{2} \right) + \frac{\sin
^{2} \theta }{\Sigma } \left( a d t - ( r^{2} + \ell^{2} + a^{2} )
d \varphi \right)^{2}
\label{Kerr_1}
\end{equation}
where
\begin{eqnarray}
\Sigma &=& r^{2} + (\ell + a \cos \theta)^{2} , \nonumber \\
\Delta&=& r^{2} - 2 M r + a^{2} - \ell^{2}+Q^2 , \label{Kerr_2}\\
\chi &=& a \sin^{2} \theta - 2 \ell \cos \theta .\nonumber
\end{eqnarray}
Here, $M$ is a parameter related to the physical mass of the gravitational source, $a$ is associated with its angular momentum per unit mass and $\ell$ denotes gravitomagnetic monopole moment
of the source which is also identified as the NUT charge. $Q$ is the electric charge. The electromagnetic field of the source can be given by the potential 1-form
\begin{equation}
A=A_\mu dx^\mu=- \frac{Qr}{\Sigma}(dt- \chi d \varphi).\label{Kerr_6}
\end{equation}
Interestingly, although the spacetime cannot be identified as a black hole, it has metric singularities at the locations
\begin{equation}
r_{\pm}=M\pm\sqrt{M^2-a^2+\ell^2-Q^2} ,\label{Kerr_8}
\end{equation}
where $\Delta=0$. It can also be seen that, the spacetime allows a family of locally non-rotating observers which rotate with coordinate angular velocity given by
\begin{equation}
\Omega=-\frac{g_{t\phi}}{g_{\phi\phi}}=\frac{\Delta \chi - a \sin^2 \theta (r^{2} + \ell^{2} + a^{2})}{\Delta \chi^2 - \sin^2 \theta (r^{2} + \ell^{2} + a^{2})^2}.\label{Kerr_9}
\end{equation}
This can be identified as the frame dragging effect which arises due to the presence of the off diagonal component $g_{t \varphi}$ of the metric where at the outermost singularity $r_{+}$, the angular velocity can be calculated as
\begin{equation}
\Omega_{+}=-\left.\left(\frac{g_{t\phi}}{g_{\phi\phi}}\right)\right|_{r=r_{+}}=\frac{a}{r_{+}^{2} + \ell^{2} + a^{2}} \, \cdot \label{Kerr_10}
\end{equation}
It is also obvious that the Killing vectors $\xi_{(t)}$ and $\xi_{(\phi)}$ generate two constants of motion namely the energy and the angular momentum. It can be shown that the Killing vector $\xi=\xi_{(t)}+\Omega_{+}\xi_{(\phi)}$ becomes null at the metric singularity where $r=r_{+}$.

\section{The motion of charged test particles}

In this section, we examine the motion of charged test particle in Kerr-Newman-Taub-NUT spacetime. To this end, we introduce the Hamilton-Jacobi equation for a charged particle
\begin{equation}
2 \frac{\partial S}{\partial \tau}=g^{\mu \nu} \left( \frac{\partial S}{\partial x^{\mu}}-q A_{\mu}\right) \left( \frac{\partial S}{\partial x^{\nu}}-q A_{\nu}\right) \label{Motion_1}
\end{equation}
where $\tau$ is an affine parameter and $q$ is the charge of the particle. Since the spacetime (\ref{Kerr_1}) admits the timelike Killing vector $\xi_{(t)}$ and spacelike Killing vector $\xi_{(\phi)}$, the solution of the Hamilton-Jacobi equation can be written as
\begin{equation}
S=- \frac{1}{2}m^2 \tau -Et +L \varphi +f(r,\theta) \label{Motion_2}
\end{equation}
where $f(r,\theta)$ is a function of the variables $r$ and $\theta$, the constants of motion $m$, $E$ and $L$ denote the mass, the energy and the angular momentum of the particle respectively. Furthermore, the separability of the Hamilton-Jacobi equation \cite{carter,frolov,aliev2} in Kerr-Newman-Tab-NUT spacetime implies that the function $f(r,\theta)$ can be expressed as a sum of two different functions which only depend on $r$ and $\theta$ independently, i.e.
\begin{equation}
f(r, \theta)=S_{r}(r)+ S_{\theta}(\theta) \label{Motion_3}.
\end{equation}
The substitution of (\ref{Motion_3}) together with the metric components $g^{\mu \nu}$ and (\ref{Motion_2}) into Hamilton-Jacobi equation (\ref{Motion_1}) results in two differential equations
\begin{equation}
\left(\frac{d S_r}{dr} \right)^2=\frac{1}{\Delta} \left\{-K-m^2r^2+\frac{1}{\Delta}\left[(r^2+a^2+\ell^2)E-aL-qQr \right]^2 \right\} \label{Motion_4}
\end{equation}
and
\begin{equation}
\left(\frac{d S_{\theta}}{d \theta} \right)^2= K-m^2(\ell+a \cos \theta)^2 -\left( \frac{\chi E-L}{\sin \theta}\right)^2 \label{Motion_5},
\end{equation}
where $K$ can be identified as the Carter separability constant. Using the expression for the canonical momenta $P_{\mu}$ such that
\begin{equation}
P_{\mu}=\frac{\partial S}{\partial x^{\mu}}=m g_{\mu \nu} \frac{d x^{\nu}}{d \tau}+q A_{\mu} \label{Motion_6}
\end{equation}
and identifying
\begin{equation}
P_{t}=-E, \qquad P_{\varphi}=L, \label{Motion_7}
\end{equation}
we obtain the following equations of motion:
\begin{equation}
\frac{d r}{d \tau}=\mp \frac{1}{\Sigma}\sqrt{\left[(r^2+\ell^2+a^2) \bar{E}-a \bar{L}-\bar{q} Q r\right]^2-\Delta\left(\frac{K}{m^2}+r^2\right)} \label{Motion_8},
\end{equation}
\begin{equation}
\frac{d \theta}{d \tau}=\mp\frac{1}{\Sigma} \sqrt{\frac{K}{m^2}-(\ell+a\cos \theta)^2-\left(\frac{\chi \bar{E}-\bar{L}}{\sin \theta}\right)^2}\label{Motion_9},
\end{equation}
\begin{eqnarray}
\frac{d t}{d \tau}&=&\frac{1}{ \Sigma \Delta \sin^{2} \theta}\left[\bar{L} \left(\Delta \chi-a\sin^{2} \theta (r^2+\ell^2+a^2)\right) \right. \label{Motion_10}\\
&+&\left.\bar{E}\left(-\Delta \chi^2+\sin^{2} \theta (r^2+\ell^2+a^2)^2\right)- \bar{q} Q r \sin^{2} \theta (r^2+\ell^2+a^2)\right] \nonumber,
\end{eqnarray}
\begin{eqnarray}
\frac{d \varphi}{d \tau}&=&\frac{1}{ \Sigma \Delta \sin^{2} \theta} \left[\bar{E}\left(-\Delta \chi+a sin^{2} \theta (r^2+\ell^2+a^2)\right)\right. \nonumber \\
&+&\left.\bar{L} \left(\Delta-a^2 sin^{2}\theta \right)- \bar{q} Q r a sin^{2}\theta\right]\label{Motion_11}
\end{eqnarray}
where we define
\begin{equation}
\bar{E}:=\frac{E}{m}, \qquad \bar{L}:=\frac{L}{m}, \qquad \bar{q} :=\frac{q }{m} \cdot \label{Motion_12}
\end{equation}
To this end, introducing a new time parameter $\lambda$ (the so-called Mino time) as in \cite{mino} such that
\begin{equation}
\frac{d\lambda}{d\tau}=\frac{1}{\Sigma}, \label{Motion_13}
\end{equation}
we can express the equations of motion in terms of Mino time:
\begin{equation}
\frac{d r}{d \lambda}=\mp \sqrt{P_r(r)} \label{Motion_14},
\end{equation}
\begin{equation}
\frac{d \theta}{d \lambda}=\mp \sqrt{P_{\theta}(\theta)}\label{Motion_15},
\end{equation}
\begin{eqnarray}
\frac{d t}{d \lambda}& = & \mp \frac{\chi(\bar{L}-\bar{E} \chi)}{\sin^2 \theta \sqrt{P_{\theta}(\theta)}} \frac{d \theta}{d \lambda} \nonumber \\
&&\mp (\bar{E} (r^2 +a^2 + \ell^2)-a \bar{L}-\bar{q}Q r)\frac{(r^2 +a^2 + \ell^2)}{\Delta \sqrt{P_r(r)}} \frac{dr}{d \lambda},\label{Motion_16}
\end{eqnarray}
\begin{eqnarray}
\frac{d \varphi}{d \lambda}&=& \mp \frac{ (\bar{L}-\bar{E} \chi)}{\sin^2 \theta \sqrt{P_{\theta}(\theta)}} \frac{d \theta}{d \lambda} \nonumber \\
&&\mp \frac{a\left[\bar{E} (r^2 +a^2 + \ell^2) -\bar{L} a- \bar{q} Q r \right]}{\Delta \sqrt{P_r(r)}} \frac{dr}{d \lambda},\label{Motion_17}
\end{eqnarray}
where
\begin{equation}
P_r(r)=\left[(r^2+\ell^2+a^2) \bar{E}-a \bar{L}-\bar{q} Q r\right]^2-\Delta\left(\frac{K}{m^2}+r^2\right)\label{Motion_18}
\end{equation}
and
\begin{equation}
P_{\theta}(\theta)=\frac{K}{m^2}-(\ell+a\cos \theta)^2-\left(\frac{\chi \bar{E}-\bar{L}}{\sin \theta}\right)^2. \label{Motion_19}
\end{equation}

\section{Analysis of the motion:}

In this part, we make an analysis of the angular and the radial motion and examine the possible orbit types as well. We also investigate the stability of spherical orbits.

\subsection{Analysis of the angular motion ($\theta$-motion)}

Writing the angular equation as
\begin{equation}
\left( \frac{d \theta}{d \lambda}\right)^2= P_{\theta}(\theta),\label{angular_1}
\end{equation}
we see that $P_\theta(\theta) \geq 0$ for the possibility of the motion where $P_{\theta}(\theta)$ can also be interpreted as the potential associated with the angular motion. Obviously, it depends on the relation between spacetime parameters $a$, $M$, $\ell$ and the energy and the angular momentum of the test particle. Associated with these parameters, either the motion is not possible (i.e $P_\theta(\theta) < 0$ ) or it is confined to an angular interval where $\theta_2 \leq \theta \leq \theta_1$ (i.e the angular motion is bound). We should also comment that depending on the values of the parameters, $\theta=\frac{\pi}{2}$ can be included in that angular interval or not. It means that, the particle can cross or cannot cross the equatorial plane depending on the spacetime parameters. From the expression of $P_\theta(\theta)$, one can easily see that if the condition
\begin{equation}
\frac{K}{m^2}\geq \ell^2+(\bar{L}-a \bar{E})^2 \label{angular_2}
\end{equation}
is satisfied, the test particle can cross the equatorial plane ($\theta=\frac{\pi}{2}$). Otherwise, the particle cannot pass through the equatorial plane but its motion is restricted to $\theta_2<\theta<\theta_1$, $\theta=\frac{\pi}{2}$ being outside of this interval. We further remark that, the physical restriction on $P_\theta (\theta)$, ($P_\theta (\theta) \geq 0$), requires that when the spacetime parameters (NUT parameter, rotation parameter) are fixed, a constraint relation between the energy and the angular momentum of the test particle can be developed as is also illustrated in \cite{hackmann4}. Nevertheless, for our case, due to the presence of NUT parameter $\ell$, it is very cumbersome to get such a relation between the energy and the angular momentum for non-vanishing spacetime parameters. It is obvious from the expression of $P_\theta(\theta)$ that the particle cannot pass through the poles (i.e $\theta=0$ and $\theta=\pi$ are not reached.) since $P_\theta(\theta)$ becomes singular at the poles for arbitrary non-zero NUT parameter $\ell$. However, in the limiting cases where $\bar{L}=-2 \ell \bar{E}$ and $\bar{L}=2 \ell \bar{E}$, the particle can approach the North pole ($\theta=0$) and the South pole ($\theta=\pi$) respectively.

\noindent In the graphs 1(a), 1(b) and 1(c), the angular potential $P_\theta(\theta)$ is plotted with respect to angular variable $\theta$ where it can be seen that there may exist one or two angular bound orbits or no motion. The physical parameters are chosen such that we get a physically acceptable angular motion for the charged test particle. It is seen that in the plot 1(a), there exist two angular intervals for which $P_\theta(\theta)\geq 0$. In the plot 1(b) however, only one physical interval seems to exist for the given spacetime parameters. In both plots, it is also understood that the test particle can cross the equatorial plane (since when $\theta=\frac{\pi}{2}$, $P_\theta (\theta) >0$). On the other hand, there exists no angular  motion in plot 1(c) since the condition $P_\theta (\theta) >0$ is not satisfied for the given spacetime parameters.

\begin{figure}
    \centering
    \begin{subfigure}[b]{0.45\textwidth}
        \includegraphics[width=\textwidth]{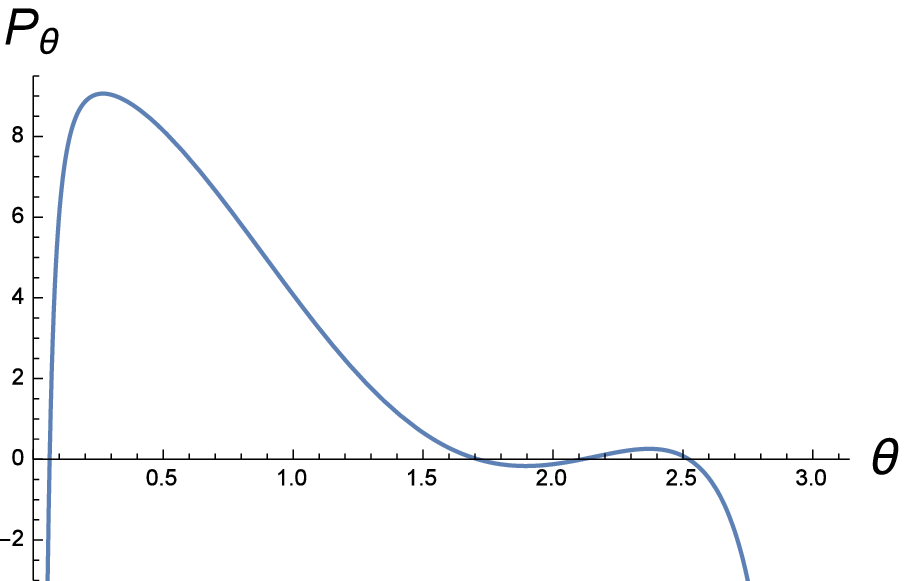}
        \caption{}
        \label{angular_potential_1}
    \end{subfigure}
    \begin{subfigure}[b]{0.45\textwidth}
        \includegraphics[width=\textwidth]{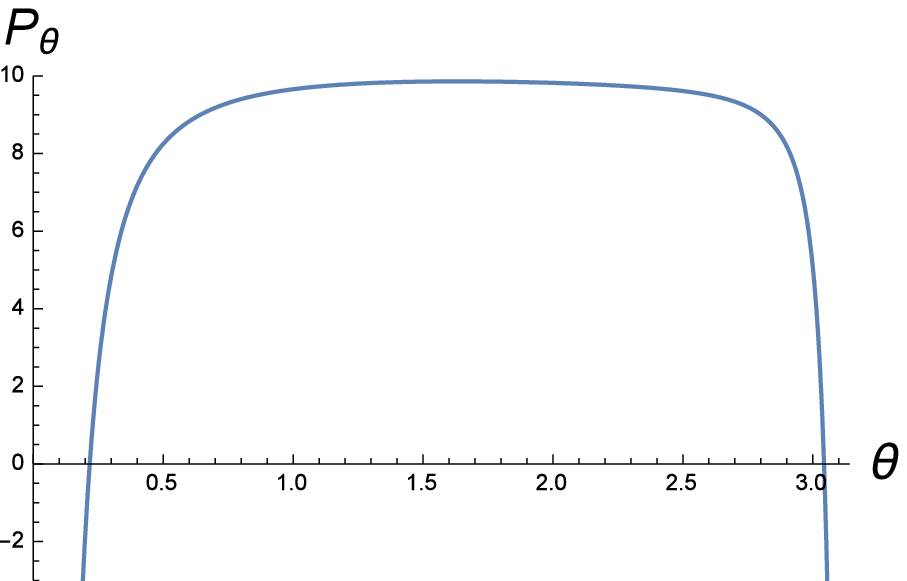}
        \caption{}
        \label{angular_potential_2}
    \end{subfigure}
    \begin{subfigure}[b]{0.45\textwidth}
        \includegraphics[width=\textwidth]{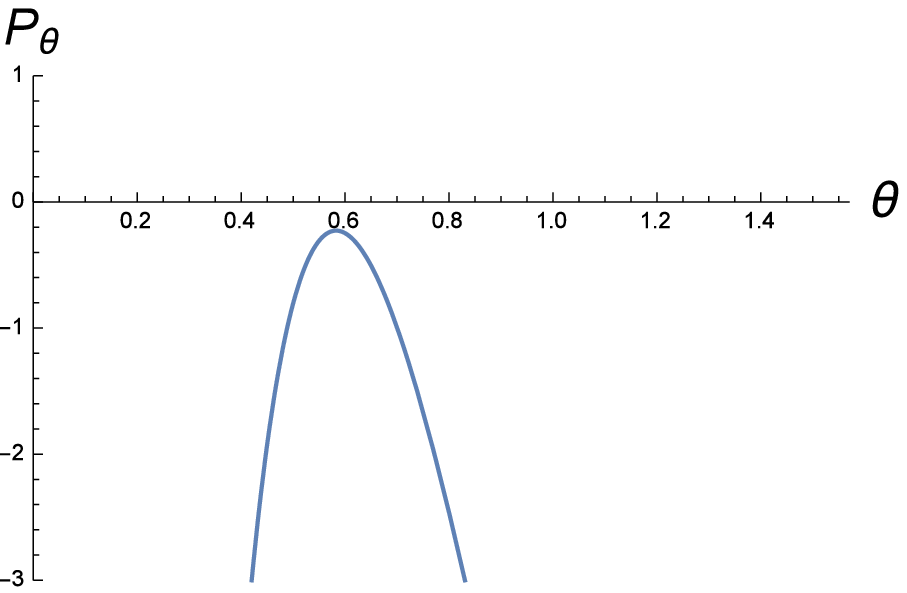}
        \caption{}
        \label{angular_potential_3}
    \end{subfigure}
    \caption{ The graphs of the angular potential $P_\theta(\theta)$ with different physical parameters. The physical parameters are chosen to obtain physically viable angular motion for the charged test particle. In figure (a), there exists two physically acceptable region for the angular motion with parameters $M=1$, $a=0.9$, $\ell=0.1$, $Q=0.4$, $\bar{q}=0.3$, $m=1$, $\bar{L}=-0.4$, $K=10$ and $\bar{E}=3$. In figure (b), there exists only one region with parameters $M=1$, $a=0.9$, $\ell=0.1$, $Q=0.4$, $\bar{q}=0.3$, $m=1$, $\bar{L}=0.5$, $K=10$ and $\bar{E}=0.96$, where the angular $\theta$-motion is possible. In figure (c), there is no physically viable angular motion for the parameters $M=1$, $a=0.9$, $\ell=0.1$, $Q=0.4$, $\bar{q}=0.3$, $m=1$, $\bar{L}=0.5$, $K=0.5$ and $\bar{E}=5$.}\label{angular_potential}
\end{figure}

\noindent In addition, it is also obvious that when
\begin{equation}
\left. P_{\theta}(\theta)\right|_{\theta=\theta_0}=0 \label{angular_3}
\end{equation}
and
\begin{equation}
\left. \frac{d P_{\theta}(\theta)}{d \theta}\right|_{\theta=\theta_0}=0 \label{angular_4}
\end{equation}
are simultaneously fulfilled, the motion is confined to $\theta=\theta_0$ plane (i.e the particle moves over a conical section). These conditions lead to the following equations:
\begin{equation}
\left.\frac{K}{m^2}\right|_{\theta=\theta_0}=(\ell+a \cos \theta_0)^2+ \left(\frac{\chi_0 \bar{E} - \bar{L}}{\sin \theta_0}\right)^2=:\frac{K_0}{m^2}, \label{angular_5}
\end{equation}
\begin{equation}
a(\ell+a \cos \theta_0)\sin \theta_0- \frac{\left(\chi_0 \bar{E} - \bar{L}\right) }{\sin \theta_0}\left[a \cos \theta_0 \bar{E} + \frac{2 \ell \bar{E} +\bar{L}\cos \theta_0 }{\sin^2 \theta_0}\right]=0, \label{angular_6}
\end{equation}
where
\begin{equation}
\chi_0 = a \sin^2 \theta_0 -2 \ell \cos \theta_0. \label{angular_7}
\end{equation}
By an analytic calculation, it can be seen that $\theta=\frac{\pi}{2}$ is not the solution of these equations for arbitrary values of the parameters $\ell$, $a$, $\bar{E}$ and $\bar{L}$. This means that for arbitrary values of gravitomagnetic monopole moment $\ell$ ($\ell\neq 0$), there exist no equatorial plane orbits (also called equatorial geodesics for the motion of uncharged test particle) as is also clearly stated in \cite{bini}. On the other hand, for $\ell=0$, $\theta=\frac{\pi}{2}$ solves the above equations provided that the Carter constant becomes
\begin{equation}
\frac{K}{m^2}= (a\bar{E}-\bar{L})^2. \label{angular_8}
\end{equation}
It can be seen that, in the vanishing of gravitomagnetic monopole moment, equatorial plane orbits can exist for arbitrary values of the parameters $a$, $\bar{E}$ and $\bar{L}$. Interestingly, the equations (\ref{angular_5}) and (\ref{angular_6}) also admit the solution $\theta_0=\frac{\pi}{2}$, if the constraint relation
\begin{equation}
\bar{L}=\frac{a(2 \bar{E}^2-1)}{2 \bar{E}} \label{angular_9}
\end{equation}
is imposed between the angular momentum and the energy of the test particle for arbitrary NUT parameter, provided that the Carter constant becomes $\frac{K_0}{m^2}=\ell^2 +\frac{a^2}{4\bar{E}^2}$ in that case. It means that, equatorial orbits can also exist for arbitrary NUT parameter if the constraint relation (\ref{angular_9}) holds between the angular momentum and the energy of the test particle as well as the spacetime rotation parameter. To our knowledge, this is a new result that has not been mentioned in previous works. Furthermore, it would be interesting to investigate the stability of the angular motion at the equatorial plane. It can be seen that, in addition to expressions given in (\ref{angular_3}) and (\ref{angular_4}), if the condition
\begin{equation}
\left.\frac{d^2 P_{\theta}(\theta)}{d \theta^2}\right|_{\theta=\theta_0} <0 \label{angular_10}
\end{equation}
is satisfied, the angular motion is stable at $\theta=\theta_0$. A straightforward calculation yields
\begin{equation}
\left.\frac{d^2 P_{\theta}(\theta)}{d \theta^2}\right|_{\theta=\frac{\pi}{2}}=-\left( 4 \ell^2 \bar{E}^2+\frac{a^2}{4 \bar{E}^2}\right)<0 \label{angular_11}
\end{equation}
where the constraint $\bar{L}=\frac{a (2 \bar{E}^2-1)}{2 \bar{E}}$ has been used. It concludes that the angular motion is stable at the equatorial plane.

\noindent Before closing this section, it would also be remarkable to examine the case with vanishing rotation parameter $a$ for $\ell \neq 0$. It can be algebraically seen that when $a=0$, the motion of the charged particle is restricted to a cone with opening angle determined by  either $
\cos \theta_0= -\frac{\bar{L}}{2 \ell \bar{E}}$ or  $\cos \theta_0= -\frac{2 \ell \bar{E}}{\bar{L}}$. It implies that in the vanishing of the rotation parameter, equatorial orbit may also exist either for the interesting case $\bar{L}=0$ (and $\ell\neq 0$) or for the case $\ell= 0$ (and $\bar{L}\neq0$) which is also discussed in \cite{bini} for the uncharged particle motion. As a further remark, we should also point out that, the presence of the charge $Q$ associated with the electromagnetic field and the charge $\bar{q}$ of the test particle does not affect the angular motion.

\subsection{Analysis of the radial motion ($r$-motion)}

First, we express the radial equation (\ref{Motion_8}) in the form
\begin{equation}
\left( \frac{ d r}{d \lambda} \right)^2 =P_r(r) \label{radial_1}
\end{equation}
where $P_r(r)$ is a fourth order polynomial in $r$ with real coefficients. As in the angular case, the possibility of the motion requires that $P_r(r)\geq0$. Then for $r$-motion, according to the roots of the polynomial $P_r(r)$, one can identify the following orbit types \cite{oneil}:\\

\noindent {\bf i.} Bound Orbit: If the particle moves in a region $r_2<r<r_1$, then the motion is bound. This can happen if $P_r(r)$ has four positive real roots or two positive real roots (with two complex roots) or two positive double roots or one triple positive root and one real positive root. In such a case, there may exist one or two bound regions.\\

\noindent {\bf ii.} Flyby Orbit: If the particle starts from $\mp \infty$ and comes to a point $r=r_1$ and goes back to infinity, then the orbit is flyby. Likewise, flyby orbits can be seen when $P_r(r)$ has four positive real roots or two positive real roots (with two complex roots) or two positive double roots or one triple positive root and one real positive root. Similarly, for this case, there may exist one or two flyby orbits.\\

\noindent {\bf iii.} Transit Orbit: If the particle starts from $\mp \infty$, crosses $r=0$ and goes to $\pm \infty$, then the orbit is transit. This is possible if $P_r(r)$ has no real roots.\\

\noindent {\bf iv.} Spherical Orbit: This is a special type of orbit, such that $P_r(r)$ has a real double root at $r=r_s$.\\

\noindent Depending on the value of the energy of the test particle, one can further examine the possible orbit types:\\

\noindent{\bf 1.} For $\bar{E}<1$: In that case, for physically acceptable motion $P_r(r)$ can have two or four real zeros since as $r\rightarrow \mp \infty$, $P_r(r)\rightarrow - \infty$. Then, there exist either one bound orbit or two bound orbits.\\

\noindent{\bf 2.} For $\bar{E}>1$: For this case, possible types of orbits can be classified according to whether $P_r(r)$ has four real zeros, two real zeros or no real zeros. When $P_r(r)$ has no real zeros (in other words all the roots are complex), then only the transit orbit is possible since in that case as $r\rightarrow \mp \infty$, $P_r(r)\rightarrow  \infty$. On the other hand, if $P_r(r)$ has two different real zeros (and two complex conjugate roots), one can get two flyby orbits. Moreover, if  $P_r(r)$ has four different real zeros, there may exist either two bound orbits or one bound, two flyby orbits or even two bound, two flyby orbits.\\

\noindent{\bf 3.} For $\bar{E}=1$: For this special case, $P_r(r)$ can possess either three real roots or one real root (with two complex conjugate roots). When $P_r(r)$ has three real roots, one can get bound and flyby orbits. In the case that $P_r(r)$ has only one real root, the possible orbit type is flyby.

\noindent In the graphs 2(a), 2(b) and 2(c), we give plots of the radial function $P_r(r)$ illustrating some possible orbit types for some particular values of the orbit parameters. As in the angular case, the parameters can be chosen to obtain a physically acceptable radial motion (i.e $P_r(r \geq0)$). For the first graph 2(a), $P_r(r)$ has no real roots so that the orbit is transit type. In graphs 2(b) and 2(c), there exist one and two bound regions respectively.

\begin{figure}
    \centering
    \begin{subfigure}[b]{0.42\textwidth}
        \includegraphics[width=\textwidth]{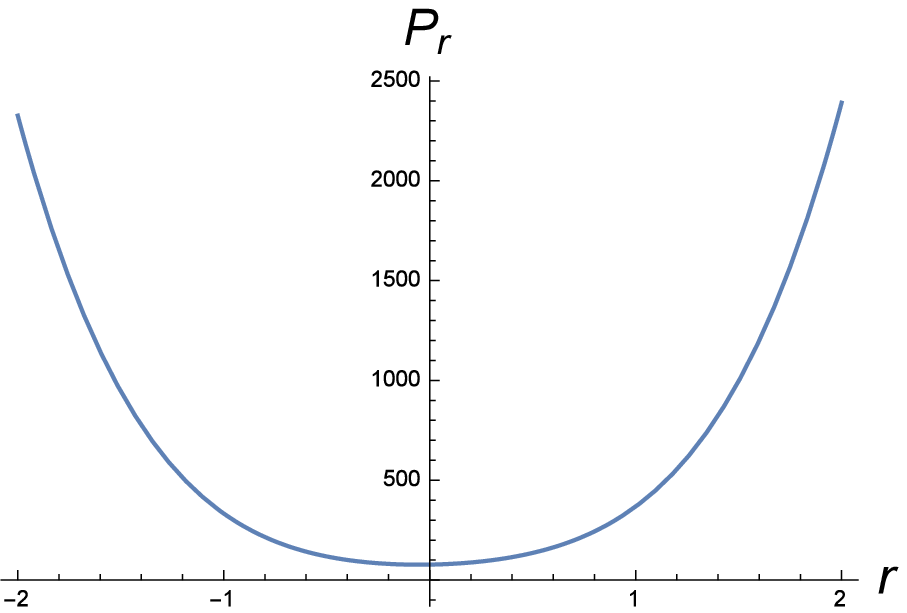}
        \caption{}
        \label{radial_potantial_1}
    \end{subfigure}
    \begin{subfigure}[b]{0.42\textwidth}
        \includegraphics[width=\textwidth]{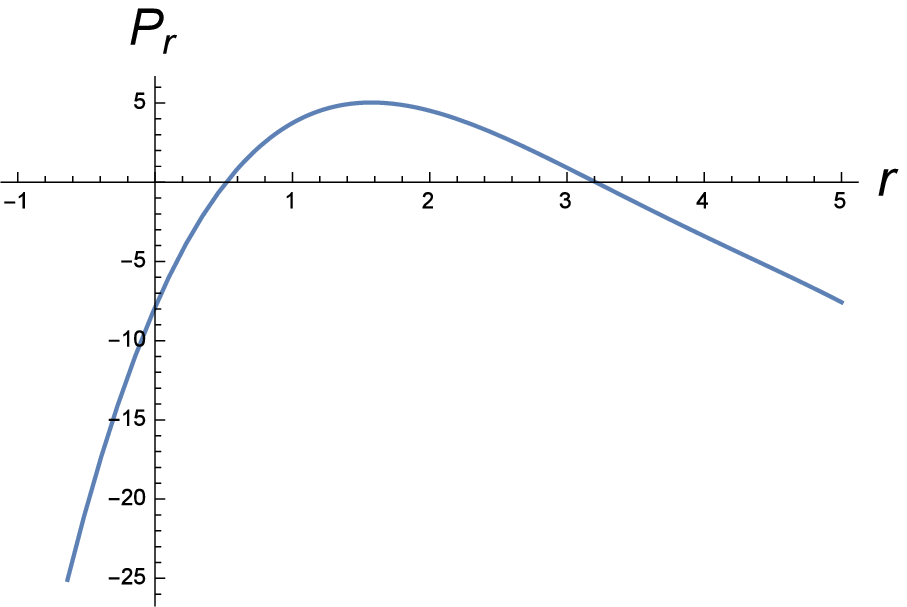}
        \caption{}
        \label{radial_potantial_2}
    \end{subfigure}
    \begin{subfigure}[b]{0.6\textwidth}
        \includegraphics[width=\textwidth]{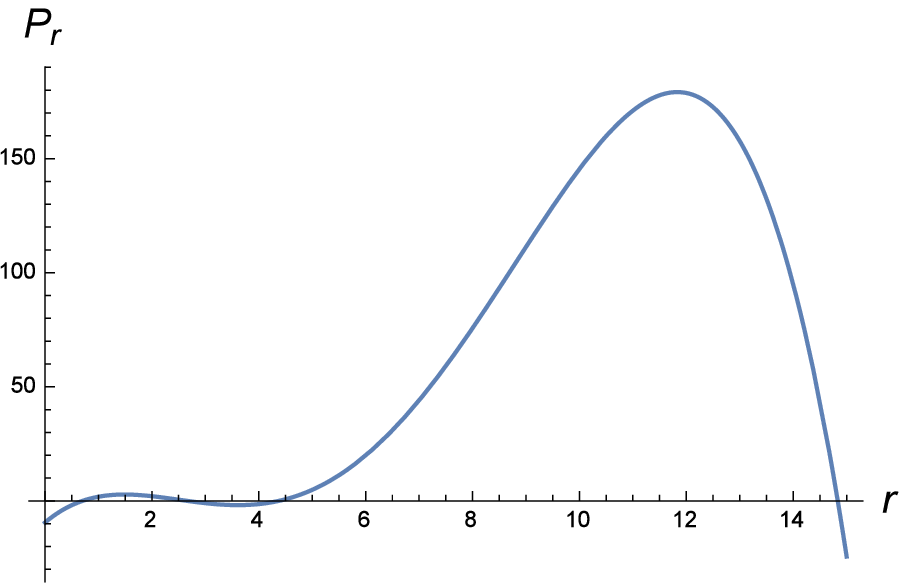}
        \caption{}
        \label{radial_potantial_3}
    \end{subfigure}
    \caption{Examples of some different orbits with respect to graph of $P_r(r)$. The parameters are chosen to illustrate the examples of transit, one bound and two bound orbits. In figure (a), there exists a transit orbit with parameters $M=1$, $a=0.9$, $K=10$, $Q=0.4$, $\bar{q}=0.3$, $\bar{L}=0.5$, $m=1$, $\ell=0.4$ and $\bar{E}=10$. In figure (b), there exists only one bound orbit with parameters $M=1$, $a=0.9$, $K=10$, $Q=0.4$, $\bar{q}=0.3$, $\bar{L}=0.5$, $m=1$, $\ell=0.4$ and $\bar{E}=0.94$. In figure (c), there are two bound orbits with parameters $M=1$, $a=0.9$, $K=10$, $Q=0.4$, $\bar{q}=0.3$, $\bar{L}=0.5$, $m=1$, $\ell=0.1$ and $\bar{E}=0.96$.}\label{radial_potantial}
\end{figure}

\noindent Alternatively, one can express the equation (\ref{radial_1}) as
\begin{equation}
\left( \frac{d r}{d \lambda} \right)^2= (r^2+\ell^2+a^2)^2 \left( \bar{E}- V_+(r) \right) \left(\bar{E}- V_-(r) \right) \label{radial_2}
\end{equation}
where
\begin{equation}
V_{\pm} (r,\bar{L}, a, \ell, \bar{q}, Q)=\frac{a \bar{L} +\bar{q} Q r \pm \sqrt{\Delta(r) \left( \frac{K}{m^2} + r^2 \right)}}{r^2 +\ell^2 +a^2} \label{radial_3}
\end{equation}
can be identified as the effective radial potentials. When $P_r(r)=0$, the $r$-motion has turning points determined by
\begin{equation}
\bar{E}=V_{\pm} \left.(r,\bar{L}, a, \ell, \bar{q}, Q)\right|_{r=r_0}=\frac{a \bar{L} +\bar{q} Q r_0 \pm \sqrt{\Delta(r_0) \left( \frac{K}{m^2} + r_0^2 \right)}}{r_0^2 +\ell^2 +a^2} \label{radial_4}
\end{equation}
where $r_0$'s denote turning points of the motion. In Figure 3, we give some examples of the graphs of effective potentials $V_{\pm}$ for two different values of the NUT parameter where in these graphs, the grey area shows the physically forbidden zone.

\begin{figure}
    \centering
    \begin{subfigure}[b]{0.48\textwidth}
        \includegraphics[width=\textwidth]{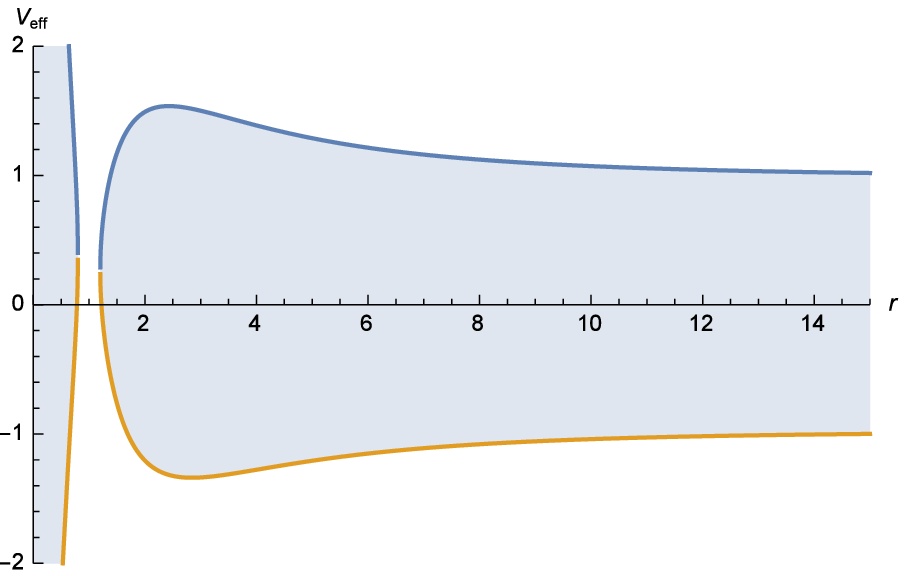}
        \caption{}
        \label{effective_potential_mp_1}
    \end{subfigure}
    \begin{subfigure}[b]{0.48\textwidth}
        \includegraphics[width=\textwidth]{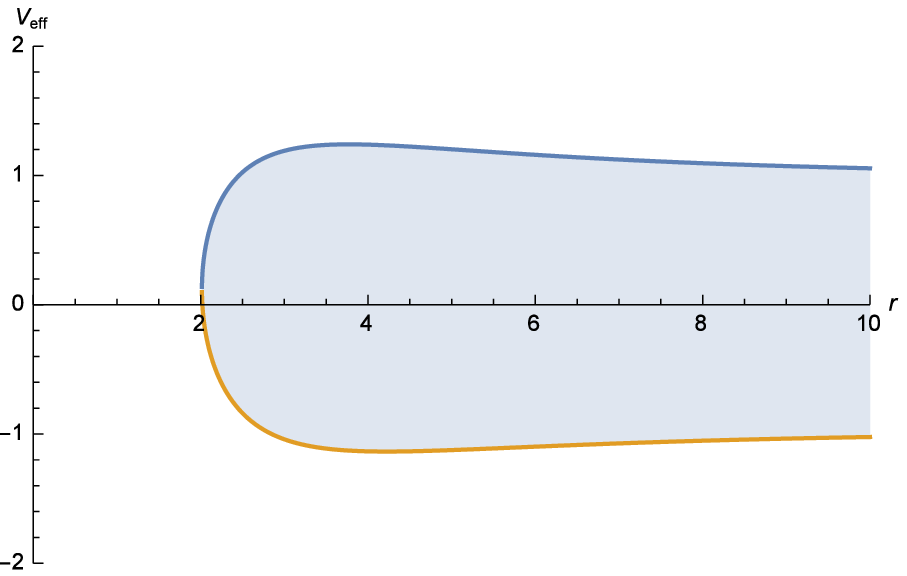}
        \caption{}
        \label{effective_potential_mp_2}
    \end{subfigure}
    \caption{Examples of effective potentials for different values of spacetime parameters and the angular momentum. In figure (a) we take $M=1$, $m=1$, $a=0.9$, $\ell=0.1$, $Q=0.4$, $q=0.3$, $\bar{L}=0.5$, $K=40$, while in figure (b) we take $M=1$, $m=1$, $a=0.9$, $\ell=1$, $Q=0.4$, $q=0.3$, $\bar{L}=0.5$, $K=40$. In both figures, the blue line represents $V_+$ while the yellow line corresponds to $V_-$ and the grey area shows the physically forbidden zone.
}\label{}
\end{figure}

\begin{figure}
    \centering
    \begin{subfigure}[b]{0.6\textwidth}
        \includegraphics[width=\textwidth]{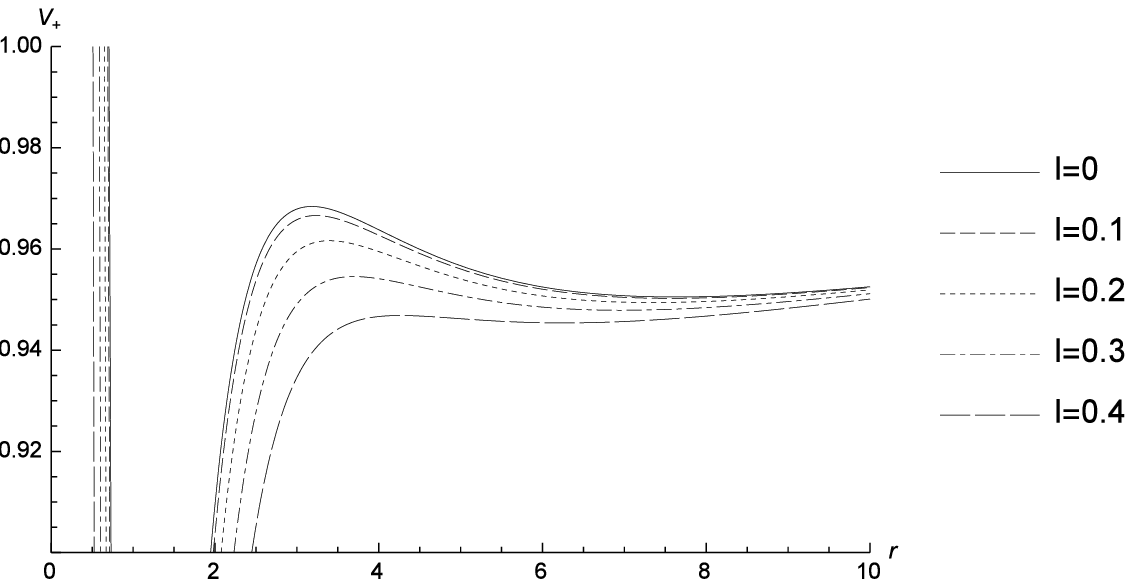}
        \caption{}
        \label{effective_potential_2}
    \end{subfigure}
    \begin{subfigure}[b]{0.6\textwidth}
        \includegraphics[width=\textwidth]{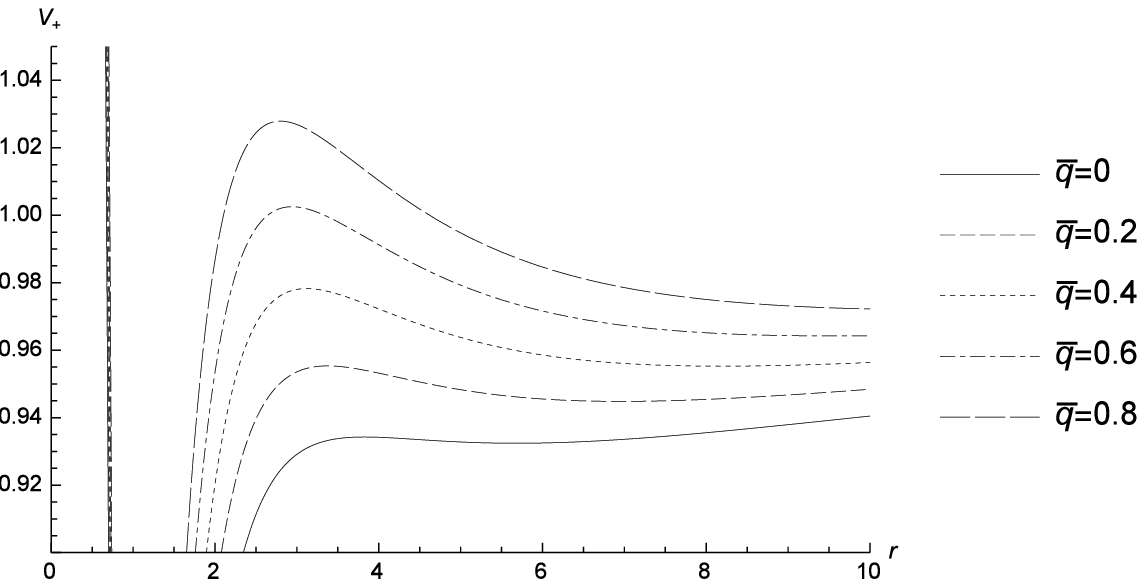}
        \caption{}
        \label{effective_potential_3}
    \end{subfigure}
        \begin{subfigure}[b]{0.6\textwidth}
        \includegraphics[width=\textwidth]{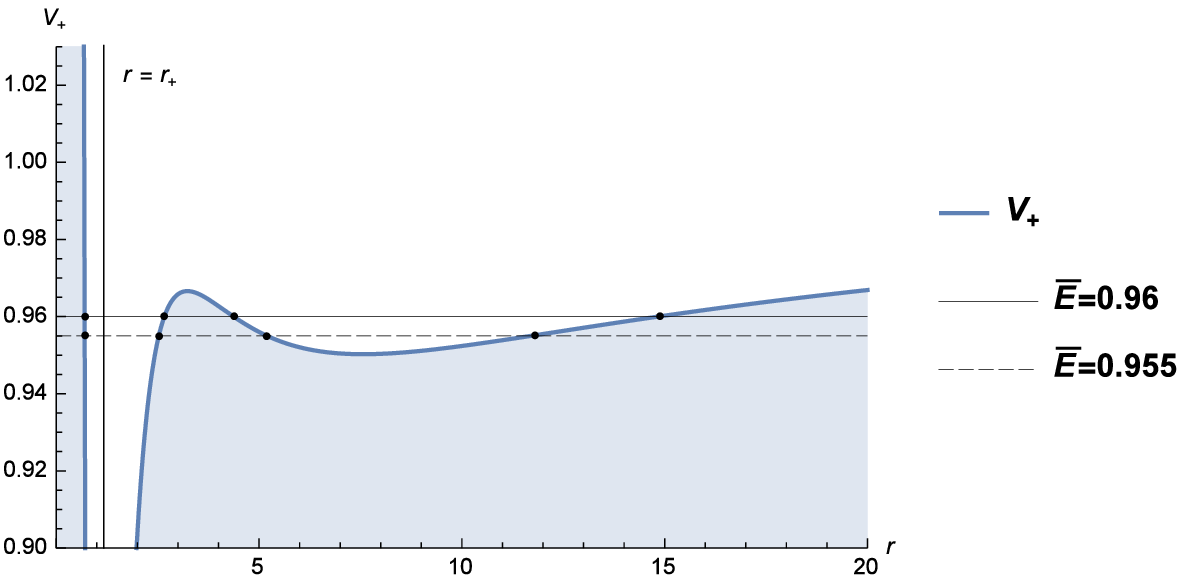}
        \caption{}
        \label{effective_potential_1}
    \end{subfigure}
    \caption{In (a), we illustrate the effective potential curves for different values of the NUT parameter by taking $M= 1$, $a= 0.9$, $Q = 0.4$, $\bar{q} = 0.3$, $\bar{L} = 0.5$, $K = 10$ and $m = 1$. In (b), the effective potential curves are plotted for different values of the charged test particle by taking $M= 1$, $a= 0.9$, $\ell= 0.1$, $Q = 0.4$, $\bar{L} = 0.5$, $K = 10$ and $m = 1$. In (c), the existence of bound orbits are revealed for $r>r_+$. For two different energy levels, there exists four turning points. The grey area shows the physically forbidden region. Here we take $M= 1$, $a= 0.9$, $\ell= 0.1$, $Q = 0.4$, $\bar{q} = 0.3$, $\bar{L} = 0.5$, $K = 10$ and $m = 1$.}\label{effective_potential}
\end{figure}

It would also be interesting to examine the variation of the effective radial potential by obtaining potential curves (for $V_+$) for different values of NUT parameter $\ell$ (the other parameters are fixed) and the charge $\bar{q}$ of the test particle. These are depicted in plots 4(a) and 4(b) respectively. As can be seen from these potential curves, the potentials have some local maxima and minima recalling that $V_+\rightarrow 1$ as $ r \rightarrow \infty $. From 4(a), for the fixed parameters $M= 1$, $a= 0.9$, $Q = 0.4$, $\bar{q} = 0.3$, $\bar{L} = 0.5$, $K = 10$ and $m = 1$, one can realize that as the value of NUT parameter $\ell$ increases, the peaks (local maxima) tend to disappear. On the other hand, one can see from the plot 4(b) that for the same parameter values (except that we take $\ell=0.1$ for this case) as the charge $\bar{q}$ of the test particle decreases in that case, the peaks have a similar behaviour. Physically, one can conclude that the effective potential can form potential well as the value of gravitomagnetic monopole moment decreases (while fixing the other parameters). The similar behaviour is seen as the value of charge of test particle increases in that case (again fixing the other parameters). This illustrates the existence of bound orbits in the region where potential well exists. In Figure 4(c), the existence of bound orbits are illustrated where the effective potential is crossed at four turning points as is also discussed in \cite{wilkins}.
\vspace{0.2cm}

\subsection{The spherical orbits and stability}

The spherical orbits are special orbits that satisfy
\begin{equation}
P_r(r=r_s)=0, \label{spherical_1}
\end{equation}
\begin{equation}
\left. \frac{dP_r}{dr}\right|_{r=r_s}=0 \label{spherical_2}
\end{equation}
at $r=r_s$. These require that
\begin{eqnarray}
&&(\bar{E_s}^2 -1)r_s^4+ 2 (M-\bar{q} Q \bar{E_s})r_s^3\nonumber\\
&&+\left( 2(\ell^2 +a^2)\bar{E_s}^2 +\bar{q}^2 Q^2-2a \bar{E_s} \bar{L_s}- \frac{K}{m^2}-(a^2-\ell^2+Q^2) \right)r_s^2 \nonumber\\
&&+ 2 \left( \frac{KM}{m^2}+a\bar{q}Q \bar{L_s} -\bar{q}Q (\ell^2+a^2) \bar{E_s} \right)r_s  \label{spherical_3}\\
&&+ (\ell^2+a^2)^2 \bar{E_s}^2 +a^2 \bar{L_s}^2- 2a(\ell^2+a^2)\bar{E_s}\bar{L_s}- \frac{K}{m^2}(a^2-\ell^2+Q^2)=0 \nonumber
\end{eqnarray}
and
\begin{eqnarray}
&&4(\bar{E_s}^2 -1)r_s^3+ 6(M-\bar{q} Q \bar{E_s})r_s^2\nonumber\\
&&+2\left( 2(\ell^2 +a^2)\bar{E_s}^2 +\bar{q}^2 Q^2-2a \bar{E_s} \bar{L_s}- \frac{K}{m^2}-(a^2-\ell^2+Q^2) \right)r_s \nonumber\\
&&+ 2 \left( \frac{KM}{m^2}+a\bar{q}Q \bar{L_s} -\bar{q}Q (\ell^2+a^2) \bar{E_s} \right)=0. \label{spherical_4}
\end{eqnarray}
Considering that $P_r(r)$ can also be written in the form
\begin{equation}
P_r(r)=\left( r-r_s \right)^2 \left( (\bar{E_s}^2-1)r^2 +\mu_1 r +\mu_2 \right), \label{spherical_5}
\end{equation}
the relations (\ref{spherical_3}) and (\ref{spherical_4}) can equivalently be expressed in the following forms:
\begin{eqnarray}
& & \left[3r_s^4+2r_s^2(\ell^2+a^2)-(\ell^2+a^2)^2 \right]\bar{E_s}^2- a^2 \bar{L_s}^2 +2 a \left( \ell^2+a^2-r_s^2 \right) \bar{E_s}\bar{L_s} \label{spherical_6}\\
& &-4r_s^3 \bar{q}Q \bar{E_s} + \left( a^2 - \ell^2+Q^2 \right) \left( \frac{K}{m^2}- r_s^2 \right)-r_s^2 \left(3r_s^2 -4r_s M-  \bar{q}^2 Q^2 + \frac{ K}{m^2} \right)=0 \nonumber
\end{eqnarray}
and
\begin{eqnarray}
& & 2r_s(\ell^2+a^2+r_s^2) \bar{E_s}^2+ a\left( \bar{q} Q -2 r_s \bar{E_s} \right) \bar{L_s}- \left(3 r_s^2 +\ell^2 +a^2 \right) \bar{q} Q \bar{E_s} \nonumber\\
&& +3r_s^2 M- \left(\frac{ K}{m^2}+a^2 -\ell^2 +Q^2 -\bar{q}^2 Q^2+2r_s^2  \right) r_s + \frac{ KM}{m^2} =0.\label{spherical_7}
\end{eqnarray}
The analytic solutions of the equations (\ref{spherical_6} ) and (\ref{spherical_7}) for the energy and angular momentum of the particle in spherical orbit yield
\begin{equation}
\bar{E_s}^{\pm}=\frac{\bar{q} Q}{2r_s}\pm \frac{ \sqrt{D_s}}{8 r_s^2 \Delta(r_s) \left( \frac{K}{m^2} +r_s^2 \right)} \label{spherical_12}
\end{equation}
and
\begin{eqnarray}
\bar{L_s}^{\pm} &=& \frac{\bar{q} Q}{2ar_s} (a^2+\ell^2 -r_s^2 ) \pm \frac{(a^2+\ell^2 +r_s^2 ) \sqrt{D_s}}{8\left( \frac{K}{m^2} +r_s^2 \right) \Delta(r_s) a r_s^2  } \label{spherical_14} \\
 && \mp \frac{1}{a} \sqrt{\Delta(r_s)\left( \frac{K}{m^2} +r_s^2 \right)} \nonumber
\end{eqnarray}
where
\begin{equation}
D_s= 16 r_s^2 \Delta(r_s) \left(\frac{K}{m^2}+r_s^2\right)\left[ (r_s-M)\left(\frac{K}{m^2}+r_s^2\right) + r_s\Delta(r_s) \right]^2.\label{spherical_13}
\end{equation}
From the expression obtained for the energy, one can identify the first term as electrostatic interaction between the charge of the test particle and the charge of the source of the spacetime.

\noindent To get a deeper insight of the analytical expressions of energy and angular momentum, we plot them as a function of gravitomagnetic monopole moment $\ell$ and the spherical radius $r_s$. In both plots, we concentrate on $\bar{L}_s^+$ and $\bar{E}_s^+$. First, looking at the graph of $\bar{L}_s^+$ vs $\ell$ for the parameter values given in Figure 5(a), it can be seen that as gravitomagnetic monopole moment increases, angular momentum also increases becoming zero at some specific value of the NUT parameter i.e at $\ell=\ell_s$. It is also interesting to  see that $\bar{L}_s\leq0$ when $0 \leq \ell \leq \ell_s$, while $\bar{L}_s>0$ when $\ell>\ell_s$. This can be physically interpreted as such that one obtains retrograde ($\bar{L}_s \leq 0$) spherical orbits for $0 \leq \ell \leq \ell_s$, while for $\ell>\ell_s$, the orbits are seen to be direct spherical orbits. In addition, we should point out that the analytical expression of $\bar{L}_s$ restrict the value of the NUT parameter $\ell$ since in the expression (\ref{spherical_14}), $D_s>0$ should also be imposed.

\begin{figure}
    \centering
    \begin{subfigure}[b]{0.4\textwidth}
        \includegraphics[width=\textwidth]{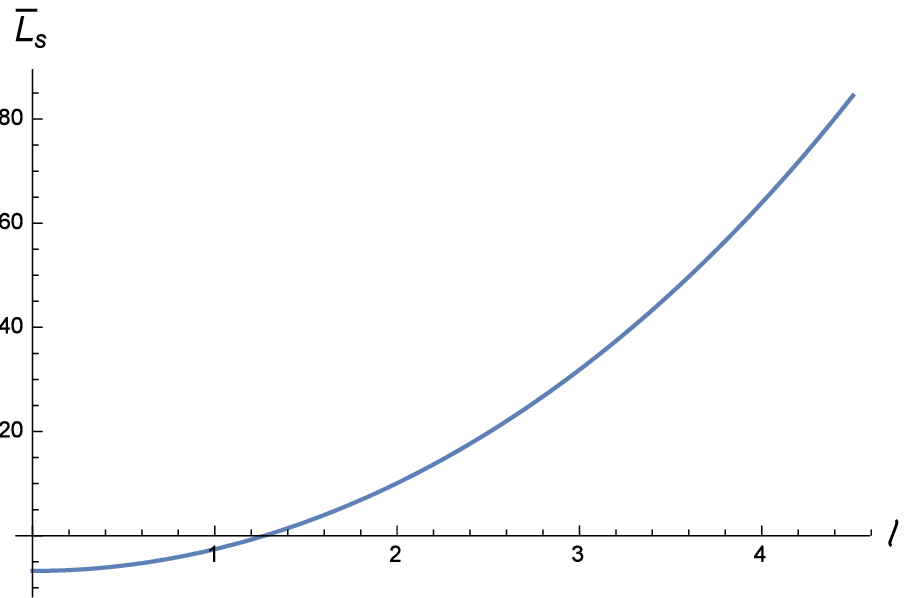}
        \caption{}
        \label{spherical_orbit_angular_momentum_1}
    \end{subfigure} \hspace{1cm}
    \begin{subfigure}[b]{0.4\textwidth}
        \includegraphics[width=\textwidth]{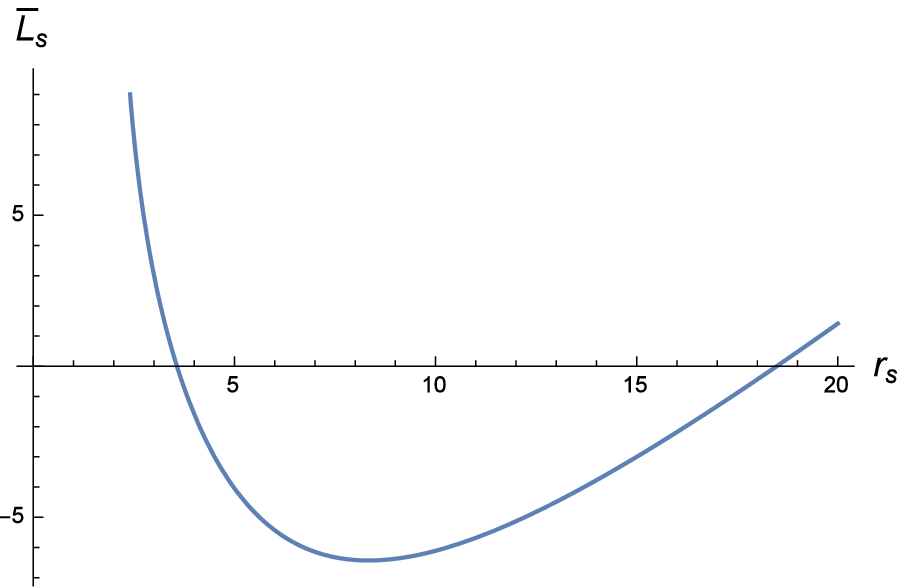}
        \caption{}
        \label{spherical_orbit_angular_momentum_2}
    \end{subfigure}
    \caption{In (a), the orbital angular momentum of the test particle in spherical orbit is plotted as a function of gravitomagnetic monopole moment $\ell$ with parameters $M= 1$, $a= 0.4$, $Q = 0.4$, $\bar{q} = 0.3$, $K = 20$, $m = 1$, $r_s=10$. In (b), the angular momentum is plotted as a function of spherical radius $r_s$ by taking the parameters $M= 1$, $a= 0.4$, $\ell= 0.4$, $Q = 0.4$, $\bar{q} = 0.3$, $K = 20$, $m = 1$.}\label{spherical_orbit_angular_momentum}
\end{figure}

\noindent For the second plot $\bar{L}_s^+$ vs $r_s$ (Figure 5(b)), it can be seen that $\bar{L}_s$ decreases to a certain extremal value and then it again increases. It can also be understood that $\bar{L}_s=0$ at some particular values of spherical radius $r_s$ i.e when $r_s=r_{s_1}$ and $r_s=r_{s_2}$ assuming that $r_{s_2}<r_{s_1}$. The graph similarly illustrates that, one has direct spherical orbits for the radial intervals where $r_s<r_{s_2}$ and $r_s>r_{s_1}$ while one obtains retrograde orbits for the interval $r_{s_2}<r_s<r_{s_1}$.

\noindent For the plot of $\bar{E}_s^+$ vs $\ell$ (Figure 6(a)), one can see that the energy of the test particle increases while $\ell$ also increases. It is also interesting to understand that, the increase of energy starts from a value where $\bar{E}<1$ and then it continues to increase to values where $\bar{E}>1$, the energy becoming unity at some specific value of NUT parameter $\ell$. It should be also added that value of $\ell$ should again be restricted since the analytic expression (\ref{spherical_12}) of $\bar{E}_s$ also suggests that $D_s\geq0$.

\begin{figure}
    \centering
    \begin{subfigure}[b]{0.4\textwidth}
        \includegraphics[width=\textwidth]{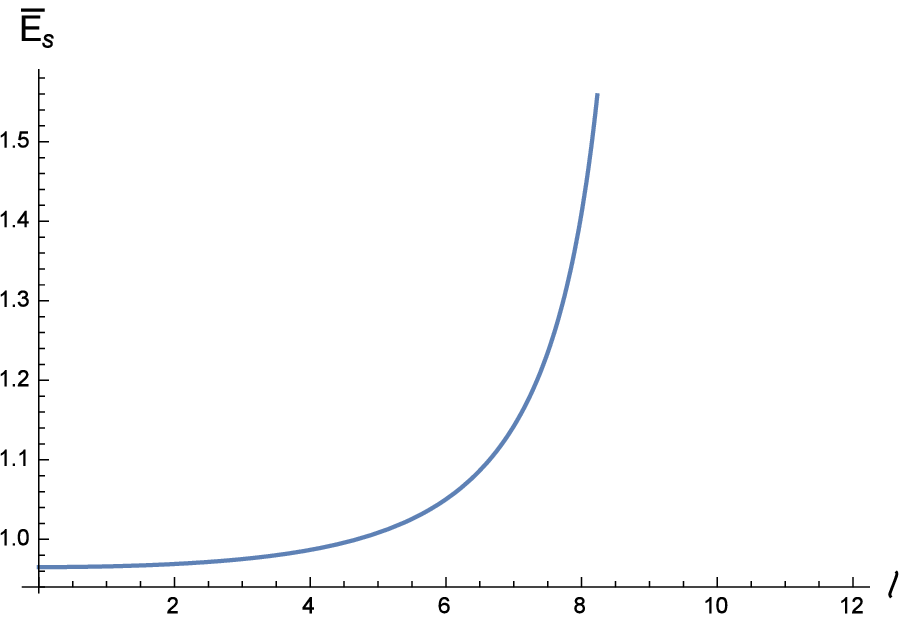}
        \caption{}
        \label{spherical_orbit_energy_1}
    \end{subfigure} \hspace{1cm}
    \begin{subfigure}[b]{0.4\textwidth}
        \includegraphics[width=\textwidth]{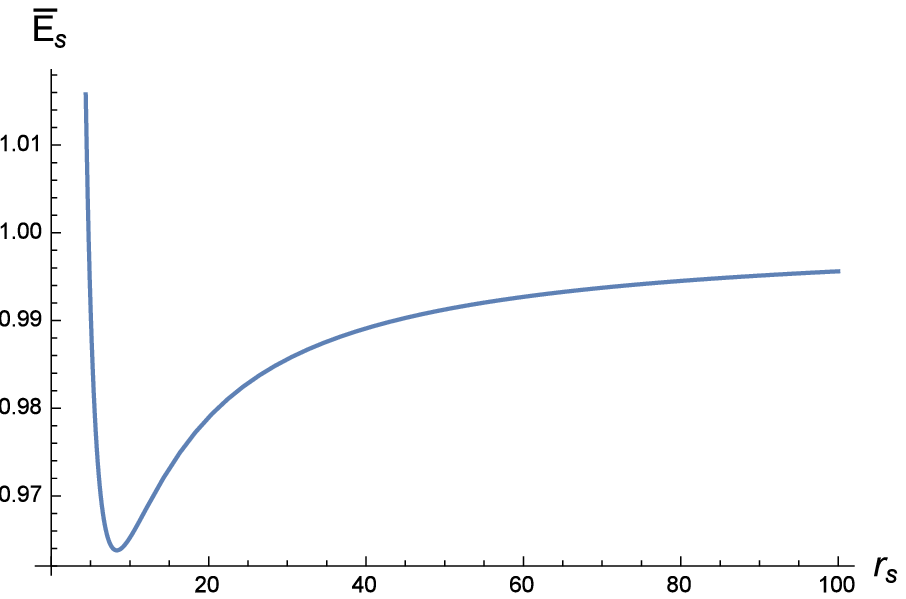}
        \caption{}
        \label{spherical_orbit_energy_2}
    \end{subfigure}
    \caption{In (a), the energy of the test particle in spherical orbit is plotted as a function of gravitomagnetic monopole moment $\ell$ with parameters $M= 1$, $a= 0.4$, $Q = 0.4$, $\bar{q} = 0.3$, $K = 20$, $m = 1$, $r_s=10$. In (b), the energy is plotted as a function of spherical radius $r_s$ by taking the parameters $M= 1$, $a= 0.4$, $\ell= 0.4$, $Q = 0.4$, $\bar{q} = 0.3$, $K = 20$, $m = 1$.}\label{spherical_orbit_energy}
\end{figure}

\noindent Finally the plot of $\bar{E}_s$ vs $r_s$ (Figure 6(b)) shows that, as the value of $r_s$ increases, the energy of test particle starts to decrease from a certain value having an extremum at some specific value of the spherical radius and then it continues to increase approaching unity (i.e $\bar{E}_s \rightarrow 1$) as $r_s\rightarrow \infty$.\\

Now we examine the stability of spherical orbits for arbitrary spacetime parameters. The stability of such orbits implies that \cite{wilkins,howes}
\begin{eqnarray}
\left.\frac{d^2P_r (r)}{dr^2}\right|_{r=r_s}<0. \label{spherical_27}
\end{eqnarray}
This further  leads to the inequality
\begin{eqnarray}
 2 \left(3 r_s^2 + \ell^2 + a^2 \right) \bar{E}^2 +2 \left(\bar{q} Q-a\bar{L} \right)\bar{E} \label{spherical_28} \\
-(5 r_s+4M)r_s -\Delta(r_s)- \frac{K}{m^2} + \bar{q}^2 Q^2<0. \nonumber
\end{eqnarray}
This means that the spherical orbits are stable  if the energy of the particle falls into the interval
\begin{equation}
\bar{E}_- <\bar{E}< \bar{E}_+, \label{spherical_29}
\end{equation}
where
\begin{equation}
\bar{E}_{\pm}= \frac{(a \bar{L}- \bar{q}Q)\pm \sqrt{D_E}}{2(3 r_s^2 +\ell^2 +a^2 )} \label{spherical_30}
\end{equation}
and
\begin{equation}
D_E= (a \bar{L} - \bar{q}Q)^2 +2 (3 r_s^2+\ell^2 +a^2 )\left(5 r_s^2 +4 M r_s + \Delta(r_s) +\frac{K}{m^2} - \bar{q}^2 Q^2\right). \label{spherical_31}
\end{equation}
Otherwise, we have an unstable spherical orbit. In Tables \ref{stability_table_1} and \ref{stability_table_2}, we investigate the stability with respect to change of the NUT parameter $\ell$ and the spherical radius $r_s$. For the values of the spacetime parameters given in tables, we have found the examples of stable orbits as the parameters $\ell$ and $r_s$ increases. However, looking at Table \ref{stability_table_1}, as the NUT parameter increases, stable spherical orbits change their class from retrograde orbits (where the rotational angular momentum $\bar{L}$ of the test particle about the axis of symmetry is in the opposite sense to that of the intrinsic angular momentum $a$ of the spacetime, i.e. $a>0, \bar{L}<0$) to direct ones (where in that case the rotational angular momentum of the test particle about the axis of symmetry is in the same sense to that of the intrinsic angular momentum of the spacetime, i.e. $a>0, \bar{L}>0$). One can see a similar effect in Table \ref{stability_table_2}, when the radius of the spherical orbit is increased. At this stage, we are unable to obtain examples of unstable spherical orbits although it is clear that there may exit unstable one(s) for the spacetime parameters that do not satisfy the inequality (\ref{spherical_28}).
\begin{table} [H]
\caption{$a=0.4M$, $Q=0.4M$, $r_s=10M$, $M=1$, $\bar{q}=0.3$,  $m=1$, $K=20$.}
\vspace{0.2cm}
\centering
\begin{tabular}{c c c c}
$\ell$ & $\bar{E}_s^+$ & $\bar{L}_s^+$  & stability of the orbit \\[0.5ex]
\hline\hline
  0.1 &  0.965109 & -6.73569 & stable \\
  0.2 & 0.965135  & -6.61086 & stable \\
  0.4 & 0.965241 & -6.1113 & stable \\
  0.8 & 0.965672 & -4.1093 & stable \\
  1.2 & 0.966415 & -0.758978 & stable \\
  2 & 0.969014  & 10.0825 & stable \\
  4 & 0.986713 & 63.9055 & stable \\[1ex]
\hline
\end{tabular}
\label{stability_table_1}
\end{table}
\begin{table} [H]
\caption{$a=0.4M$, $Q=0.4M$, $\ell=0.4 M$, $M=1$, $\bar{q}=0.3$,  $m=1$, $K=20$.}
\vspace{0.2cm}
\centering
\begin{tabular}{c c c c}
$r_s$ & $\bar{E}_s^+$ & $\bar{L}_s^+$  & stability of the orbit \\[0.5ex]
\hline\hline
  5 &  0.991365 & -4.04398 & stable \\
 10 & 0.965241  & -6.1113 & stable \\
  12 & 0.96844 & -5.13413 & stable \\
  13 & 0.970084 & -4.49053 & stable \\
 15 & 0.973125 & -3.0001& stable \\
20 & 0.978958  & 1.41363 & stable \\[1ex]
\hline
\end{tabular}
\label{stability_table_2}
\end{table}

\section{Analytical solutions}

In this section we present analytical solutions of the equations of motion (\ref{Motion_14})-(\ref{Motion_17}). We see that the solutions can be expressed in terms of Jacobian elliptic functions $F(y,k)$ and Weierstrass $\wp$, $\sigma$ and $\zeta$ functions.
\subsection{$\theta$-motion}
Making a transformation $x=\cos \theta$, (\ref{Motion_15}) can be cast into the following form:
\begin{equation}
\left(\frac{d x}{d \lambda}\right)^{2}=C_{0}+C_{1} x+C_{2} x^{2}+C_{3} x^{3}+C_{4} x^{4}=:P_{\theta}(x) \label{ang_sol_1}
\end{equation}
where
\begin{equation}
C_{0}=\frac{K}{m^2} -\ell^2- \left(\bar{L} -a \bar{E} \right)^2  , \label{ang_sol_2}
\end{equation}
\begin{equation}
C_{1}=2 \ell a (2 \bar{E}^2-1)-4 \ell \bar{E} \bar{L} , \label{ang_sol_3}
\end{equation}
\begin{equation}
C_{2} = - C_{0} - 4\ell^2 \bar{E}^2 +a^2 \left(\bar{E}^2-1  \right)- \bar{L}^2 , \label{ang_sol_4}
\end{equation}
\begin{equation}
C_{3}=2 \ell a (1-2 \bar{E}^2) \label{ang_sol_5}
\end{equation}
and
\begin{equation}
C_{4}=a^2 (1-\bar{E}^2) . \label{ang_sol_6}
\end{equation}
We also recall that $-1 \leq x \leq 1$. In general, the solution of the fourth order polynomial equation (\ref{ang_sol_1}) can be expressed in terms of elliptic functions.
We see that the transformation (for $\bar{E} \neq 1$)
\begin{equation}
x=\frac{\alpha_3}{\left(4y-\frac{\alpha_2}{3}\right)} +x_{\theta} \label{ang_sol_7}
\end{equation}
brings the polynomial equation (\ref{ang_sol_1}) into the standard Weierstrass form of the differential equation
\begin{equation}
\left( \frac{d y}{d\lambda}\right)^2= 4 y^3- g_2 y- g_3 \label{ang_sol_13}
\end{equation}
whose solution can be given by Weierstrass $\wp$ function
\begin{equation}
y(\lambda)=\wp\left(\lambda-\lambda_0; g_2,g_3\right), \label{ang_sol_16}
\end{equation}
where $\lambda_0$ describes the initial Mino time.
Here the constants read
\begin{equation}
\alpha_1=C_3 + 4 C_4 x_{\theta}, \label{ang_sol_9}
\end{equation}
\begin{equation}
\alpha_2=C_2 + 3 C_3 x_{\theta} + 6 C_4 x_{\theta}^2, \label{ang_sol_10}
\end{equation}
\begin{equation}
\alpha_3=C_1 + 2 C_2 x_{\theta}+3C_3 x_{\theta}^2 +4C_4 x_{\theta}^3,\label{ang_sol_11}
\end{equation}
\begin{equation}
g_2= \frac{1}{4} \left( \frac{\alpha_2^2}{3} -\alpha_1 \alpha_3 \right)\label{ang_sol_14}
\end{equation}
and
\begin{equation}
g_3= \frac{1}{8} \left( \frac{\alpha_1 \alpha_2 \alpha_3}{6} - \frac{C_4 \alpha_3^2 }{2} - \frac{\alpha_2^3 }{27} \right). \label{ang_sol_15}
\end{equation}
Hence the solution of $\theta-$equation can be obtained as
\begin{equation}
\theta(\lambda)=\arccos\left( \frac{\alpha_3}{4 \wp (\lambda-\lambda_0; g_2,g_3)- \frac{\alpha_2}{3}}+x_{\theta} \right). \label{ang_sol_17}
\end{equation}
For that solution, one can assume that there exist at least one real root (in fact there can exist at least two real roots for a fourth order polynomial).

\noindent Alternatively, there may exist orbits for which the polynomial $P_{\theta}(x)$ has four real roots. For the orbits $\bar{E}^2 <1$, ($C_4 >0$), one can assume $P_{\theta}(x)$ has real roots $x_1$, $x_2$, $x_3$, $x_4$ ordered as $x_4 < x_3 < x_2 < x_1$. In this case, there exist three intervals namely $x \leq x_4$, $x_3 \leq x \leq x_2$ and $x \geq x_1$ where $P_{\theta}(x) \geq 0$ \cite{wang}. For the intervals $x \leq x_4$ or $x \geq x_1$, one can consider the transformation \footnote{Equivalently, one can apply a similar transformation on the variables such that $ \frac{x-x_3}{x-x_4}=\frac{x_2-x_3}{x_2-x_4}y_{\theta}^2$  and obtain the same analytical result for the solution of the angular equation where in that case the transformation is valid when $x_3 \leq x \leq x_2$.}
\begin{equation}
\frac{x-x_1}{x-x_2} = \frac{x_1 -x_4}{x_2 -x_4}y_{\theta}^2. \label{ang_sol_18}
\end{equation}
Then with this transformation, the solution yields
\begin{equation}
F(y_{\theta},k_{\theta})= \sqrt{C_4 (x_1-x_3)(x_2-x_4)} \, \frac{(\lambda-\lambda_0)}{2} \label{ang_sol_21}
\end{equation}
with
\begin{equation}
k_{\theta}^2=\frac{(x_1-x_4)(x_2-x_3)}{(x_1-x_3)(x_2-x_4)}.\label{ang_sol_22}
\end{equation}
Here $F(y_{\theta},k_{\theta})$ describes incomplete Jacobian elliptic function of the first kind.

\noindent For the orbits where $\bar{E}^2 >1$, ($C_4<0$), as for the case $\bar{E}^2 <1$, we can still assume four real roots for the polynomial $P_\theta(x)$ ordered as $\bar{x}_4 < \bar{x}_3 <\bar{x}_2 <\bar{x}_1$ (obviously, the roots $\bar {x}_i$'s are different from the roots $x_i$'s, $i=1,2,3,4$). In this case, there exist two intervals $\bar{x}_2 \leq x\leq \bar{x}_1$ and  $\bar{x}_4 \leq x \leq \bar{x}_3$ for which $P_\theta(x)\geq 0$. Concentrating on the interval  $\bar{x}_2 \leq x \leq \bar{x}_1$, one may consider the transformation \footnote{Equivalently, one can apply a similar transformation on the variables such that $ \frac{x-\bar{x}_4}{x-\bar{x}_1}=\frac{\bar{x}_3-\bar{x}_4}{\bar{x}_3-\bar{x}_1}y_{\theta}^2$  and obtain the same analytical result for the solution of the angular equation where the transformation is valid when $\bar{x}_4 \leq x \leq \bar{x}_3$.}
\begin{equation}
\frac{x-\bar{x}_2}{x-\bar{x}_3} = \frac{\bar{x}_1 -\bar{x}_2}{\bar{x}_1 -\bar{x}_3}y_{\theta}^2. \label{ang_sol_23}
\end{equation}
leading to the solution
\begin{equation}
F(y_{\theta},\tilde{k}_{\theta})= \sqrt{-C_4(\bar{x}_1-\bar{x}_3)(\bar{x}_2-\bar{x}_4)}  \, \frac{(\lambda- \lambda_0)}{2} ,\label{ang_sol_24}
\end{equation}
with
\begin{equation}
\tilde{k}_{\theta}^2=\frac{(\bar{x}_1-\bar{x}_2)(\bar{x}_3-\bar{x}_4)}{(\bar{x}_1-\bar{x}_3)(\bar{x}_2-\bar{x}_4)}.\label{ang_sol_25}
\end{equation}

\noindent For $\bar{E}=1$, $P_\theta(x)$ becomes a third order polynomial such that the transformation
\begin{equation}
x=\frac{1}{\bar{C}_3} \left( 4y-\frac{\bar{C}_2}{3}\right) \label{ang_sol_26}
\end{equation}
obviously brings the equation (\ref{ang_sol_1}) into standard Weierstrass form
\begin{equation}
\left( \frac{d y}{d\lambda}\right)^2=4 y^3-\bar{g}_2 y-\bar{g}_3 \label{ang_sol_27}
\end{equation}
where in that case
\begin{equation}
\bar{g}_2=\frac{1}{4} \left( \frac{\bar{C}_2^2}{3}-\bar{C}_1 \bar{C}_3\right),\label{ang_sol_28}
\end{equation}
\begin{equation}
\bar{g}_3=\frac{1}{8} \left( \frac{\bar{C}_1 \bar{C}_2 \bar{C}_3}{6}-\frac{\bar{C}_0 \bar{C}_3^2}{2}-\frac{\bar{C}_2^3}{27}\right)\label{ang_sol_29}
\end{equation}
whose solution can similarly be expressed in terms of Weierstrass $\wp$ function as
\begin{equation}
y(\lambda)=\wp(\lambda-\lambda_0;\bar{g}_2,\bar{g}_3) \label{ang_sol_30}
\end{equation}
leading to $\theta$-solution
\begin{equation}
\theta(\lambda)=\arccos \left( \frac{1}{\bar{C}_3}(4\wp(\lambda-\lambda_0;\bar{g}_2,\bar{g}_3)-\frac{\bar{C}_2}{3})\right).\label{ang_sol_31}
\end{equation}
We further remark that, here
\begin{equation}
\bar{C}_0=\left. C_0 \right|_{\bar{E}=1}, \quad \bar{C}_1=\left. C_1 \right|_{\bar{E}=1}, \quad \bar{C}_2=\left. C_2 \right|_{\bar{E}=1}, \quad \bar{C}_3=\left. C_3 \right|_{\bar{E}=1}. \label{ang_sol_32}
\end{equation}

\noindent In addition, one can also consider the special case where $\bar{E}=1$ and $a=0$ (which corresponds to Taub-NUT Reissner-Nordstr\"{o}m
spacetime) for which the angular solution reads
\begin{equation}
\theta(\lambda)=\arccos \left[ \frac{2 \ell \bar{L}}{\frac{K}{m^2}+3 \ell^2}+\sqrt{\tilde{C}} \sin \left( \sqrt{\frac{K}{m^2}+3 \ell^2} \ (\lambda-\lambda_0)\right)\right] \label{ang_sol_33}
\end{equation}
where we identify
\begin{equation}
\tilde{C}=\frac{1}{\left( \frac{K}{m^2}+3 \ell^2\right)^2}  \left[ \left(\frac{K}{m^2}- \ell^2-\bar{L}^2 \right) \left(\frac{K}{m^2}+3 \ell^2\right)+4 \ell^2 \bar{L}^2 \right] \label{ang_sol_34}
\end{equation}
provided that $\tilde{C}>0$.

\subsection{$r$-motion}

On the other hand, the transformation (\ref{Motion_13}) on the time variable brings the radial equation (\ref{Motion_8}) into
\begin{equation}
\left(\frac{d r}{d \lambda}\right)^{2}=N_{0}+N_{1} r+N_{2} r^{2}+N_{3} r^{3}+N_{4} r^{4}=:P_r(r), \label{rad_sol_1}
\end{equation}
where
\begin{equation}
N_4=\bar{E}^2-1, \label{rad_sol_2}
\end{equation}
\begin{equation}
N_3=2(M-\bar{E} \bar{q} Q) , \label{rad_sol_3}
\end{equation}
\begin{equation}
N_2 = - C_{0} +a^2( \bar{E}^2-1)+ 2 \ell^2 \bar{E}^2+Q^2 (\bar{q}^2 -1)- \bar{L}^2 , \label{rad_sol_4}
\end{equation}
\begin{equation}
N_1=2\bar{q} Q \left( a\bar{L}-(\ell^2+a^2)\bar{E}\right)+\frac{2MK}{m^2}\label{rad_sol_5}
\end{equation}
and
\begin{equation}
N_0=\left( (\ell^2+a^2) \bar{E}-a\bar{L} \right)^2 - \left( a^2-\ell^2 +Q^2 \right)\frac{K}{m^2} . \label{rad_sol_6}
\end{equation}
If one performs the transformation (for $\bar{E}^2 \neq 1$)
\begin{equation}
r=\frac{\beta_3}{\left(4v-\frac{\beta_2}{3} \right)}+ r_1 \label{rad_sol_7}
\end{equation}
the equation (\ref{rad_sol_1}) can be brought into the standard Weierstrass form
\begin{equation}
\left(\frac{d v}{d \lambda} \right)^2=4v^3-h_2 v-h_3, \label{rad_sol_8}
\end{equation}
where
\begin{equation}
\beta_1=N_3 + 4 N_4 r_1,\label{rad_sol_9}
\end{equation}
\begin{equation}
\beta_2=N_2 + 3 N_3 r_1 + 6 N_4r_1^2, \label{rad_sol_10}
\end{equation}
\begin{equation}
\beta_3=N_1 + 2 N_2 r_1+3N_3 r_1^2 +4N_4 r_1^3, \label{rad_sol_11}
\end{equation}
with
\begin{equation}
h_2=\frac{1}{12} \left(\beta_2^2-3\beta_1 \beta_3 \right), \qquad h_3=\frac{1}{8}\left( \frac{\beta_1 \beta_2 \beta_3}{6}-\frac{N_4 \beta_3^2}{2} -\frac{\beta_2^3}{27}\right),\label{rad_sol_12}
\end{equation}
whose solution can again be given by Weierstrass $\wp$ function
\begin{equation}
v(\lambda)=\wp (\lambda-\lambda_0; h_2,h_3), \label{rad_sol_13}
\end{equation}
so that the solution for $r$ can be written as
\begin{equation}
r=\frac{\beta_3}{\left(4\wp (\lambda-\lambda_0; h_2,h_3)-\frac{\beta_2}{3} \right)}+ r_1, \label{rad_sol_14}
\end{equation}
where we have assumed that $P_r(r)$ has at least two real roots (Here $r_1$ is assumed to be one real root of $P_r(r)$).

\noindent As in the angular motion, if one considers that the radial polynomial $P_r(r)$ has four distinct real roots $r_1$, $r_2$, $r_3$, $r_4$ (for $\bar{E}^2>1$) ordered as $r_4<r_3<r_2<r_1$, one can alternatively express the solutions in terms of Jacobian elliptic functions. If one performs the similar transformations done in the angular case, one can end up with the following solutions:

\noindent For the orbits where $\bar{E}^2>1$, the solution reads
\begin{equation}
F(y_r,k_r)=\sqrt{N_4\left(r_1-r_3\right)(r_2-r_4)} \frac{\left(\lambda-\lambda_0\right)}{2}, \label{rad_sol_15}
\end{equation}
where we have affected the transformation
\begin{equation}
\frac{r-r_1}{r-r_2}=\frac{r_1-r_4}{r_2-r_4} y_r^2 \label{rad_sol_16}
\end{equation}
with
\begin{equation}
k_r^2= \frac{(r_1-r_4)(r_2-r_3)}{(r_1-r_3)(r_2-r_4)}, \label{rad_sol_17}
\end{equation}
while for the orbits where $\bar{E}^2<1$, the solution can similarly be expressed as
\begin{equation}
F(y_r,\bar{k}_r)=\sqrt{-N_4\left(\bar{r}_1-\bar{r}_3\right)(\bar{r}_2-\bar{r}_4)} \frac{\left(\lambda-\lambda_0\right)}{2}, \label{rad_sol_18}
\end{equation}
where in that case, the transformation
\begin{equation}
\frac{r-\bar{r}_2}{r-\bar{r}_3}=\frac{\bar{r}_1-\bar{r}_2}{\bar{r}_1-\bar{r}_3} y_r^2 \label{rad_sol_19}
\end{equation}
is valid with
\begin{equation}
\bar{k}_r^2= \frac{(\bar{r}_1-\bar{r}_2)(\bar{r}_3-\bar{r}_4)}{(\bar{r}_1-\bar{r}_3)(\bar{r}_2-\bar{r}_4)}. \label{rad_sol_20}
\end{equation}
Here $\bar{r}_1$, $\bar{r}_2$, $\bar{r}_3$ and $\bar{r}_4$ correspond to real roots of the polynomial $P_r(r)$ with $\bar{E}^2<1$ and it is assumed to be ordered as $\bar{r}_4<\bar{r}_3<\bar{r}_2<\bar{r}_1$.

\noindent Interestingly, for the special case where $P_r(r)$ has a double real root such that $r_1=r_2=r_s$, the radial polynomial can be put into the form (with $\bar{E}^2 \neq 1$) \cite{chandrasekhar}
\begin{equation}
P_r(r)=(r-r_s)^2 \left\{(\bar{E}^2-1)r^2 +2r r_s \left( \bar{E}^2-1 + \frac{M-\bar{q}Q\bar{E}}{r_s}\right) + \frac{N_0}{r_s^2}\right\}. \label{rad_sol_21}
\end{equation}
At this stage, we can affect the transformation
\begin{equation}
\rho=\frac{1}{r-r_s}, \label{rad_sol_22}
\end{equation}
to obtain (\ref{rad_sol_1}) in the following form:
\begin{equation}
\left( \frac{d \rho}{d \lambda} \right)^2= \alpha + \beta \rho +\gamma \rho ^2, \label{rad_sol_23}
\end{equation}
where
\begin{equation}
\alpha=N_4=\bar{E}^2-1, \label{rad_sol_24}
\end{equation}
\begin{equation}
\beta= 4r_s (\bar{E}^2-1) +2 (M-\bar{E}\bar{q} Q) \label{rad_sol_25}
\end{equation}
and
\begin{equation}
\gamma=3 r_s^2 (\bar{E}^2-1) +2r_s (M-\bar{E}\bar{q} Q) + \frac{N_0}{r_s^2}.\label{rad_sol_26}
\end{equation}
The solution of the equation (\ref{rad_sol_23}) can be obtained for three different cases, namely for the cases $\gamma >0$, $\gamma=0$ and $\gamma<0$. For $\gamma=0$, the solution can be expressed as
\begin{equation}
\lambda-\lambda_0=\mp \frac{2 \sqrt{\alpha+\beta \rho}}{\beta}, \label{rad_sol_27}
\end{equation}
for $\gamma >0$, the solution can be given by
\begin{equation}
\lambda-\lambda_0=\mp \frac{1}{\sqrt{\gamma}} \ln \left| \rho+\frac{\beta}{2 \gamma}+\sqrt{\rho^2+\frac{\beta}{\gamma}\rho+\frac{\alpha}{\gamma}}\right| , \label{rad_sol_28}
\end{equation}
while for $\gamma<0$ the solution reads
\begin{equation}
\lambda-\lambda_0=\mp \frac{1}{\sqrt{-\gamma}} \arcsin \left( \frac{ 2 \gamma \rho+\beta}{\sqrt{\beta^2-4 \gamma  \alpha}}\right). \label{rad_sol_29}
\end{equation}

\noindent Interestingly for $\bar{E} =1$, $P_r(r)$ turns into a third order polynomial such that the transformation
\begin{equation}
r=\frac{\bar{N}_3}{\left( 4v-\frac{\bar{N}_2}{3}\right)}\label{rad_sol_30}
\end{equation}
brings the radial equation into standard Weierstrass-$\wp$ form
\begin{equation}
\left( \frac{d v}{d\lambda}\right)^2=4 v^3-\bar{h}_2 v-\bar{h}_3 \label{rad_sol_31}
\end{equation}
where in that case we identify
\begin{equation}
\bar{h}_2=\frac{1}{4} \left( \frac{\bar{N}_2^2}{3}-\bar{N}_1 \bar{N}_3\right),\label{rad_sol_32}
\end{equation}
\begin{equation}
\bar{h}_3=\frac{1}{8} \left( \frac{\bar{N}_1 \bar{N}_2 \bar{N}_3}{6}-\frac{\bar{N}_0 \bar{N}_3^2}{2}-\frac{\bar{N}_2^3}{27}\right)\label{rad_sol_33}
\end{equation}
with
\begin{equation}
\bar{N}_0=\left. N_0 \right|_{\bar{E}=1}, \quad \bar{N}_1=\left. N_1 \right|_{\bar{E}=1}, \quad \bar{N}_2=\left. N_2 \right|_{\bar{E}=1}, \quad \bar{N}_3=\left. N_3 \right|_{\bar{E}=1}. \label{rad_sol_34}
\end{equation}
\noindent Then, for $\bar{E}=1$, the solution reads
\begin{equation}
r(\lambda)= \frac{\bar{N}_3}{4\wp(\lambda-\lambda_0;\bar{h}_2,\bar{h}_3)-\frac{\bar{N}_2}{3}}.\label{rad_sol_35}
\end{equation}

\subsection{$t$-motion}

To obtain the solution of the equation (\ref{Motion_16}), we recall that it can be written in differential form as (taking the $+$ sign only)
\begin{equation}
dt=dI_{\theta,1}^{(t)} +dI_{\theta,2}^{(t)}+ d I_r^{(t)} \label{t_motion_1}
\end{equation}
where the integration yields
\begin{equation}
t-t_0= I_{\theta,1}^{(t)}+I_{\theta,2}^{(t)}+I_r^{(t)}. \label{t_motion_2}
\end{equation}
Here
\begin{eqnarray}
I_{\theta,1}^{(t)}= a \bar{L}  \int_{\theta_{0}}^{\theta (\lambda)} \frac{d\theta}{ \sqrt{P_{\theta}(\theta)}} =a \bar{L} (\lambda-\lambda_{0}) \label{t_motion_3}
\end{eqnarray}
where we have integrated (\ref{Motion_15}), taking the $+$ sign only. The second expression $I_{\theta,2}^{(t)}$ can be written as
\begin{eqnarray}
I_{\theta,2}^{(t)}&=& -2 \ell \bar{L}\int_{\theta_{0}}^{\theta (\lambda)} \frac{\cos \theta}{\sin^2 \theta \sqrt{P_{\theta} (\theta)}}\, d \theta - a^2 \bar{E} \int_{\theta_{0}}^{\theta (\lambda)} \frac{\sin^2 \theta}{ \sqrt{P_{\theta} (\theta)}}\, d \theta  \label{t_motion_4}\\
&&
+ 4a \ell \bar{E} \int_{\theta_{0}}^{\theta (\lambda)} \frac{\cos \theta}{ \sqrt{P_{\theta} (\theta)}}\, d \theta
-4 \ell^2 \bar{E} \int_{\theta_{0}}^{\theta (\lambda)} \frac{\cos^2 \theta}{\sin^2 \theta \sqrt{P_{\theta} (\theta)}}\, d \theta .  \nonumber
\end{eqnarray}
The integration $I_{\theta,2}^{(t)}$ can be accomplished first by taking $x=\cos \theta$ and next by making further transformation
\begin{equation}
x=\frac{\alpha_3}{\left( 4y- \frac{\alpha_2}{3}\right)}+ x_{\theta}, \label{t_motion_5}
\end{equation}
where $x_{\theta}$ is one real root of the polynomial equation (\ref{ang_sol_1}) and $\alpha_2$ and $\alpha_3$ are defined as in (\ref{ang_sol_10}) and (\ref{ang_sol_11}) respectively. With the additional transformation $ \wp(s)=y$, integrations with respect to variable $s$ yield
\begin{eqnarray}
I_{\theta,2}^{(t)}&=&\left(a^2 \bar{E}^2 (x_{\theta}^2 -1)+ 4 a \ell \bar{E} x_{\theta}- \frac{2 x_{\theta} \ell}{1-x_{\theta}^2} \left(2 \ell \bar{E} x_{\theta }+ \bar{L} \right)\right)(\lambda-\lambda_{0})\nonumber \\
& &-2 \ell  \sum_{i=1}^2 \sum_{j=1}^2 \frac{( \bar{L} G_i+ 2 \ell \bar{E}\bar{G}_i) }{\wp^{\prime} (a_{ij})} \left( \zeta(a_{ij})(\lambda- \lambda_{0} )+ \ln \frac{\sigma (s- a_{ij})}{\sigma (s_{0}- a_{ij})} \right) \nonumber \\
&& +a \bar{E}  \left(\frac{ax_{\theta}}{2}+ \ell \alpha_3\right) \sum_{i=1}^2 \frac{1}{\wp^{\prime} (b_{1i})} \left( \zeta(b_{1i})(\lambda- \lambda_{0} )+ \ln \frac{\sigma (s- b_{1i})}{\sigma (s_{0}- b_{1i})} \right) \nonumber \\
&& - \frac{a^2 \bar{E} \alpha_3^2}{16} \sum_{i=1}^2 \frac{1}{{\wp^{\prime}}^2 (b_{1i})} (\lambda- \lambda_{0} )\left( \wp(b_{1i})+ \frac{\wp^{\prime \prime}(b_{1i})}{\wp^{\prime}(b_{1i})} \right) \label{t_motion_6}\\
&& - \frac{a^2 \bar{E} \alpha_3^2}{16} \sum_{i=1}^2 \frac{1}{{\wp^{\prime}}^2 (b_{1i})} \left( \zeta(s- b_{1i} )+ \frac{\wp^{\prime \prime}(b_{1i})}{\wp^{\prime}(b_{1i})} \ln \frac{\sigma (s- b_{1i})}{\sigma (s_{0}- b_{1i})}- \zeta_{0}^{(i)} \right). \nonumber
\end{eqnarray}
Here we identify $ \wp(a_{ij})=a_i$ and $ \wp(b_{1i})=b_1=\frac{\alpha_2}{12}$, $(i,j=1,2)$ with
\begin{equation}
a_1=\frac{\bar{c}}{4 (1-x_{\theta})}, \qquad  a_2=-\frac{\bar{c}}{4 (1+x_{\theta})}, \label{t_motion_7}
\end{equation}
where
\begin{equation}
\bar{c}=\alpha_3+ \frac{\alpha_2}{3} (1-x_{\theta}). \label{t_motion_8}
\end{equation}
Also we define
\begin{equation}
G_1=\frac{\alpha_3}{8(1-x_{\theta}^2)}\left( \frac{(1-3 x_{\theta}^2)}{(1-x_{\theta})}+\frac{1}{3\bar{c}}\left(3\alpha_3 x_{\theta}-\alpha_2\right) \right), \label{t_motion_9}
\end{equation}
\begin{equation}
G_2=\frac{\alpha_3}{8(1-x_{\theta}^2)}\left( \frac{(1-3 x_{\theta}^2)}{(1+x_{\theta})}-\frac{1}{3\bar{c}}\left(3\alpha_3 x_{\theta}-\alpha_2\right) \right), \label{t_motion_10}
\end{equation}
\begin{equation}
\bar{G}_1=\frac{\alpha_3}{4(1-x_{\theta}^2)}\left( \frac{ x_{\theta}}{(1-x_{\theta})}+ \frac{1}{3\bar{c}} (  x_{\theta} \alpha_2 -\alpha_3)) \right) \label{t_motion_11}
\end{equation}
and
\begin{equation}
\bar{G}_2=\frac{\alpha_3}{4(1-x_{\theta}^2)}\left( \frac{ x_{\theta}}{(1+x_{\theta})}- \frac{1}{3\bar{c}} (  x_{\theta} \alpha_2 -\alpha_3)) \right). \label{t_motion_12}
\end{equation}
Next, we consider the radial integral
\begin{eqnarray}
I_r^{(t)}=\int_{r_{0}}^{r (\lambda)}\left(\bar{E} (r^2 +a^2 + \ell^2)-a \bar{L}-\bar{q}Q r\right)\frac{(r^2 +a^2 + \ell^2)}{\Delta \sqrt{P_r(r)}}dr. \label{t_motion_13}
\end{eqnarray}
The radial integrals can be evaluated by a similar transformation such that
\begin{equation}
r=\frac{\beta_3}{(4v-\frac{\beta_2}{3})}+r_1 \label{t_motion_14}
\end{equation}
where $r_1$ is again assumed to be one real root of $P_r(r)$ and $\beta_2$ and $\beta_3$ are introduced in (\ref{rad_sol_10}) and (\ref{rad_sol_11}) respectively. Next taking $v=\wp(s)$, the integration yields
\begin{eqnarray}
I_r^{(t)}&=&\left[\bar{E}(r_1^2+\ell^2+a^2)^2-\bar{q} Q  (r_1^2+\ell^2+a^2)r_1-a \bar{L} (r_1^2+1)\right]\frac{(\lambda-\lambda_{0})}{\Delta(r_1)}\nonumber \\
&&+ \sum_{i=1}^{3}\sum_{j=1}^{2} \frac{(\bar{E} \omega_i -\bar{q} Q \bar{\omega}_i)}{\wp^{\prime} (v_{ij})} \left( \zeta(v_{ij})(\lambda-\lambda_{0})+ \ln \frac{\sigma(s-v_{ij})}{\sigma(s_{0}-v_{ij})}\right)\nonumber \\
&&- \frac{\bar{E}\beta_3^2}{16} \sum_{j=1}^{2} \frac{1}{{\wp^{\prime}}^2 (v_{3j})} \left\{(\lambda-\lambda_{0})\left(\wp(v_{3j})+\frac{\wp^{\prime \prime}(v_{3j})}{\wp^{\prime}(v_{3j})} \right) \right.  \nonumber\\
&& \left. + \frac{\wp^{\prime \prime}(v_{3j})}{\wp^{\prime}(v_{3j})}\ln \frac{\sigma(s-v_{3j})}{\sigma(s_{0}-v_{3j})} + \zeta(s-v_{3j})- \zeta_{0}^{(j)}\right\} \label{t_motion_15} \\
&&+ \sum_{i=1}^{2}\sum_{j=1}^{2} \frac{1}{{\wp^{\prime}} (v_{ij})} \left( \zeta(v_{ij})(\lambda-\lambda_{0})+ \ln \frac{\sigma(s-v_{ij})}{\sigma(s_{0}-v_{ij})} \right) \times \nonumber \\
&& \left[\left(2 \bar{E} (\ell^2 +a^2 )-a\bar{L} \right)\tilde{\omega}_i - \bar{q} Q (\ell^2+a^2)\hat{\omega}_i + (\ell^2+a^2)\left(  \bar{E} (\ell^2 +a^2 )-a\bar{L}\right)\check{\omega}_i \right].\nonumber
\end{eqnarray}
Here $ \wp(v_{ij})=v_i$ and $ \wp(v_{3j})=v_3=\frac{\beta_2}{12}$, $(i,j=1,2)$ with
\begin{equation}
v_1=\frac{1}{ 4\Delta(r_1)}\left( \frac{\Delta (r_1) \beta_2}{3} -r_1 +M \beta_3 - \sqrt{(r_1-M \beta_3)^2 - \Delta (r_1) \beta_3^2}\right) \label{t_motion_16}
\end{equation}
and
\begin{equation}
v_2=\frac{1}{4 \Delta(r_1)}\left( \frac{\Delta (r_1) \beta_2}{3} -r_1 +M \beta_3 + \sqrt{(r_1-M \beta_3)^2 - \Delta (r_1) \beta_3^2}\right). \label{t_motion_17}
\end{equation}
We also identify
\begin{equation}
\omega_1=-\frac{\left[r_1(r_1 -M \beta_3 ) -\beta_3 \Delta(r_1)  +2 r_1  \Delta(r_1) (v_2-v_1) \right]^4}{ 16\Delta^3(r_1) (v_2-v_1)  \left[r_1-M \beta_3 + 2 \Delta(r_1)(v_2-v_1)\right]^2}, \label{t_motion_18}
\end{equation}
\begin{equation}
\omega_2=\frac{\left[r_1(r_1 -M \beta_3 ) -\beta_3 \Delta(r_1)  -2 r_1  \Delta(r_1) (v_2-v_1) \right]^4}{ 16\Delta^3(r_1) (v_2-v_1)  \left[r_1-M \beta_3 - 2 \Delta(r_1)(v_2-v_1)\right]^2}, \label{t_motion_19}
\end{equation}
\begin{equation}
\bar{\omega}_1= -\frac{\left[r_1(r_1 -M \beta_3 ) -\beta_3 \Delta(r_1)  +2 r_1  \Delta(r_1) (v_2-v_1) \right]^3}{ 16\Delta^3(r_1) (v_2-v_1)  \left[r_1-M \beta_3 + 2 \Delta(r_1)(v_2-v_1)\right]}, \label{t_motion_20}
\end{equation}
\begin{equation}
\bar{\omega}_2= \frac{\left[r_1(r_1 -M \beta_3 ) -\beta_3 \Delta(r_1)  -2 r_1  \Delta(r_1) (v_2-v_1) \right]^3}{ 16\Delta^3(r_1) (v_2-v_1)  \left[r_1-M \beta_3 - 2 \Delta(r_1)(v_2-v_1)\right]}, \label{t_motion_21}
\end{equation}
\begin{equation}
\omega_3= \beta_3 r_1-\frac{(r_1-M \beta_3)}{2}, \qquad \bar{\omega}_3= \frac{\beta_3}{4}, \label{t_motion_22}
\end{equation}
\begin{equation}
\tilde{\omega}_1= -\frac{\left[r_1(r_1 -M \beta_3 ) -\beta_3 \Delta(r_1)  +2 r_1  \Delta(r_1) (v_2-v_1) \right]^2}{ 16\Delta^3(r_1) (v_2-v_1)}, \label{t_motion_23}
\end{equation}
\begin{equation}
\tilde{\omega}_2= \frac{\left[r_1(r_1 -M \beta_3 ) -\beta_3 \Delta(r_1)  -2 r_1  \Delta(r_1) (v_2-v_1) \right]^2}{ 16\Delta^3(r_1) (v_2-v_1)}, \label{t_motion_24}
\end{equation}
\begin{equation}
\hat{\omega}_1= \frac{1}{\Delta(r_1)(v_1-v_2)} \left( v_1- \frac{\beta_2}{12}\right) \left( r_1\left( v_1- \frac{\beta_2}{12}\right)+ \frac{\beta_3}{4} \right), \label{t_motion_25}
\end{equation}
\begin{equation}
\hat{\omega}_2= \frac{1}{\Delta(r_1)(v_2-v_1)} \left( v_2- \frac{\beta_2}{12}\right) \left( r_1\left( v_2- \frac{\beta_2}{12}\right)+ \frac{\beta_3}{4} \right), \label{t_motion_26}
\end{equation}
\begin{equation}
\check{\omega}_1= \frac{1}{\Delta(r_1)(v_1-v_2)} \left( v_1- \frac{\beta_2}{12}\right)^2\label{t_motion_27}
\end{equation}
and
\begin{equation}
\check{\omega}_2= \frac{1}{\Delta(r_1)(v_2-v_1)} \left( v_2- \frac{\beta_2}{12}\right)^2. \label{t_motion_28}
\end{equation}

\subsection{$\varphi$-motion}

Similarly, the equation (\ref{Motion_17}) can be written in differential form as (again taking the $+$ sign only)
\begin{equation}
d\varphi=dI_{\theta}^{(\varphi)} + d I_r^{(\varphi)} \label{azimuthal_1}
\end{equation}
upon integration which yields
\begin{equation}
\varphi-\varphi_0= I_{\theta}^{(\varphi)} +  I_r^{(\varphi)}. \label{azimuthal_2}
\end{equation}
The angular integral
\begin{equation}
I_{\theta}^{(\varphi)} =\int \frac{\left(\bar{L}-\bar{E}(a \sin^2 \theta-2\ell \cos \theta) \right)}{\sin^2 \theta \sqrt{P_{\theta}(\theta)}}d \theta \label{azimuthal_3}
\end{equation}
can be accomplished by making the transformation
\begin{equation}
\cos \theta=\frac{\alpha_3}{\left( 4\wp(s)- \frac{\alpha_2}{3}\right)}+ x_{\theta} \label{azimuthal_4}
\end{equation}
such that it produces the solution
\begin{eqnarray}
I_{\theta}^{(\varphi)} &=& \left( \frac{(\bar{L} +2 \ell \bar{E} x_{\theta})}{(1-x_{\theta}^2)} -a \bar{E}\right)(\lambda-\lambda_{0}) \label{azimuthal_5} \\
&& +  \sum_{i=1}^2 \sum_{j=1}^2 \frac{( \bar{L} \bar{G}_i+ 2 \ell \bar{E}G_i) }{\wp^{\prime} (a_{ij})} \left( \zeta(a_{ij})(\lambda- \lambda_{0} )+ \ln \frac{\sigma (s- a_{ij})}{\sigma (s_{0}- a_{ij})} \right) \nonumber
\end{eqnarray}
where $G_i$ and $\bar{G}_i$ are defined through (\ref{t_motion_9})-(\ref{t_motion_12}). On the other hand, the radial integral
\begin{equation}
 I_r^{(\varphi)}= \int_{r_{0}}^{r(\lambda)} \frac{a\left(\bar{E} (r^2+a^2 +\ell^2)-a\bar{L} -\bar{q} Q r \right)}{\Delta \sqrt{P_r(r)}}dr \label{azimuthal_6}
\end{equation}
can be calculated by affecting the transformation
\begin{equation}
r=\frac{\beta_3}{(4\wp(s)-\frac{\beta_2}{3})}+r_1 \label{azimuthal_7}
\end{equation}
that gives
\begin{eqnarray}
&&I_r^{(\varphi)}=\frac{a(\bar{E}(r_1^2  +a^2 + \ell^2)-\bar{q} Q r_1 -a\bar{L})}{\Delta(r_1)}(\lambda- \lambda_{0})\label{azimuthal_8} \\
&+& \sum_{i=1}^2 \sum_{j=1}^2 \frac{a(\bar{E}\tilde{\omega}_i -\bar{q} Q \hat{\omega}_i +(\bar{E}(a^2 + \ell^2) -a\bar{L}) \check{\omega}_i)}{\wp^{\prime} (v_{ij})} \left( \zeta(v_{ij})(\lambda-\lambda_{0})+ \ln \frac{\sigma(s-v_{ij})}{\sigma(s_{0}-v_{ij})} \right)\nonumber
\end{eqnarray}
where $\tilde{\omega}_i$, $\hat{\omega}_i$ and $\check{\omega}_i$ are identified through (\ref{t_motion_23})-(\ref{t_motion_28}).

\section{Discussion of the orbits and observables}

In this part, using the analytical solutions of the equations of motion that we have obtained in the previous chapter, we plot some possible orbit types for fixed energy, angular momentum and spacetime parameters. Furthermore, at the end of the section, we calculate the observables of the bound orbits and express them in terms of the elliptic functions.

\begin{figure}
    \centering
    \begin{subfigure}[b]{0.5\textwidth}
        \includegraphics[width=\textwidth]{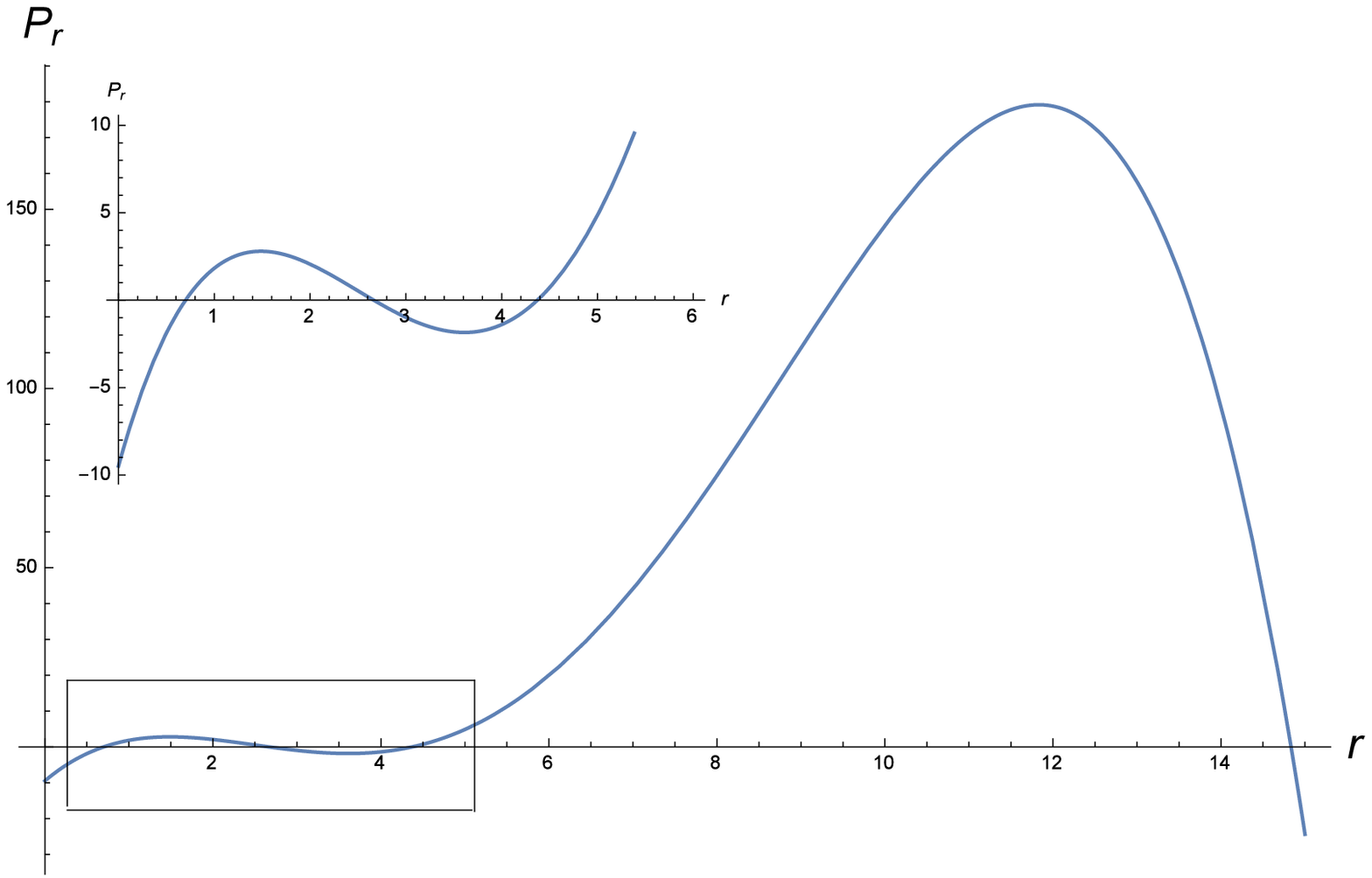}
        \caption{The graph of $P_r(r)$}
        \label{new_3D_bound_1_1}
    \end{subfigure}
     \begin{subfigure}[b]{0.45\textwidth}
        \includegraphics[width=\textwidth]{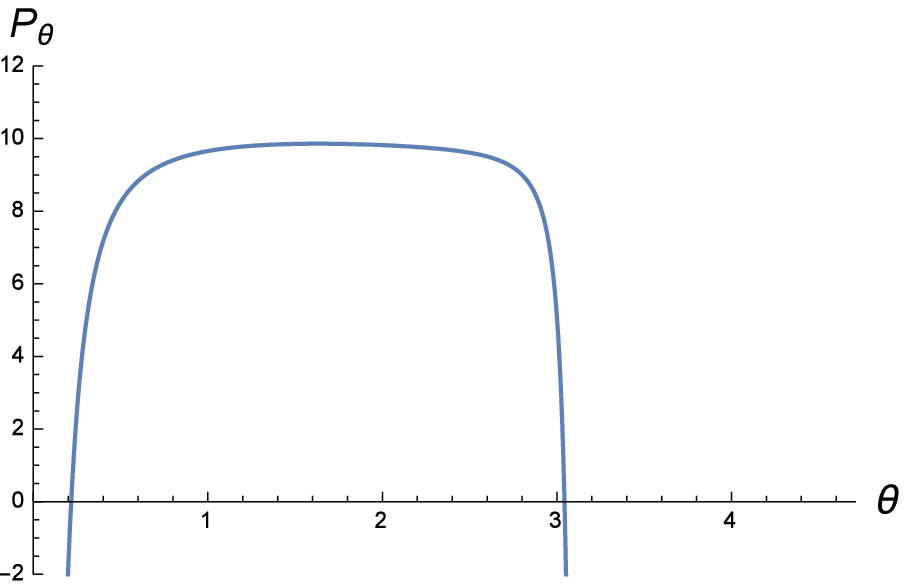}
        \caption{The graph of $P_{\theta}(\theta)$}
        \label{new3D_bound_1_angular_potential_1}
    \end{subfigure}
    \begin{subfigure}[b]{0.4\textwidth}
        \includegraphics[width=\textwidth]{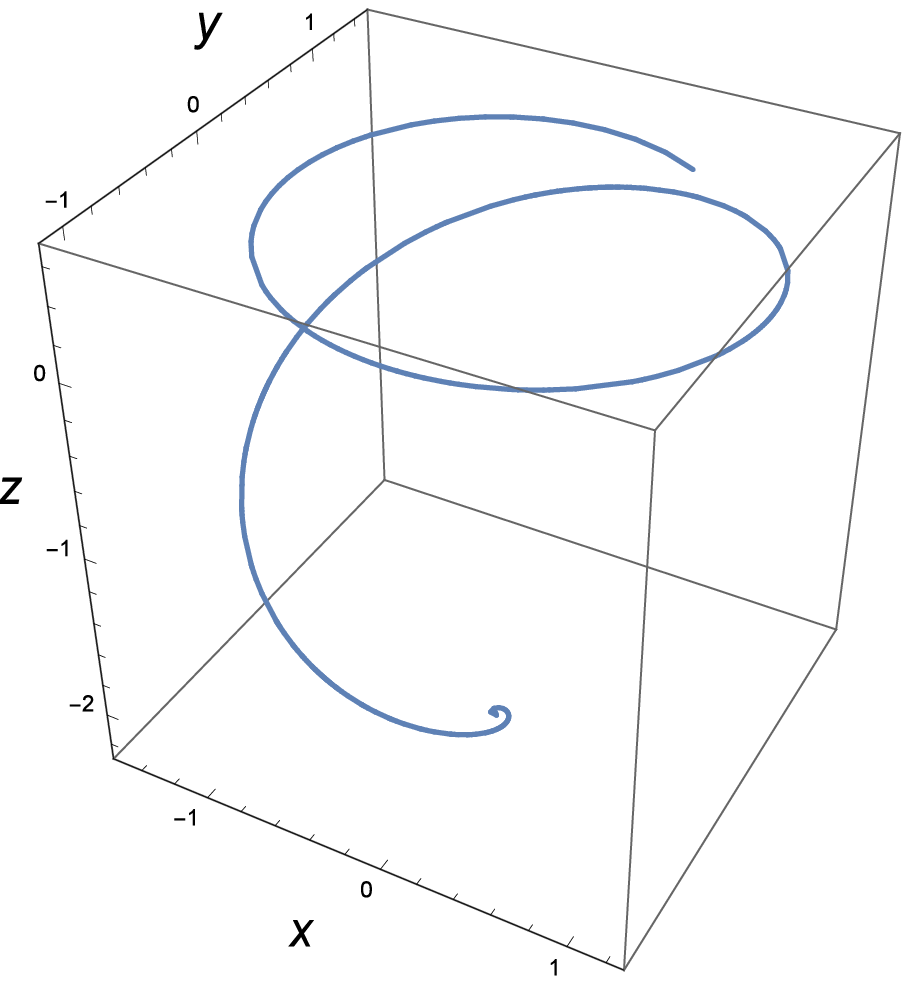}
        \caption{3D-bound orbit with $\bar{E}<1$}
        \label{new_3D_bound_1_2}
    \end{subfigure}\hspace{0.5cm}
    \begin{subfigure}[b]{0.3\textwidth}
        \includegraphics[width=\textwidth]{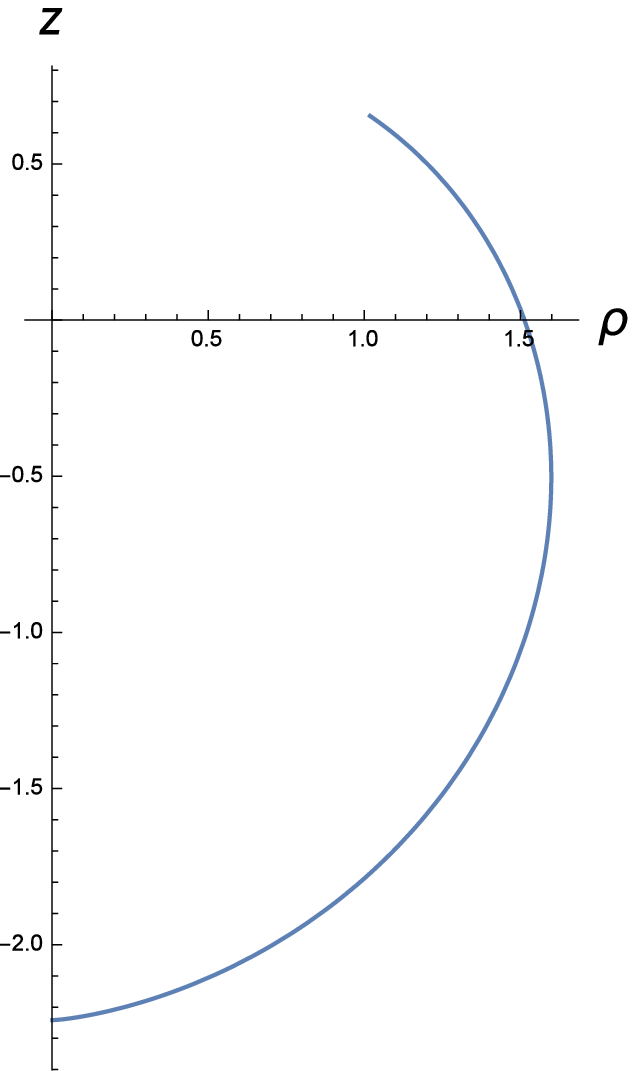}
        \caption{$r$-$\theta$ plane}
        \label{new_3D_bound_1_3}
    \end{subfigure} \hspace{0.5cm}
        \begin{subfigure}[b]{0.4\textwidth}
        \includegraphics[width=\textwidth]{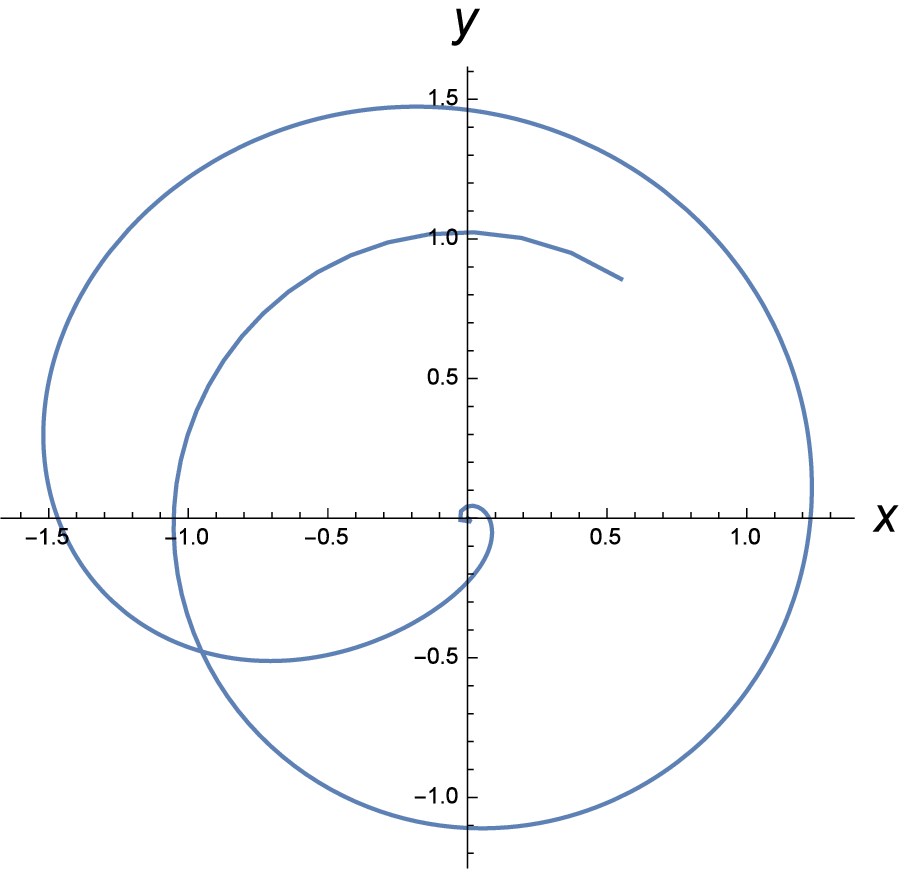}
        \caption{Projection onto $xy$-plane}
        \label{new_3D_bound_1_4}
    \end{subfigure}
    \caption{The plots are obtained for the parameters $M=1$, $a=0.9$, $K=10$, $Q=0.4$, $\bar{q}=0.3$, $\bar{L}=0.5$, $\bar{E}=0.96$, $\ell=0.1$ and $m=1$. Here $\rho^2=x^2+y^2$. There exist two bound orbits.}\label{3D_bound_1}
\end{figure}

\begin{figure}
    \centering
    \begin{subfigure}[b]{0.5\textwidth}
        \includegraphics[width=\textwidth]{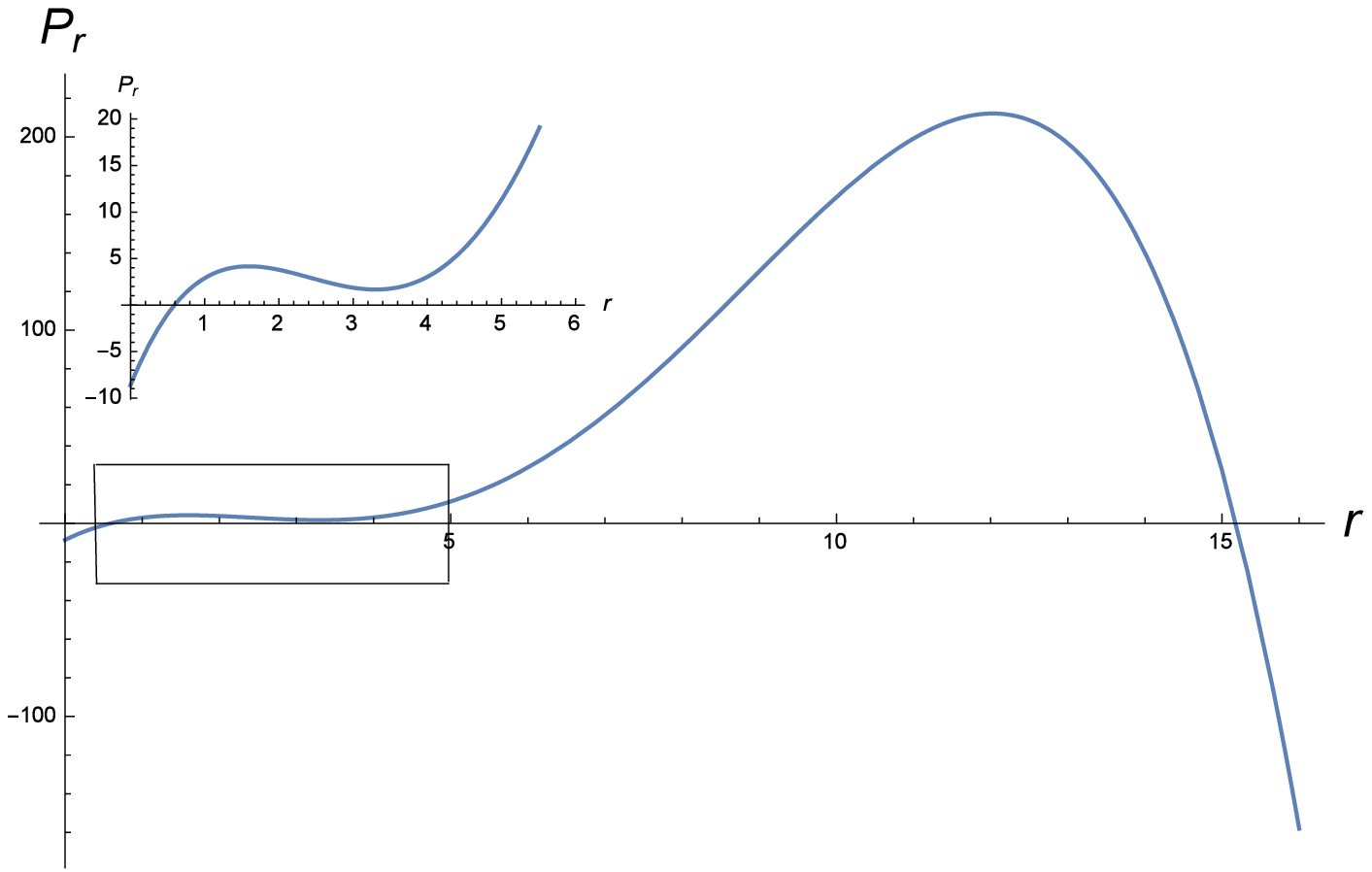}
        \caption{The graph of $P_r(r)$}
        \label{new_3D_bound_3_1}
    \end{subfigure}
    \begin{subfigure}[b]{0.45\textwidth}
        \includegraphics[width=\textwidth]{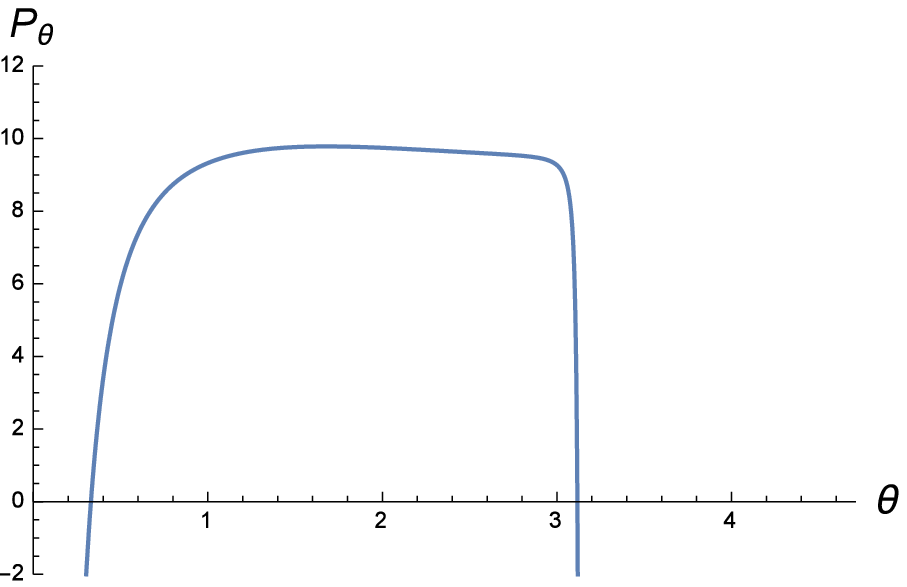}
        \caption{The graph of $P_{\theta}(\theta)$}
        \label{new3D_bound_3_angular_potential_3}
    \end{subfigure}
    \begin{subfigure}[b]{0.4\textwidth}
        \includegraphics[width=\textwidth]{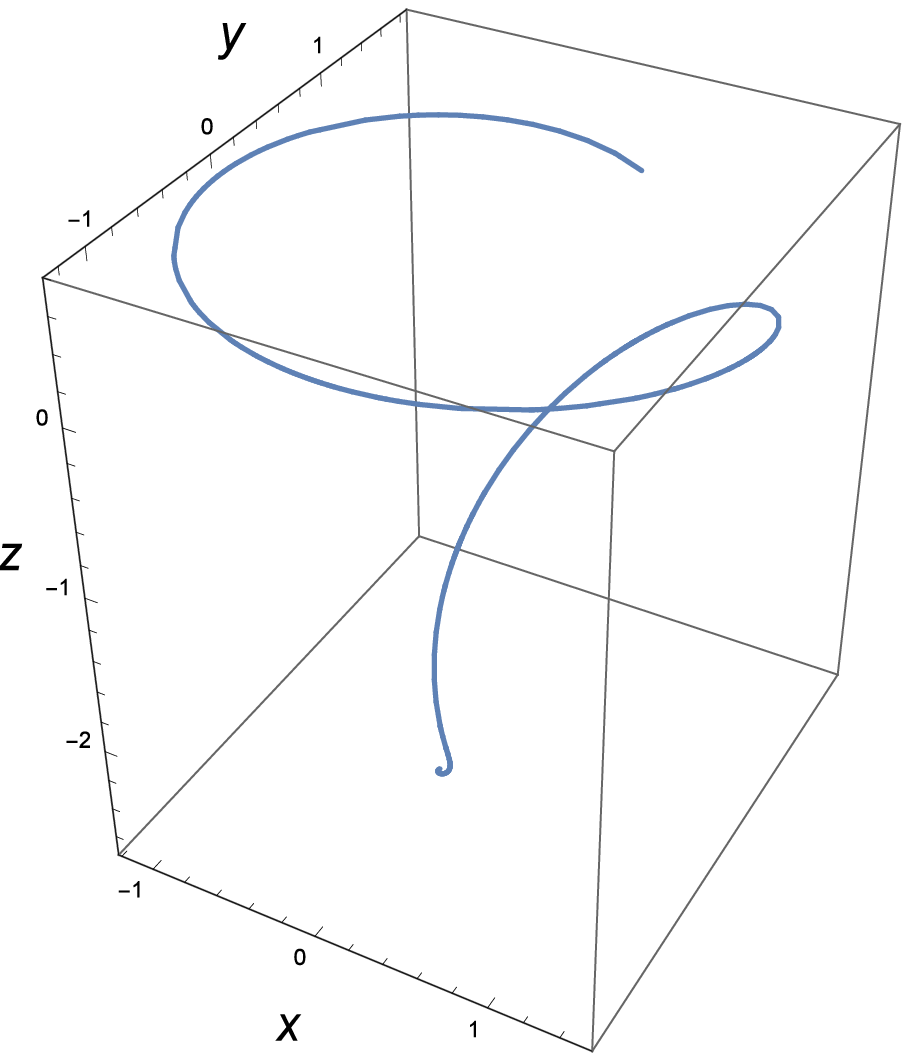}
        \caption{3D-bound orbit with $\bar{E}<1$}
        \label{new_3D_bound_3_2}
    \end{subfigure}\hspace{0.5cm}
    \begin{subfigure}[b]{0.3\textwidth}
        \includegraphics[width=\textwidth]{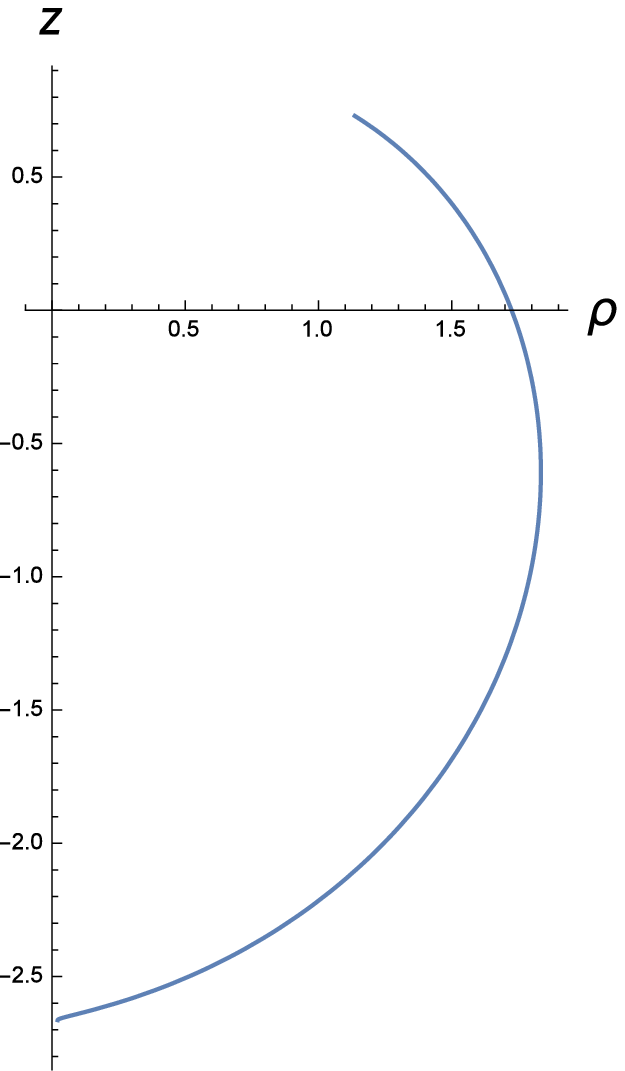}
        \caption{$r$-$\theta$ plane}
        \label{new_3D_bound_3_3}
    \end{subfigure}\hspace{0.5cm}
        \begin{subfigure}[b]{0.4\textwidth}
        \includegraphics[width=\textwidth]{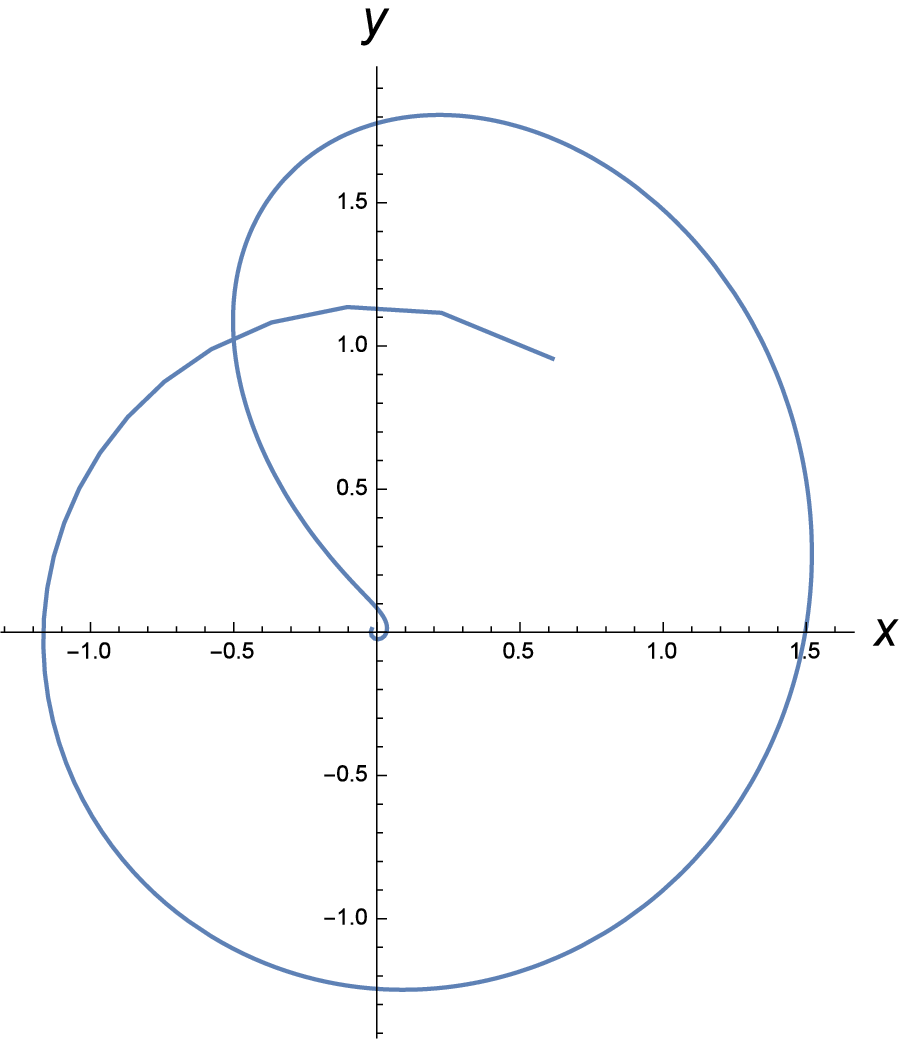}
        \caption{Projection onto $xy$-plane}
        \label{new_3D_bound_3_4}
    \end{subfigure}
    \caption{The plots are obtained for the parameters $M=1$, $a=0.9$, $K=10$, $Q=0.4$, $\bar{q}=0.3$, $\bar{L}=0.5$, $\bar{E}=0.96$, $\ell=0.3$ and $m=1$. Here $\rho^2=x^2+y^2$. There exists one bound orbit.}\label{3D_bound_2}
\end{figure}

First, we examine the possible orbit types with respect to value of the energy parameter as well as the value of the NUT parameter. In Figures \ref{3D_bound_1} and \ref{3D_bound_2}, orbits are plotted for $\bar{E}<1$. It can be seen that for $\bar{E}<1$, there may exist one bound or two bound orbits. In Figure \ref{3D_bound_1}, for the value of NUT parameter $\ell=0.1$, we obtain two bound orbits ($P_r(r)$ has four real zeros ordered as $r_4<r_3<r_2<r_1$), while in Figure \ref{3D_bound_2}, for $\ell=0.3$ (i.e NUT parameter is slightly increased) there exists one bound orbit ($P_r(r)$ has two real zeros). A further investigation of the bound orbits in Figure \ref{3D_bound_1} illustrates that, there exist two bound regions for $r_2<r<r_1$ and $r_4<r<r_3$ where in the region $r_2<r<r_1$, $r_+<r_2<r_1$ i.e. the bound orbit exists outside the outer singularity at $r=r_+$. On the other hand, looking at the bound region $r_4<r<r_3$, one can see that $0<r_4<r_-$ while $r_3>r_+$. Meanwhile, in Figure \ref{3D_bound_2}, there exist only one bound region $0<r_2<r<r_1$ where $r_2<r_-$ and $r_1>r_+$. The bound region of Figure \ref{3D_bound_2} is identical to the second bound region of Figure \ref{3D_bound_1} in the sense that in both regions, $r_{min}<r_-$ and $r_{max}>r_+$ (Here, $r_{min}$ and $r_{max}$ correspond to minimum and maximum values of $r$-coordinate for the bound region). This can be further interpreted as, if the test particle starts its motion at $r(\lambda=0)=r_{in}\geq r_{min}$ (where $r_{in}$ describes the initial radial position of the particle), it can pass through the spacetime singularities at $r=r_-$ and $r=r_+$ and can enter the region where $r_+<r \leq r_{max}$. It can also be seen from the graphs of $P_\theta(\theta)$ in Figures \ref{3D_bound_1} and \ref{3D_bound_2}, the test particle can cross the equatorial plane where at $\theta=\frac{\pi}{2}$, $P_\theta(\theta)>0$.

\begin{figure}
    \centering
    \begin{subfigure}[b]{0.5\textwidth}
        \includegraphics[width=\textwidth]{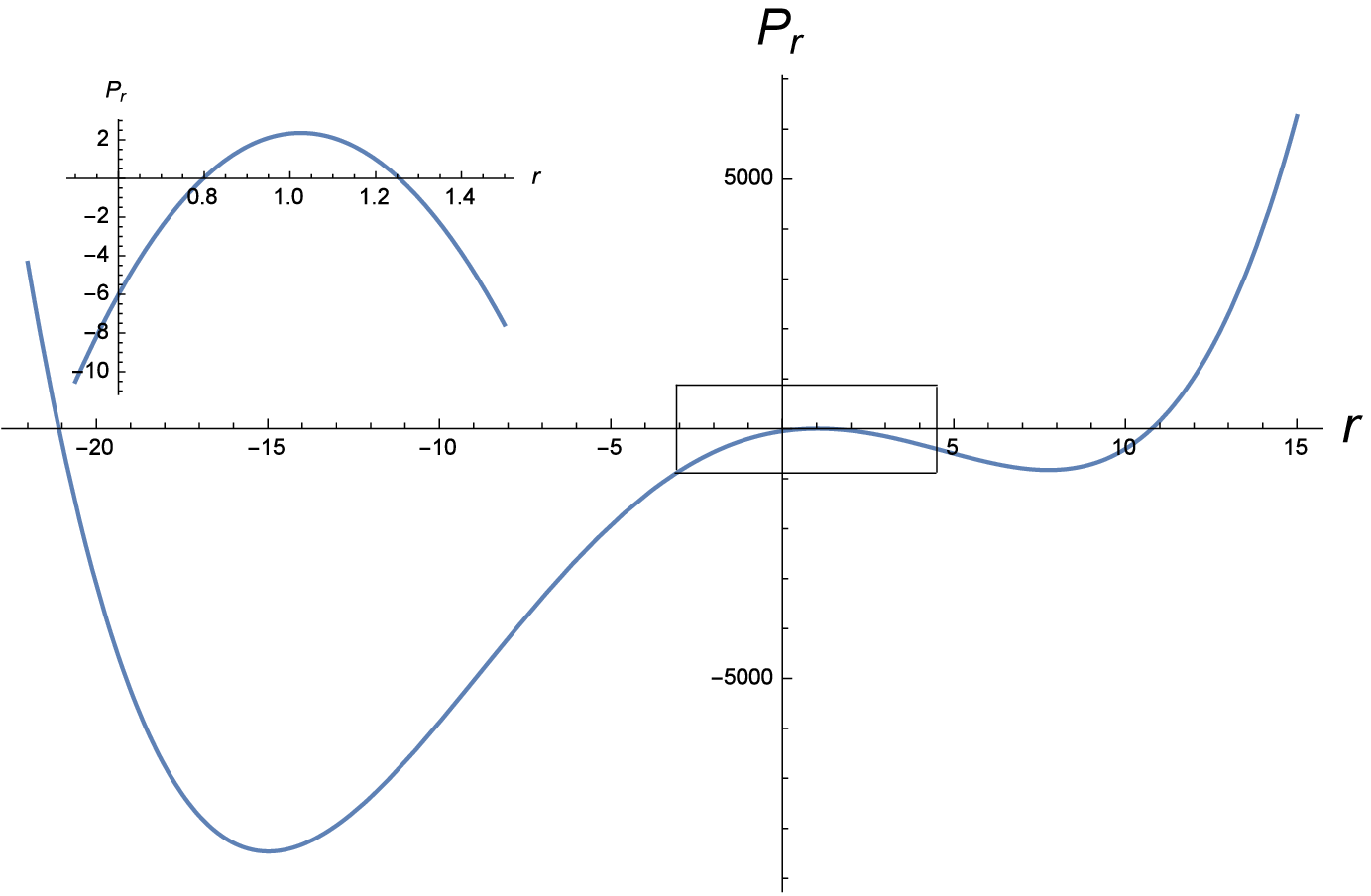}
        \caption{The graph of $P_r(r)$}
        \label{new_3D_flyby_1_1}
    \end{subfigure}\hspace{0.5cm}
     \begin{subfigure}[b]{0.45\textwidth}
        \includegraphics[width=\textwidth]{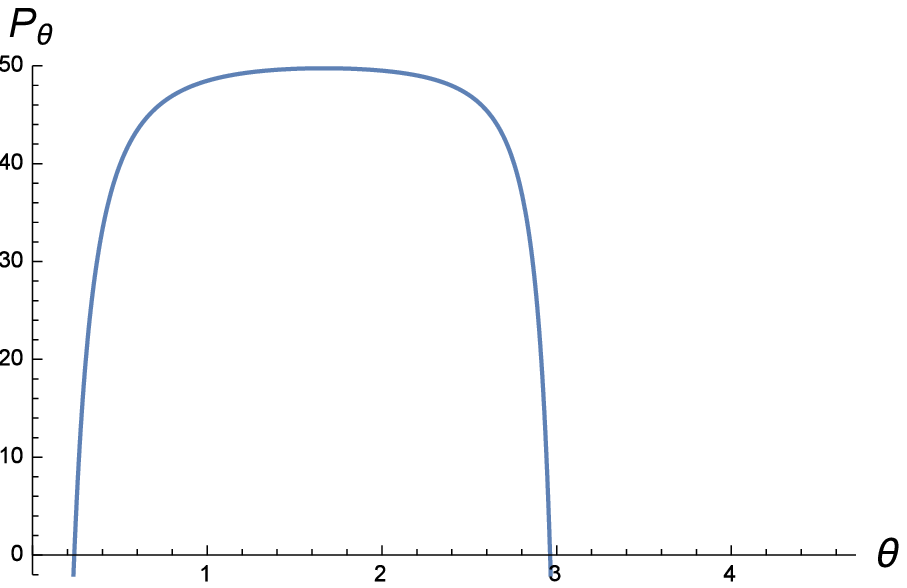}
        \caption{The graph of $P_{\theta}(\theta)$}
        \label{new3D_flyby_1_angular_potential_1}
    \end{subfigure}\hspace{0.5cm}
    \begin{subfigure}[b]{0.2\textwidth}
        \includegraphics[width=\textwidth]{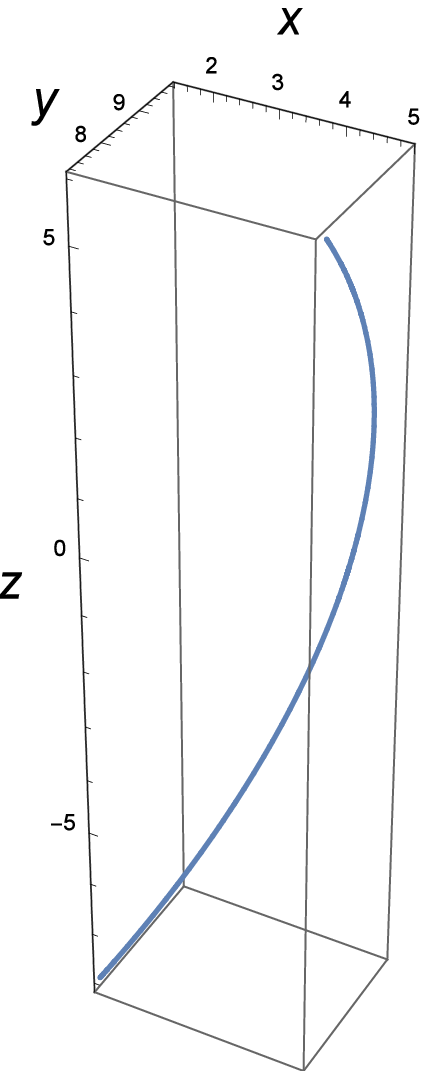}
        \caption{3D-flyby orbit for\\ $r>r_1>r_+$ with $\bar{E}>1$.}
        \label{new_3D_flyby_1_2}
    \end{subfigure}\hspace{1cm}
    \begin{subfigure}[b]{0.23\textwidth}
        \includegraphics[width=\textwidth]{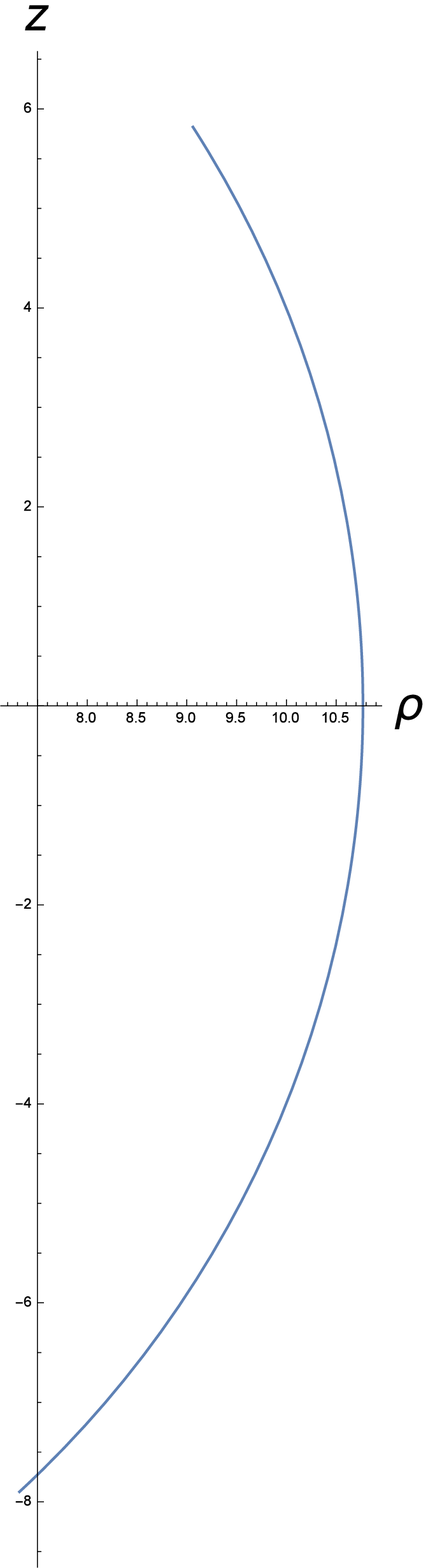}
        \caption{$r$-$\theta$ plane}
        \label{new_3D_flyby_1_3}
    \end{subfigure}\hspace{0.8cm}
        \begin{subfigure}[b]{0.4\textwidth}
        \includegraphics[width=\textwidth]{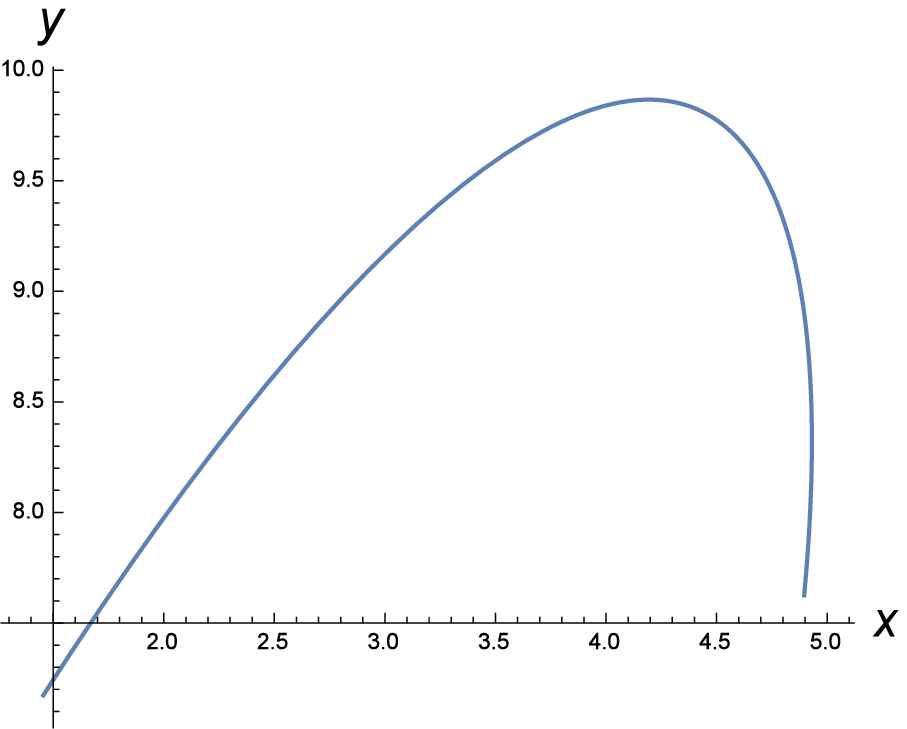}
        \caption{Projection onto $xy$-plane}
        \label{new_3D_flyby_1_4}
    \end{subfigure}
    \caption{The plots are obtained for the parameters $M=1$, $a=0.9$, $K=50$, $Q=0.4$, $\bar{q}=0.3$, $\bar{L}=1.5$, $\bar{E}=1.1$, $\ell=0.1$ and $m=1$. Here $\rho^2=x^2+y^2$ and $r_1=10.768$ ($r_1>r_+$) is one real root of $P_r(r)$. There exists one bound orbit and two flyby orbits.}\label{3D_flyby_1}
\end{figure}

On the other hand, for $\bar{E}>1$, one can get one bound and two flyby orbits ($P_r(r)$ has four real zeros) or two flyby orbits ($P_r(r)$ has two real zeros) or transit orbits ($P_r(r)$ has no real zeros). Looking at the $P_r(r)$ graph of Figure \ref{3D_flyby_1}, it can be seen that there exist one bound and two flyby orbits. A close investigation illustrates that, flyby orbits exist in the region where $r>r_1>r_+$ and $r<r_4<0<r_-$ (here $r_1$ being the largest real root of $P_r(r)$ while $r_4$ being the smallest real root of $P_r(r)$) while a bound orbit can exist in the region $0<r_3<r<r_2$ where $r_3<r_-$ and $r_2>r_+$. The examination of the flyby orbits illustrates that the particle can escape to $+ \infty $ if it starts at $r=r_{min}\geq r_1>r_+$ while it can escape to $-\infty$ if it starts at $r=r_{min}$ where $r_{min}<0<r_-$ for this case. In Figure \ref{3D_flyby_1}, the plot of 3D-flyby orbit is obtained for a specific value of the NUT parameter for the region where $r\geq r_1>r_+$ ($r_1$ corresponds to the largest root of $P_r(r)$). In this plot, it is also obvious from the graph of $P_\theta(\theta)$ that the particle can cross the equatorial plane.

\begin{figure}
    \centering
    \begin{subfigure}[b]{0.45\textwidth}
        \includegraphics[width=\textwidth]{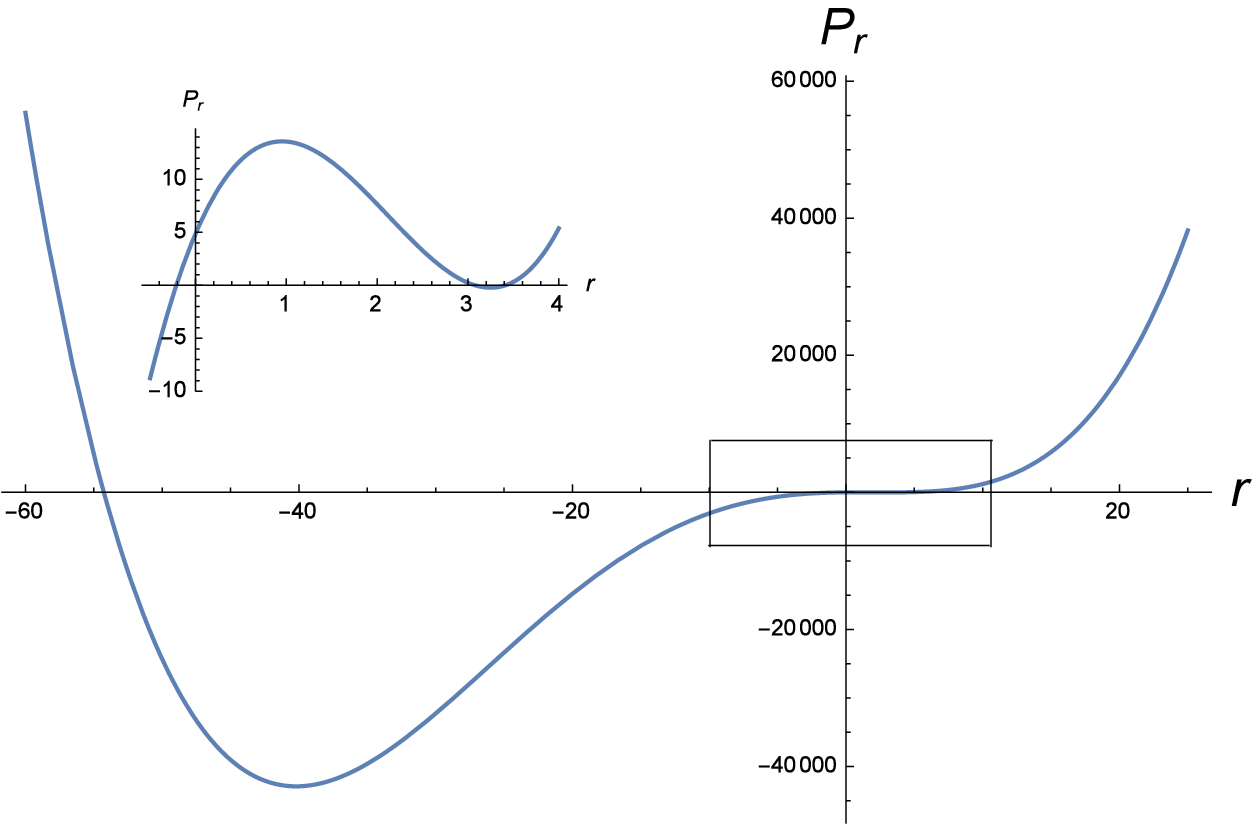}
        \caption{The graph of $P_r(r)$}
        \label{new_3D_bound_energy_greater_than_1_1_1}
    \end{subfigure}
    \begin{subfigure}[b]{0.45\textwidth}
        \includegraphics[width=\textwidth]{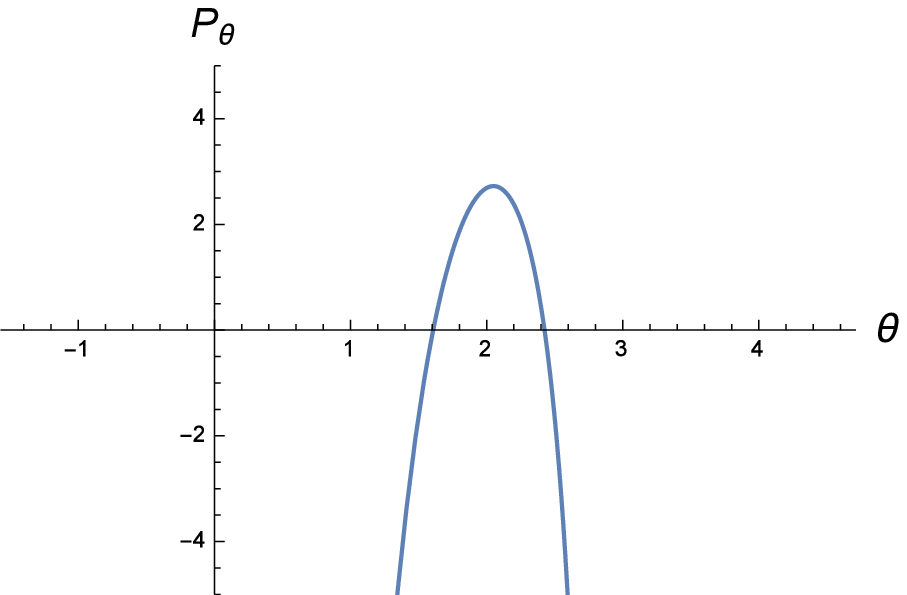}
        \caption{The graph of $P_{\theta}(\theta)$}
        \label{new3D_bound_energy_greater_than_1_angular_potential_1}
    \end{subfigure}
    \begin{subfigure}[b]{0.4\textwidth}
        \includegraphics[width=\textwidth]{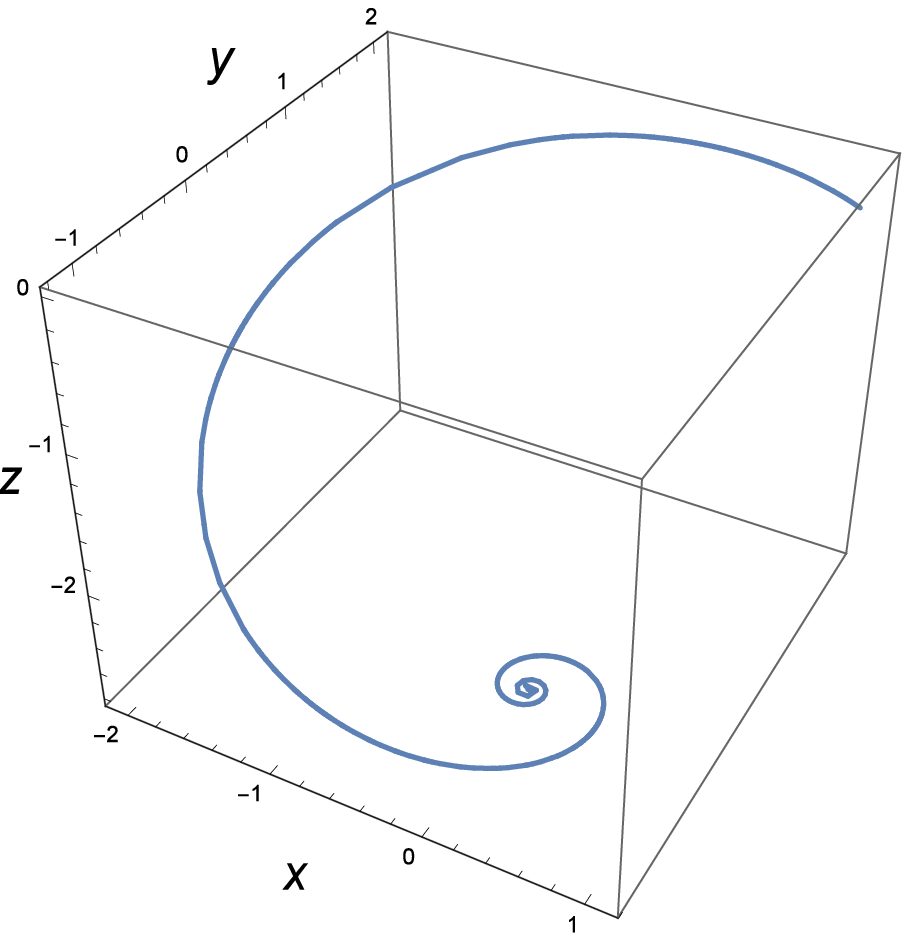}
        \caption{3D-bound orbit with $\bar{E}>1$}
        \label{new_3D_bound_energy_greater_than_1_1_2}
    \end{subfigure}\hspace{0.5cm}
    \begin{subfigure}[b]{0.3\textwidth}
        \includegraphics[width=\textwidth]{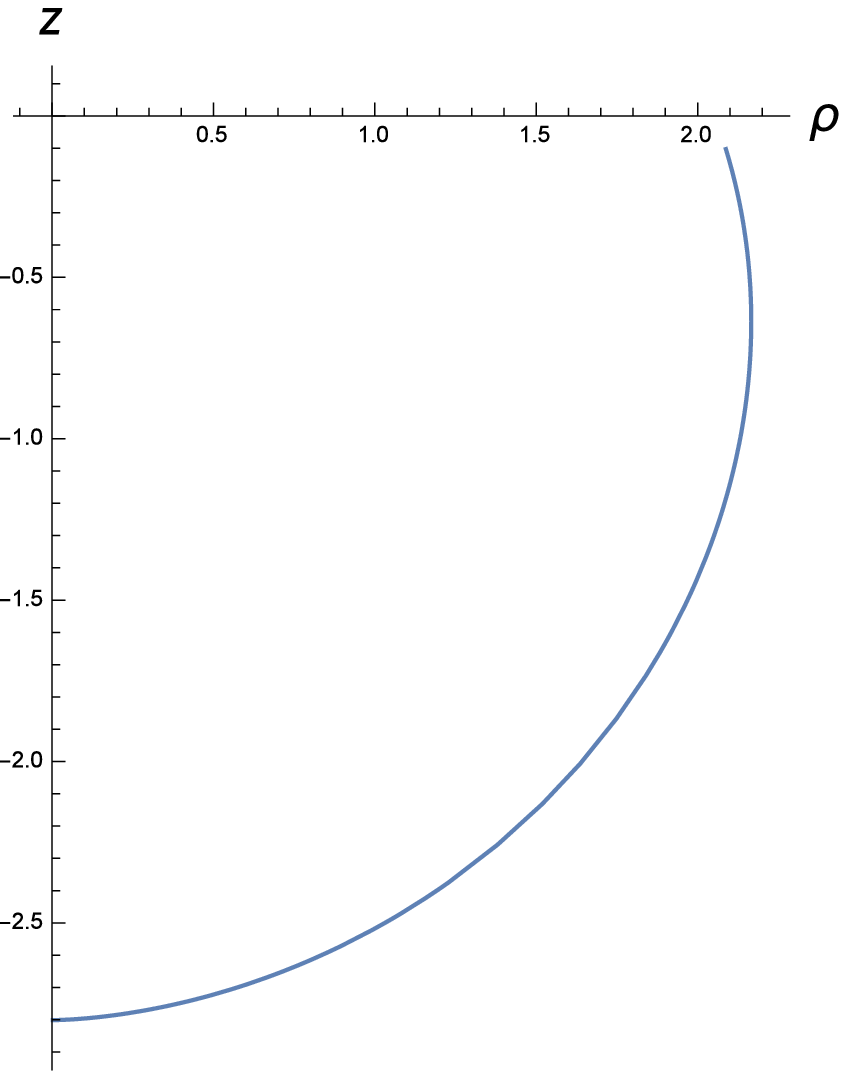}
        \caption{$r$-$\theta$ plane}
        \label{new_3D_bound_energy_greater_than_1_1_3}
    \end{subfigure}\hspace{0.5cm}
        \begin{subfigure}[b]{0.4\textwidth}
        \includegraphics[width=\textwidth]{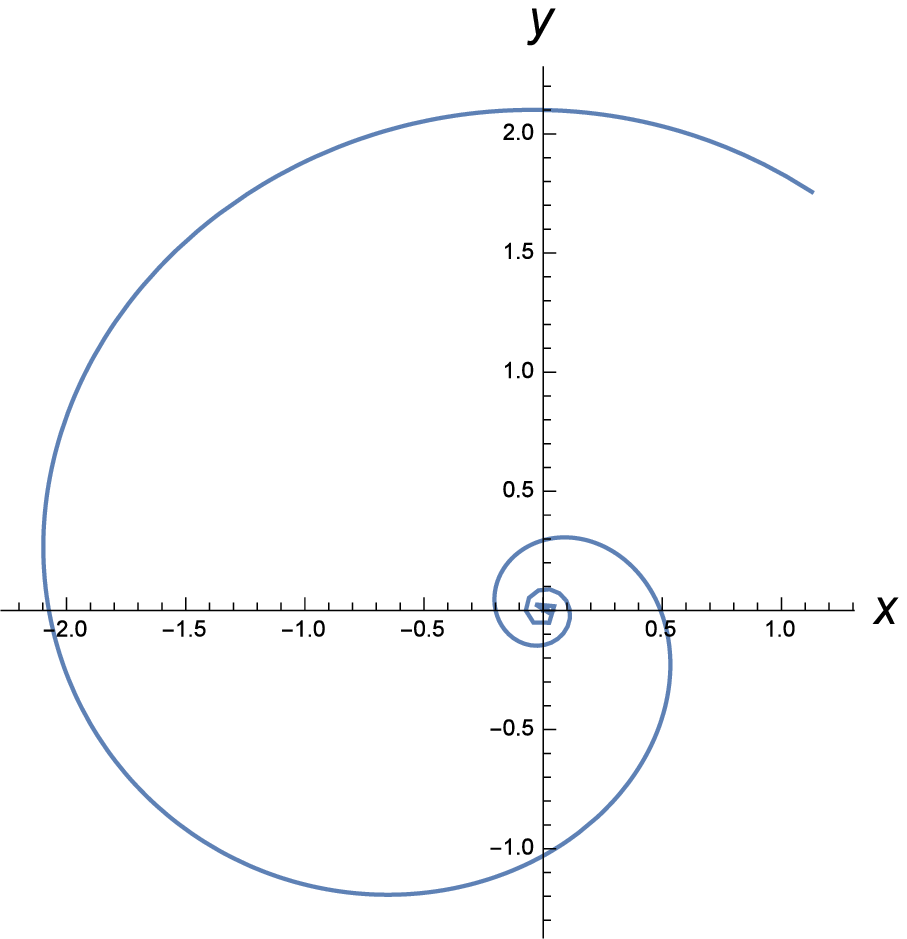}
        \caption{Projection onto $xy$-plane}
        \label{new_3D_bound_energy_greater_than_1_1_4}
    \end{subfigure}
    \caption{The plots are obtained for the parameters $M=1$, $a=0.9$, $K=10$, $Q=0.1$, $\bar{q}=0.3$, $\bar{L}=4$, $\bar{E}=1.02$, $\ell=1$ and $m=1$. Here $\rho^2=x^2+y^2$. There exists one bound orbit and two flyby orbits.}\label{3D_bound_3}
\end{figure}

Meanwhile in Figure \ref{3D_bound_3}, a 3D-bound orbit (with $\bar{E}>1$) can be obtained similarly for a specific value of gravitomagnetic monopole moment $\ell$. In this figure, the bound orbit exists in the region $r_{min} \leq r \leq r_{max}$ where in that case $r_{min}<0<r_-$ and $r_{max}>r_+$. However, for this plot, it is also clear from the graph of $P_\theta(\theta)$ that the test particle cannot cross the equatorial plane.

\begin{figure}
    \centering
    \begin{subfigure}[b]{0.45\textwidth}
        \includegraphics[width=\textwidth]{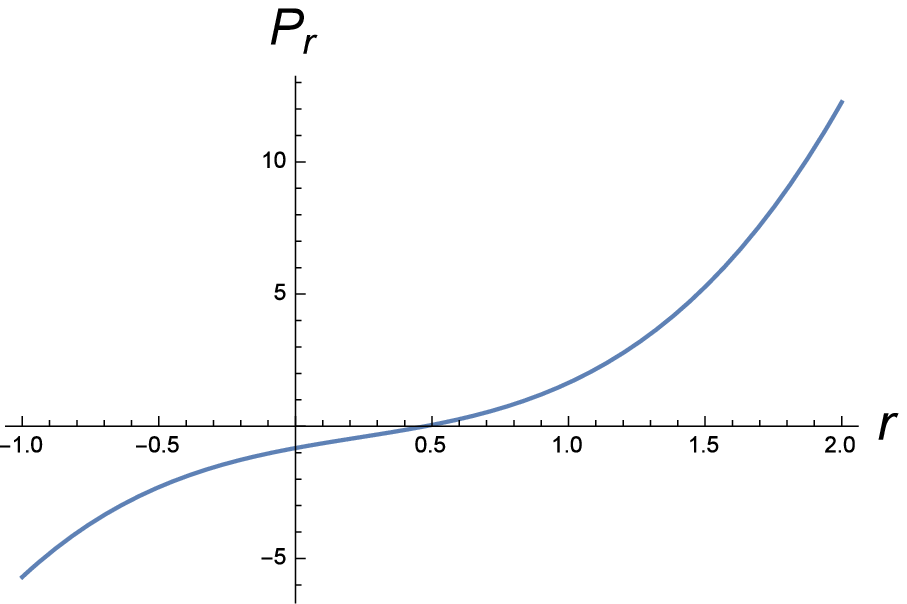}
        \caption{The graph of $P_r(r)$}
        \label{new_3D_energy_equal_to_1_1_1}
    \end{subfigure}
    \begin{subfigure}[b]{0.45\textwidth}
        \includegraphics[width=\textwidth]{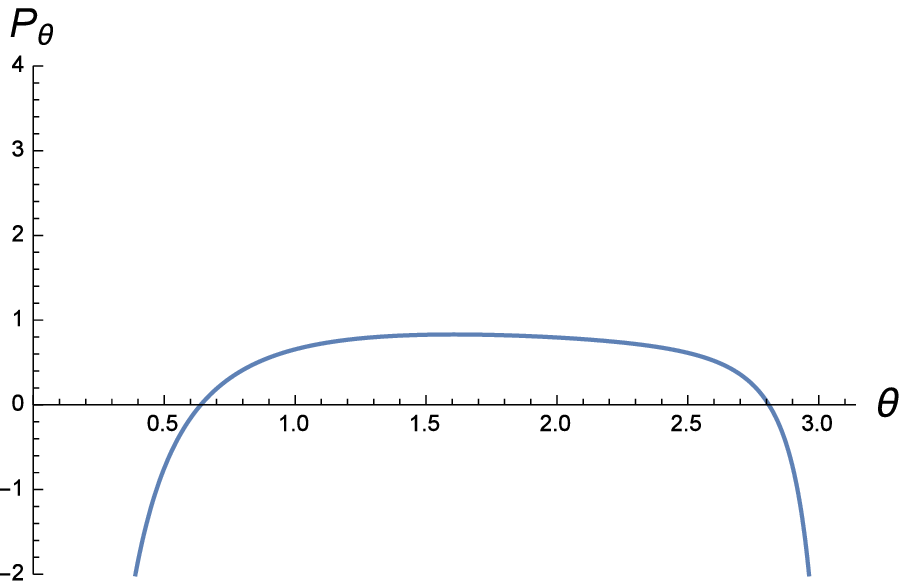}
        \caption{The graph of $P_{\theta}(\theta)$}
        \label{new3D_angular_potential_for_energy_equal_to_1_1}
    \end{subfigure}
    \begin{subfigure}[b]{0.45\textwidth}
        \includegraphics[width=\textwidth]{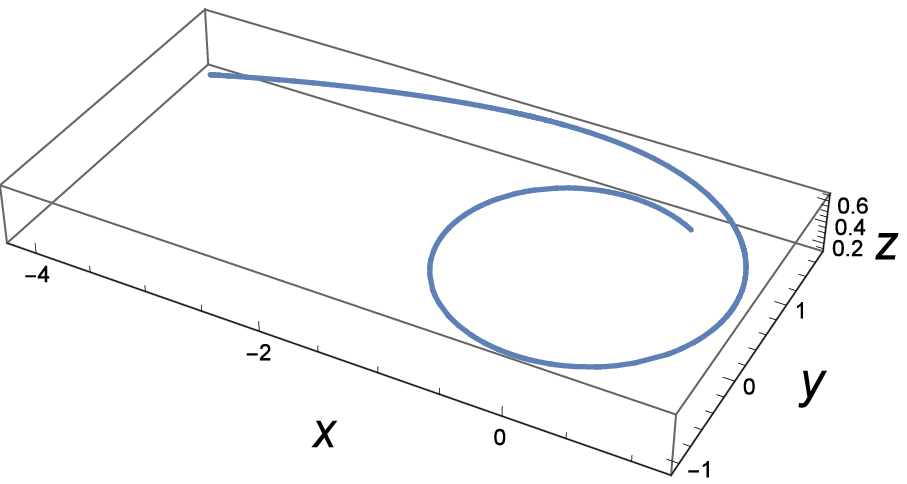}
        \caption{3D-flyby orbit with $\bar{E}=1$}
        \label{new_3D_energy_equal_to_1_1_2}
    \end{subfigure}
    \begin{subfigure}[b]{0.55\textwidth}
        \includegraphics[width=\textwidth]{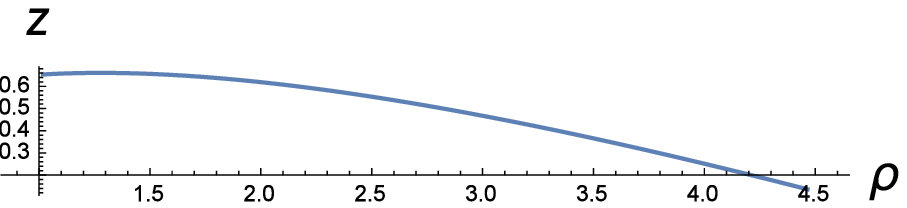}
        \caption{$r$-$\theta$ plane}
        \label{new_3D_energy_equal_to_1_1_3}
    \end{subfigure}\hspace{0.5cm}
        \begin{subfigure}[b]{0.50\textwidth}
        \includegraphics[width=\textwidth]{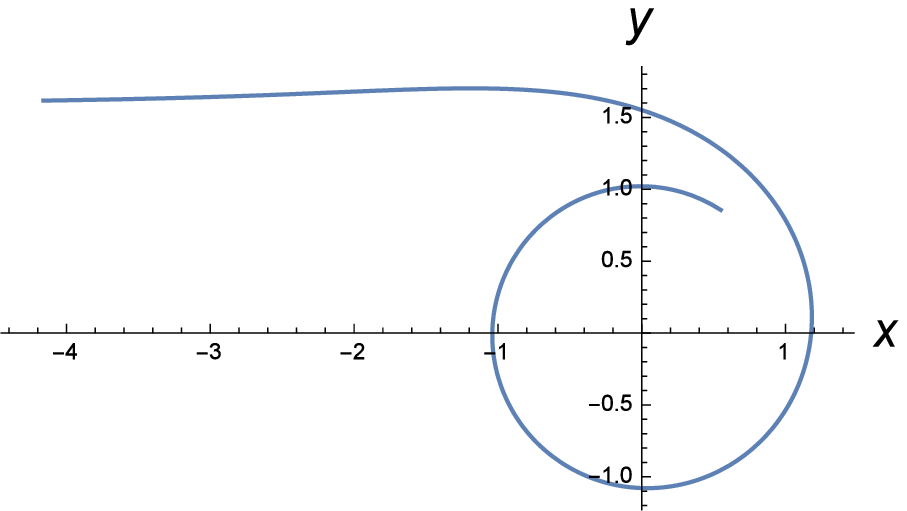}
        \caption{Projection onto $xy$-plane}
        \label{new_3D_energy_equal_to_1_1_4}
    \end{subfigure}
    \caption{The plots are obtained for the parameters $M=1$, $a=0.9$, $K=1$, $Q=0.4$, $\bar{q}=0.3$, $\bar{L}=0.5$, $\bar{E}=1$, $\ell=0.1$ and $m=1$. Here $\rho^2=x^2+y^2$. There exists one flyby orbit.}\label{3D_flyby_3}
\end{figure}

\begin{figure}
    \centering
    \begin{subfigure}[b]{0.45\textwidth}
        \includegraphics[width=\textwidth]{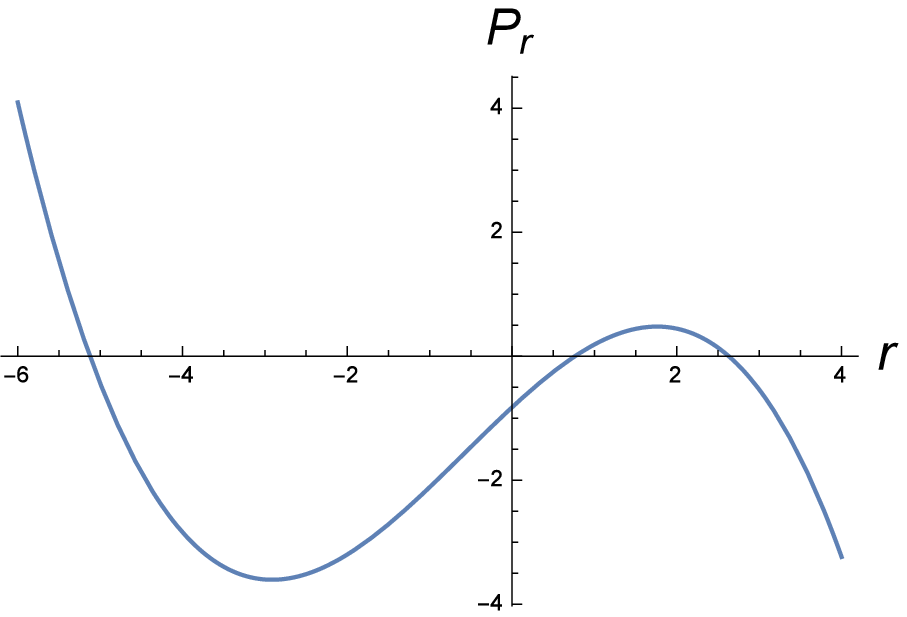}
        \caption{The graph of $P_r(r)$}
        \label{new_3D_energy_equal_to_1_2_1}
    \end{subfigure}
    \begin{subfigure}[b]{0.45\textwidth}
        \includegraphics[width=\textwidth]{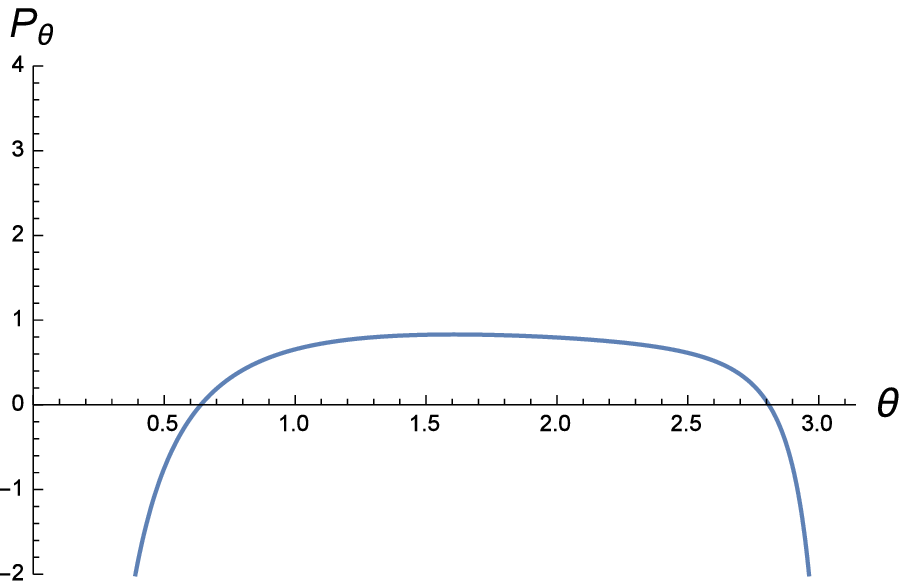}
        \caption{The graph of $P_{\theta}(\theta)$}
        \label{new3D_angular_potential_for_energy_equal_to_1_2}
    \end{subfigure}
    \begin{subfigure}[b]{0.4\textwidth}
        \includegraphics[width=\textwidth]{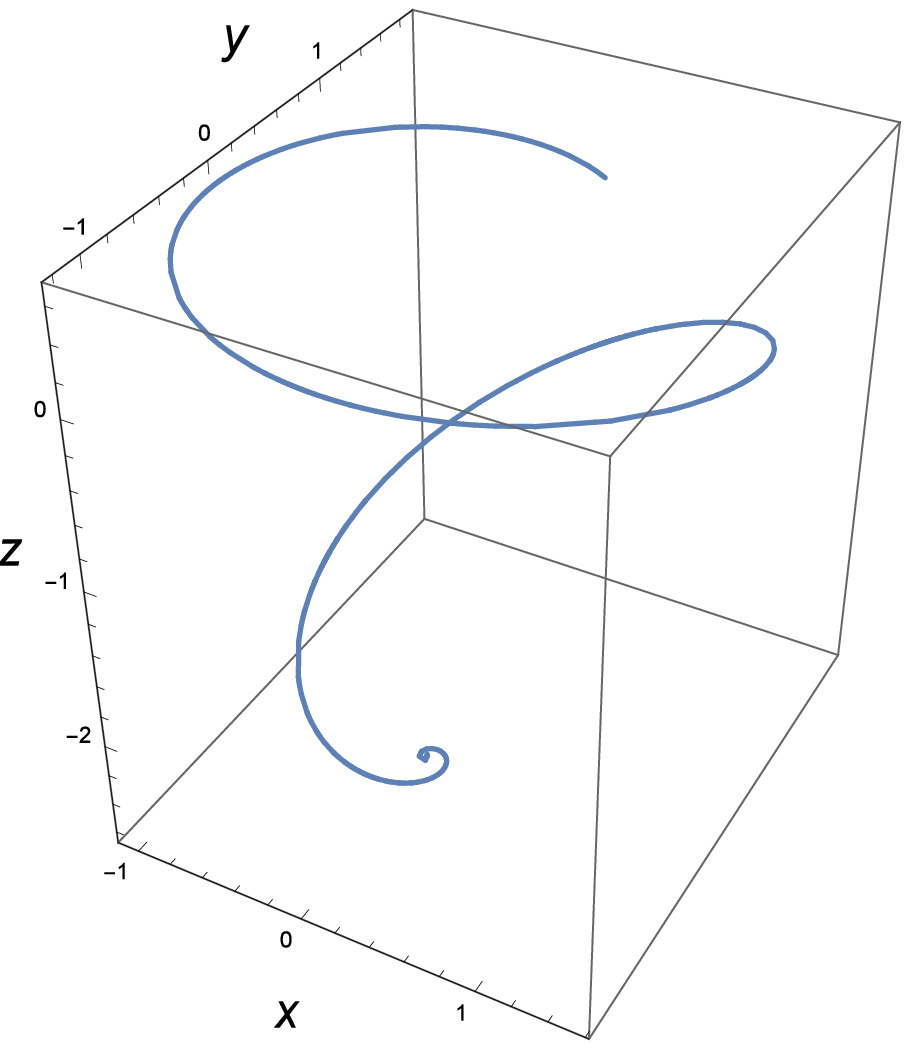}
        \caption{3D-bound orbit with $\bar{E}=1$}
        \label{new_3D_energy_equal_to_1_2_2}
    \end{subfigure}\hspace{0.45cm}
    \begin{subfigure}[b]{0.3\textwidth}
        \includegraphics[width=\textwidth]{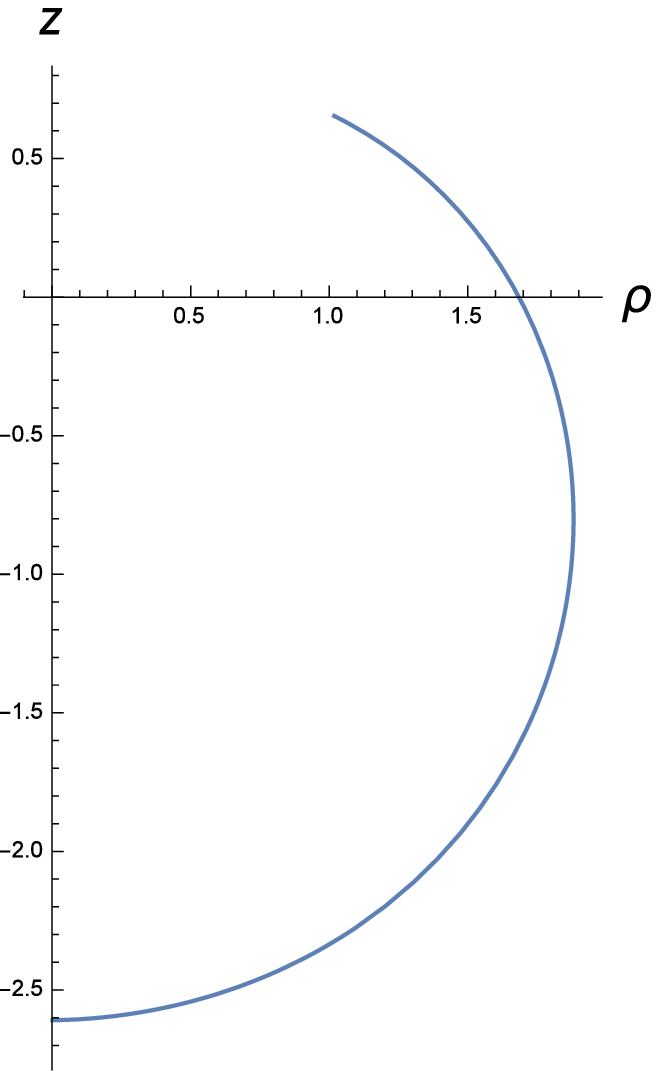}
        \caption{$r$-$\theta$ plane}
        \label{new_3D_energy_equal_to_1_2_3}
    \end{subfigure}\hspace{0.5cm}
        \begin{subfigure}[b]{0.4\textwidth}
        \includegraphics[width=\textwidth]{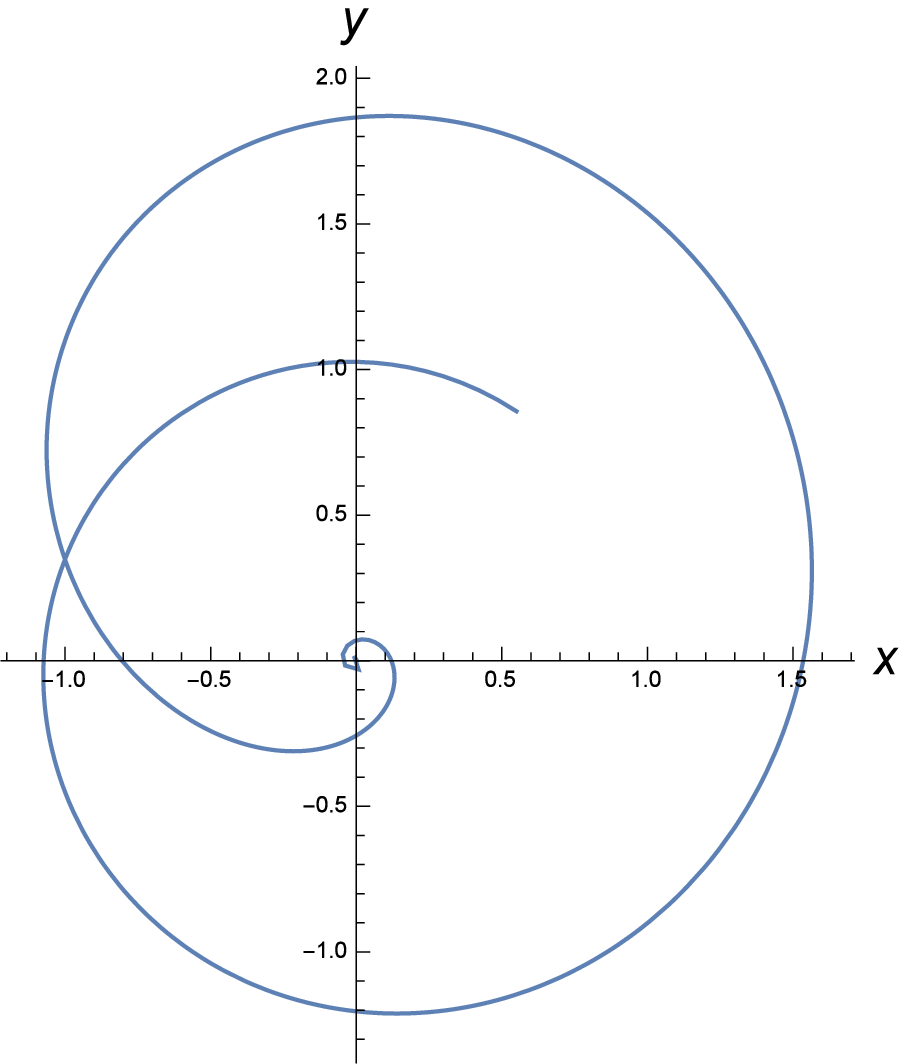}
        \caption{Projection onto $xy$-plane}
        \label{new_3D_energy_equal_to_1_2_4}
    \end{subfigure}
    \caption{The plots are obtained for the parameters $M=1$, $a=0.9$, $K=1$, $Q=0.4$, $\bar{q}=2.6$, $\bar{L}=0.5$, $\bar{E}=1$, $\ell=0.1$ and $m=1$. Here $\rho^2=x^2+y^2$. There exists one bound and one flyby orbit.}\label{3D_bound_5}
\end{figure}

Finally, for $\bar{E}=1$ the possible orbit types can be one bound and one flyby ($P_r(r)$ has three real zeros) or one flyby orbit ($P_r(r)$ has only one real zero). In Figure \ref{3D_flyby_3}, we give an example of flyby orbit, while in Figure \ref{3D_bound_5} we exhibit 3D-bound orbit for $\bar{E}=1$ (for the fixed value of the NUT parameter but different values of charge $\bar{q}$ of the test particle). We should also remark that in both plots the equatorial plane is crossed.

We should further point out that, in all the plots except Figure \ref{3D_bound_3}, the angular motion of the particle is confined to the interval $\theta_2 \leq \theta \leq \theta_1$, where $0 <\theta_2<\frac{\pi}{2}$ while $\frac{\pi}{2} < \theta_1 <\pi $. However, in the Figure \ref{3D_bound_3}, looking at the graph of $P_\theta(\theta)$ and also the projection on $r-\theta$ plane, one can deduce that the particle is similarly restricted to move in the angular region $\theta_2 \leq \theta \leq \theta_1$ where in that case both $\theta_1$ and $\theta_2$ lie in the interval $ \frac{\pi}{2} < \theta_i < \pi$, ($i=1,2$).\\

\noindent{\bf Observables of the bound orbit:}

\noindent Here, to obtain the observables for a charged particle in Kerr-Newman-Taub-NUT spacetime, we assume that the particle makes a bound motion in $r$ and $\theta$ coordinates. Considering that motion in $r$-coordinate is bounded in the interval $ r_2 \leq r \leq r_1$, one can calculate the fundamental period $\Lambda_r$ for the radial motion as
\begin{equation}
\Lambda_r=2 \int_{r_2}^{r_1} \frac{dr}{\sqrt{P_r(r)}}=2\int_{v_0}^{\infty} \frac{dv}{\sqrt{P_3(v)}} \label{observable_1}
\end{equation}
where $P_3 (v)= 4 v^3 - h_2 v - h_3$ with $h_2$ and $h_3$ introduced in (\ref{rad_sol_12}). The integral can be accomplished via the transformation
\begin{equation}
\xi_r=\frac{1}{\kappa_r} \left( \frac{e_2-e_3}{v-e_3}\right)^{1/2} \label{observable_2}
\end{equation}
where $e_1$, $e_2$ and $e_3$ correspond to the roots of the polynomial $P_3 (v)= 0$ with $\kappa_r^2=\frac{e_2-e_3}{e_1-e_3}$. We also choose $v_0=e_1$. Then one obtains the radial period as
\begin{equation}
\Lambda_r=\frac{2}{\sqrt{e_1-e_3}} K(\kappa_r) \label{observable_3}
\end{equation}
where $K(\kappa_r)$ denotes the complete elliptic function with modulus $\kappa_r$.

\noindent Similarly, if one considers that the angular motion is also bounded in the angular interval $\theta_2 \leq  \theta \leq \theta_1$, one can evaluate the fundamental period $\Lambda_{\theta}$ for the angular motion in the form
\begin{equation}
\Lambda_{\theta}=2 \int_{\theta_2}^{\theta_1} \frac{d\theta}{\sqrt{P_{\theta}(\theta)}}= \frac{2}{\sqrt{\bar{e}_1 -\bar{e}_3}} K(\kappa_{\theta}), \label{observable_4}
\end{equation}
where in this case $\bar{e}_1$, $\bar{e}_2$,  $\bar{e}_3$ correspond to roots of the polynomial $P_3 (y)= 4 y^3 - g_2 y - g_3$. Here, $g_2$ and $ g_3$ are expressed in (\ref{ang_sol_15}) and (\ref{ang_sol_16}) respectively and $\kappa_{\theta}^2 =\frac{\bar{e}_2-  \bar{e}_3}{\bar{e}_1 -  \bar{e}_3}$. Similarly $K(\kappa_{\theta})$ describes the complete elliptic function with modulus $\kappa_{\theta}$. Then one can also evaluate the corresponding angular frequencies
\begin{equation}
\Upsilon_r=\frac{2 \pi}{\Lambda_r}=\frac{\pi \sqrt{e_1 -e_3}}{K(\kappa_r)} \label{observable_5}
\end{equation}
and
\begin{equation}
\Upsilon_{\theta}=\frac{2 \pi}{\Lambda_{\theta}}=\frac{\pi \sqrt{\bar{e}_1 -\bar{e}_3}}{K(\kappa_{\theta})} \label{observable_6}
\end{equation}
for the radial and $\theta$-motion respectively.

\noindent Furthermore, one can obtain the angular frequencies $\Upsilon_{\varphi} $ and $\Upsilon_t $ for the $\varphi$-motion and $t$-motion respectively  from the solutions of $\varphi(\lambda)$ and $t(\lambda)$. By using the arguments outlined in \cite{drasco}, one notices that the solutions $\varphi(\lambda)$ and $t(\lambda)$ can be both expressed in the following forms
\begin{equation}
\varphi(\lambda)= \Upsilon_{\varphi} \left( \lambda - \lambda_0\right) + \varphi^{(r)} (\lambda)+\varphi^{(\theta)} (\lambda) \label{observable_7}
\end{equation}
and
\begin{equation}
t(\lambda)= \Upsilon_{t} \left( \lambda -\lambda_0\right) + t^{(r)} (\lambda)+t^{(\theta)} (\lambda), \label{observable_8}
\end{equation}
where $\Upsilon_{\varphi} $ and $\Upsilon_t $ correspond to frequencies in Mino time for $\varphi$-motion and $t$-motion respectively. From the solutions, one can obtain
\begin{eqnarray}
\Upsilon_{\varphi}&=& \frac{(\bar{L} +2 \ell \bar{E} x_{\theta})}{(1-x_{\theta}^2)} -a \bar{E}+ \frac{a(\bar{E}(r_1^2 + a^2 + \ell^2)-\bar{q} Q r_1 -a\bar{L})}{\Delta(r_1)} \label{observable_9}\\
&+& \sum_{i=1}^2 \sum_{j=1}^2  \left( \zeta(a_{ij}) \frac{( \bar{L} \bar{G}_i+ 2 \ell \bar{E}G_i) }{\wp^{\prime} (a_{ij})}
+\zeta(v_{ij}) \frac{a\left[\bar{E}\tilde{\omega}_i -\bar{q} Q \hat{\omega}_i +(\bar{E}(a^2 + \ell^2) -a\bar{L}) \check{\omega}_i\right]}{\wp^{\prime} (v_{ij})} \right) \nonumber
\end{eqnarray}
and
\begin{equation}
\Upsilon_{t}=\Upsilon_{t}^{(r)}+\Upsilon_{t}^{(\theta)}, \label{observable_10}
\end{equation}
where
\begin{eqnarray}
\Upsilon_{t}^{(r)}&=&\left[\bar{E}(r_1^2+\ell^2+a^2)^2-\bar{q} Q  (r_1^2+\ell^2+a^2)r_1-a \bar{L} (r_1^2+1)\right]\frac{1}{\Delta(r_1)}\nonumber \\
& &+ \sum_{i=1}^{3}\sum_{j=1}^{2} \frac{(\bar{E} \omega_i -\bar{q} Q \bar{\omega}_i)}{\wp^{\prime} (v_{ij})}  \zeta(v_{ij})
- \frac{\bar{E}\beta_3^2}{16} \sum_{j=1}^{2} \frac{1}{{\wp^{\prime}}^2 (v_{3j})}  \left(\wp(v_{3j})+\frac{\wp^{\prime \prime}(v_{3j})}{\wp^{\prime}(v_{3j})} \right) \nonumber \\
& &+ \sum_{i=1}^{2}\sum_{j=1}^{2} \left[ \left(2 \bar{E} (\ell^2 +a^2 )-a\bar{L} \right)\tilde{\omega}_i \right. \label{observable_11} \\
& & \left. - \bar{q} Q (\ell^2+a^2)\hat{\omega}_i + (\ell^2+a^2)\left(  \bar{E} (\ell^2 +a^2 )-a\bar{L}\right)\check{\omega}_i \right] \frac{\zeta(v_{ij})}{{\wp^{\prime}} (v_{ij})} \nonumber
\end{eqnarray}
and
\begin{eqnarray}
\Upsilon_{t}^{(\theta)}&=& a \bar{L} + a^2 \bar{E}^2 (x_{\theta}^2 -1)+ 4 a \ell \bar{E} x_{\theta}- \frac{2 x_{\theta} \ell}{1-x_{\theta}^2} \left(2 \ell \bar{E} x_{\theta }+ \bar{L} \right) \nonumber \\
& &-2 \ell  \sum_{i=1}^2 \sum_{j=1}^2 \frac{( \bar{L} G_i+ 2 \ell \bar{E}\bar{G}_i) }{\wp^{\prime} (a_{ij})}  \zeta(a_{ij}) \label{observable_12} \\
&& +a \bar{E}  \left(\frac{ax_{\theta}}{2}+ \ell \alpha_3\right) \sum_{i=1}^2 \frac{1}{\wp^{\prime} (b_{1i})}  \zeta(b_{1i}) \nonumber \\
&& - \frac{a^2 \bar{E} \alpha_3^2}{16} \sum_{i=1}^2 \frac{1}{{\wp^{\prime}}^2 (b_{1i})} \left( \wp(b_{1i})+ \frac{\wp^{\prime \prime}(b_{1i})}{\wp^{\prime}(b_{1i})} \right). \nonumber
\end{eqnarray}

\noindent Finally, as illustrated in \cite{drasco}, the angular frequencies obtained using Mino time $\lambda$ can be related to the angular frequencies $ \Omega_r$, $\Omega_{\theta}$ and $\Omega_{\varphi}$ calculated with respect to a distant observer time as
\begin{equation}
\Omega_r=\frac{\Upsilon_{r}}{\Upsilon_{t}}, \qquad \Omega_{\theta}=\frac{\Upsilon_{\theta}}{\Upsilon_{t}}, \qquad \Omega_{\varphi}=\frac{\Upsilon_{\varphi}}{\Upsilon_{t}}. \label{observable_13}
\end{equation}
 Considering that these frequencies are not equal, it enables us to evaluate the precession of the orbital ellipse and Lense-Thirring effect for angular motions $\varphi$ and $\theta$. These can be given by
 \begin{equation}
\Omega_{perihelion}=\Omega_{\varphi}-\Omega_{r}, \qquad  \Omega_{LT}=\Omega_{\varphi}-\Omega_{\theta}. \label{observable_14}
\end{equation}

\noindent Although we couldn't provide a numerical value for perihelion precision and Lense-Thirring effect, we can deduce that the NUT parameter and the charge of the test particle definitely influence these observables.

\section{Conclusion}
In this work, we have examined the motion of a charged test particle in Kerr-Newman-Taub-NUT spacetime. We have analyzed the angular and radial parts of the orbital motion by discussing possible orbit types that may be observed. In the analysis of the angular part, we can see that the NUT parameter has a significant effect for the existence of equatorial orbits such that unlike in the background of Kerr and Kerr-Newman spacetimes where equatorial orbits does exist for any spacetime parameter, one cannot obtain equatorial orbits in Kerr-Taub-NUT and Kerr-Newman-Taub-NUT spacetimes for arbitrary spacetime parameters, the energy and the angular momentum of the test particle. Meanwhile, equatorial orbits can exist either for vanishing NUT parameter ($\ell=0$) or if the specific relation (\ref{angular_9}) is imposed between the energy and the orbital angular momentum of the test particle for the case of non-vanishing NUT parameter. Next by making an analysis of radial part of the motion, we have classified the orbit types according to the number of real roots of the radial polynomial as well as the value of the energy of the test particle (whether it is smaller or greater than unity or equal to unity). One can conclude that for $\bar{E}<1$ either one bound or two bound regions can be seen. It means that the particle cannot escape to infinity for the energy value $\bar{E}<1$. For $\bar{E}>1$ and $\bar{E}=1$, one can observe bound and flyby orbits with respect to the change in the values of the spacetime parameters and the orbital angular momentum of the test particle. In the radial part of the motion, we have also discussed the effective potential. In the plots of the effective potential, we have seen that the existence of the NUT parameter and the charge of the test particle affects the existence of the bound orbits such that as the value of the NUT parameter decreases and the value of the charge of the test particle increases, potential well occurs which further leads to formation of bound orbits. Furthermore, we have discussed spherical orbits by calculating the energy and angular momentum of the test particle in such an orbit by also investigating the stability. We have observed that, as the NUT parameter varies, the stable spherical orbits change their class from retrograde orbits (where $a>0, \bar{L}<0$) to direct ones (where $a>0, \bar{L}>0$).

Next, we have obtained the analytical solutions of the orbit equations in terms of Weierstrass $\wp$, $\sigma$, and $\zeta$ functions. The results that we have obtained, reduce to the solutions given in \cite{hackmann4} for the case $\ell=0$ and $Q=0$ (of course by taking cosmic string parameter to be unity in their paper). Furthermore, our solutions go over into the results presented in \cite{grunau4} and \cite{hackmann6} for the cases $\ell=0$, $a=0$ and $\ell=0$ respectively (in the vanishing of magnetic charge in their work).

Using the analytical solutions of the equations of motion, we have obtained the plots of some orbit types in the region where $r>r_+$, for fixed spacetime parameters as well as the energy and the angular momentum of the test particle. We remark that as the values of the NUT parameter and the charge of the test particle vary, one can observe different orbit types as can be seen from the plots of the orbits. In all the plots, we have seen that a bound orbit may exist for all possible values of the energy of the test particle (i.e for $\bar{E}<1$, $\bar{E}>1$ and $\bar{E}=1$) for some particular values of the spacetime parameters. We have also observed that, the test particle may or may not cross the equatorial plane depending on the spacetime parameters, the energy and the angular momentum of the particle.

In addition, we have theoretically calculated the perihelion precision and Lense-Thirring effect for a bound orbit which has the paramount importance for the detection of the NUT parameter in astrophysical observations. It can be seen from the relations of the observables that the NUT parameter and the charge of the test particle do strictly affect the perihelion precision and Lense-Thirring phenomena.

For future work, it would also be remarkable to examine the equatorial orbits of the charged test particles in Kerr-Newman-Taub-NUT background with the relation (\ref{angular_9}) imposed. Furthermore, one can similarly investigate the effect of the cosmological constant on the orbital motion of a test particle in a spacetime where both the NUT and rotation parameters exist \cite{aliev1}. In addition, a numerical investigation of the observables for the bound orbits can be made to expose the existence of the NUT parameter in astrophysical phenomena. These are devoted to future research.

\section*{Acknowledgement}
We would like to thank Mehmet Ergen for fruitful discussions while obtaining the plots of the orbits.

\end{document}